\documentclass[useAMS,usenatbib]{mn2e}
\usepackage{amsmath}
\usepackage{amsfonts}
\usepackage{epsfig}
\usepackage{graphicx}
\numberwithin{equation}{section}

\newcommand{\p}{\partial}
\newcommand{\case}{\textstyle\frac}
\newcommand{\bv}{{\mbox{\boldmath $v$}}}
\newcommand{\br}{{\mbox{\boldmath $r$}}}
\newcommand{\bw}{{\mbox{\boldmath $w$}}}
\newcommand{\bI}{{\mbox{\boldmath $I$}}}
\newcommand{\eps}{\epsilon}
\newcommand{\ccase}{\textstyle\frac}
\newcommand {\veps}{{\varepsilon}}
\newcommand{\bl}{{\mbox{\boldmath $l$}}}

\title{Gravitational Loss-Cone Instability in Stellar Systems\\
\ with Retrograde Orbit Precession}
\author[E. V. Polyachenko et al.]
       {E.~V.~Polyachenko,$^1$\thanks{E-mail: epolyach@inasan.ru}, V. L. Polyachenko,$^1$
        I. G. Shukhman,$^2$\thanks{E-mail: shukhman@iszf.irk.ru}\\
       $^1$Institute of Astronomy, Russian Academy of Sciences, 48 Pyatnitskya St., Moscow 119017, Russia\\
       $^2$Institute of Solar-Terrestrial Physics, Russian Academy of Sciences, Siberian Branch, P.O. Box 4026, Irkutsk 664033, Russia}

\date{Accepted \qquad
      Received }

\pagerange{\pageref{firstpage}--\pageref{lastpage}}
\pubyear{2006}

\begin{document}
\maketitle

\label{firstpage}

\begin{abstract}

We study spherical and disk clusters in a near-Keplerian potential
of galactic centers or massive black holes. In such a potential
orbit precession is commonly retrograde, i.e. direction of the
orbit precession is opposite to the orbital motion. It is assumed
that stellar systems consist of nearly radial orbits. We show that
if there is a loss cone at low angular momentum (e.g., due to
consumption of stars by a black hole), an instability similar to
loss-cone instability in plasma may occur. The gravitational
loss-cone instability is expected to enhance black hole feeding
rates. For spherical systems, the instability is possible for the
number of spherical harmonics $l \ge 3$. If there is some amount
of counter-rotating stars in flattened systems, they generally
exhibit the instability independently of azimuthal number $m$. The
results are compared with those obtained recently by Tremaine for
distribution functions monotonically increasing with angular
momentum.

The analysis is based on simple characteristic equations
describing small perturbations in a disk or a sphere of stellar
orbits highly elongated in radius. These characteristic equations
are derived from the linearized Vlasov equations (combining the
collisionless Boltzmann kinetic equation and the Poisson
equation), using the action\,--\,angle variables. We use two
techniques for analyzing the characteristic equations: the first
one is based on preliminary finding of neutral modes, and the
second one employs a counterpart of the plasma
Penrose\,--\,Nyquist criterion for disk and spherical
gravitational systems.

\end{abstract}

\section{Introduction}
Mechanisms of ``fuel'' supply for galactic nuclear activity
usually assume exposure of stars and gas clouds to a
non-axisymmetric gravitational potential. The bar mode instability
and tidal action from nearby galaxies are most commonly considered
to be responsible for formation of the potential \citep{Sul90}. We
believe, however, that the nuclear activity results from processes
in the immediate vicinity of central objects. It is unlikely that
large-scale instabilities, such as the global bar mode, can
provide for precise targeting of the star or gas flow towards the
center.

As an example of local mechanism that can maintain the activity we
consider an instability in the stellar environment of the galactic
center. This instability has a well-known prototype in plasma
physics: the loss-cone instability in the simplest plasma traps
similar to  mirror machines (see the pioneer paper by \cite{RP},
and, e.g., \cite{Mih}), which is due to peculiar anisotropy in the
velocity distribution of plasma particles. The anisotropy is
caused by departure of particles with sufficiently small velocity
component transverse to the symmetry axis of the system. The
presence of this ``loss cone'' produces deformation of the plasma
distribution function (DF) in transverse velocities, giving it
unstable  (beam-like) character.

Similar deformation in the DF in angular momentum takes place in
clusters in case of deficiency of stars with low angular momentum
due to their absorption by the galactic nucleus, black hole, or to
other reasons. Then the deformation can have a ``beam-like''
character: the DF becomes an increasing function of angular
momentum, $\p f_0/\p L>0$. In this case, provided an additional
condition discussed later is met, the deformation can trigger the
instability which we shall call the \textsl{gravitational
loss-cone instability}.

We have mentioned the principal possibility of the instability in,
e.g., \cite{PS80}, and \cite{FP84}. However, the search of a
specific example of the gravitational loss-cone instability has
been unsuccessful until one of the authors \citep{P91b} found the
desired instability in a disk model. The delay and initial
difficulty in finding it might be due to the unusual character of
this instability, which is directly related to relatively very
slow precession motion of stellar orbits. Thus typical frequencies
and growth rates here are anomalously small, if measured in, for
example, orbital frequencies of stars. Yet it does not mean
slowness in absolute units taking into account swift growth of all
typical frequencies when going from the galactic periphery to its
center.

\cite{T05} have studied the \textsl{secular} instability, which is
identical to the gravitational loss-cone instability. He examined
the disk and sphere models of a low-mass stellar system
surrounding a massive central object. In such ``near-Keplerian''
systems, the gravitational force is dominated by a central point
mass. For the disk models with arbitrary orbits, Tremaine has
found unstable solutions. Note that for disks of nearly radial
orbits the instability was proved by \cite{P91b}, and we will give
here another proof based on the general integral equation for
eigen modes \citep{EP05}. Stability of spherical systems for
arbitrary orbits has also been probed into by \cite{T05}, who
found no evidence of instability for $l\le 2$ modes. The study for
$l\ge 3$ modes in general case is difficult, but it becomes
feasible if one restricts consideration to nearly radial orbits.
In this paper we show that the loss-cone instability occurs just
from $l=3$ (see Sec.\,4 and Sec.\,5).

For massive black holes in galactic centers, the collisional
diffusion and subsequent partial absorption of nearest stars is
most often considered as a mechanism providing for nuclear
activity (e.g., \cite{LS77,SM78}). The very existence of the
collisionless (collective) mechanism may initiate revision of the
dominating viewpoint regarding the nature of the activity. In this
paper, however, we present a mere demonstration of the existence
of gravitational loss-cone instability in simplest models. An
exception is some preliminary estimations of efficiency of the
proposed collective mechanism in the Sec.\,5.

\begin{figure*}
 \begin{center}
 \includegraphics[width=120mm, draft=false]{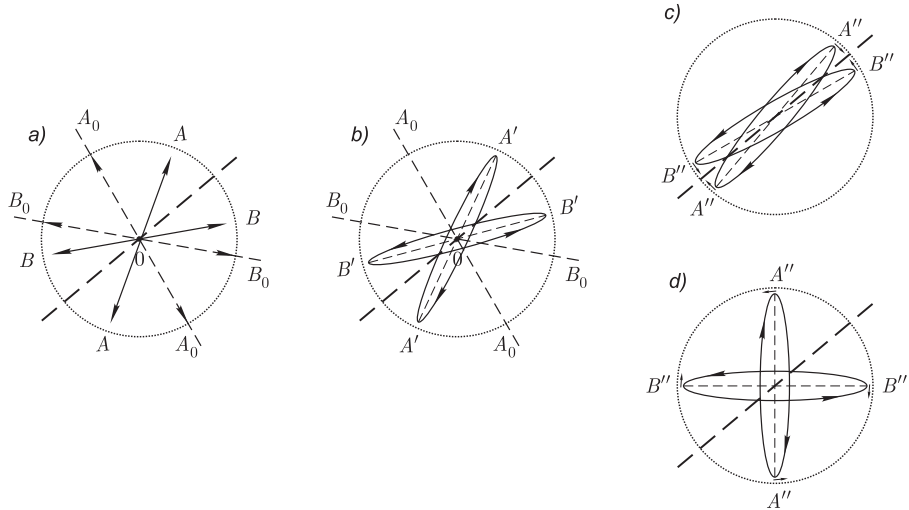}
 \end{center}
\caption{\small Diagram explaining the physical mechanism of radial orbit
instability: \textit{a} -- two typical radial orbits $A_0$ and $B_0$ in
equilibrium state, $A$ and $B$ are their positions in perturbed state from both
sides of the perturbed potential minimum (bold dashes) at initial instant of
evolution; \textit{b} -- stars at the orbits $A$ (now $A'$) and $B$ (now $B'$)
have gained small angular momentum ($L<0$ at $A'$ and $L>0$ at $B'$); \textit{c}
-- a case of prograde orbit precession (the direction of precession is indicated
by arrows): further merging of large axes of orbits takes place, i. e. radial
orbit instability develops; \textit{d} -- retrograde precession: return of large
axes to the initial equilibrium position, i. e. neutral oscillations. These
oscillations can be unstable in the presence of loss cone.}
 \label{fig1}
 \end{figure*}

Existence of the above-mentioned additional condition for the
instability originates from fundamental distinctions between
gravitating and plasma systems. In gravitating systems, particles
have only one kind of ``charge'', and they attract each other.
This fact ultimately leads to the Jeans instability substituting
Langmuir oscillations in plasma \citep[e.g.][]{FP84}. In systems
with nearly radial orbits we are going to study, there is a
specific form of the Jeans instability called the radial orbit
instability \citep{PS81, FP84}. It develops only in systems with
prograde orbit precession (see Fig.\,\ref{fig1}). Conversely, the
gravitational loss-cone instability can occur only when orbit
precession is \textsl{retrograde}. This retrograde precession is
the additional condition of the instability.

As is well-known, the radial orbit instability is suppressed if the dispersion of
orbit precession velocity exceeds some critical value \citep{P92}. We shall
obtain a similar result for the gravitational loss-cone instability (see
Sec.\,4).

Below we study the gravitational loss-cone instability in two
models representing active stellar subsystems with nearly radial
orbits. As noted above, the instability in disks was studied
earlier in \cite{P91b, T05}. Nevertheless, we believe that it
merits more detailed consideration here, although the main goal of
this paper is to study instability in spherical systems. In
\cite{P91b}, instability has been determined using the
Penrose\,--\,Nyquist general criterion (see, e.g.,
 \cite{Pen} or \cite{Mih}).

In this paper we apply a more illustrative method based on finding neutral modes
and subsequent application of the perturbation theory. This allows us to obtain
the frequencies and growth rates of perturbations corresponding to small
deviations from the neutral modes into the unstable domain. For unstable modes
remote from the neutral ones, their complex frequencies are found numerically. At
first we shall carry out this instability analysis for a simpler disk model
(Sec.\,3), then we shall use it for more complicated spherical geometry
(Sec.\,4).

Another reason for revisiting disks is that a principal characteristic
equation in \cite{P91b} was taken ready-made, without derivation. Here we
present a detailed derivation of this equation. In addition, we
justify the use of a suitable \textsl{rotating spoke} approximation: a spoke
consists of stars with fixed energy $E=E_0$ and low values of
angular momentum $L$. The approximation is then applied to the spherical
model.

The paper is organized as follows. The characteristic equations
derived in Sec. 3.1 and 4.1\,--\,4.4 are then applied to studying
the gravitational loss-cone instability of disks (Sections 3.2 and
3.3) and spheres (Sections 4.5\,--\,4.8). In particular, in
Sec.\,4.8 we obtain general instability criterion for spherical
systems analogous to the well-known Penrose\,--\,Nyquist criterion
for plasma, and establish its correspondence to our neutral-mode
approach. The instability of various DF functions is discussed in
terms of this criterion. The study is prefaced with an overview of
the orbit precession in the axial and centrally-symmetric
gravitational fields (Sec.\,2). Appendix A is devoted to
derivation of a basic integral equation for spherical systems in
terms of the action-angle formalism, and in Appendix B we prove
the instability criterion theorem.

\section{Some remarks on the orbit precession}

In low-frequency perturbations of stellar clusters we are
interested in, with typical frequencies, $\omega$, of order of the
mean precession velocity of near-radial orbits, these latter
participate as a whole (in contrast to high-frequency
perturbations, for which $\omega$ is of order of orbital
frequencies; they depend on individual stars). A detailed
justification of these statements (which are fairly obvious) can
be found in our papers \cite{P92}, \cite{EP04}, \cite{PP04} and in
\cite{LB79}. For perturbations of interest, precessing orbits
replace individual stars. So a preliminary overview of some useful
data on precession becomes very desirable.

\begin{figure}
 \begin{center}
 \includegraphics[width=50mm, draft=false]{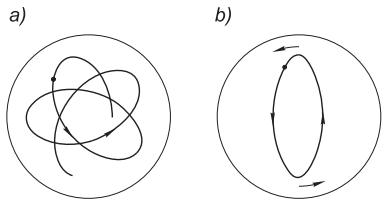}
 \end{center}
\caption{\small Typical stellar orbit in the plane of axially-symmetric disk
or in the spherical cluster
\textit{a} -- ``rosette trajectory'' in the inertial system of
reference \textit{b} -- closed precessing orbit in a system of
reference rotating with angular velocity $\Omega = \Omega_{\rm
pr}$.}
 \label{fig2}
 \end{figure}

In spherical potentials, star orbits are rosettes that generally
are not closed \citep{LL76} (see Fig.\,\ref{fig2}a). It is
possible, however, to find a rotating (with angular velocity,
$\Omega$) reference frame in which the orbit is a closed oval.
Therefore, star movement along the rosette are quick oscillations
in a closed oval, which in turn slowly rotates (or
\textit{precesses}) with the rate $\Omega_{\rm pr}=\Omega$
(Fig.\,\ref{fig2}b). The latter is the orbit precession rate.

There are only two potentials \citep{Arn89} in which any orbits
are closed ellipses: (i) for quadratic potential, $\Phi_0(r)
\propto r^2$, the ellipses are symmetric with respect to the
center, and (ii) for the point mass potential,
$\Phi_0(r)=-GM_c/r$, in which the orbits (Keplerian ellipses) are
asymmetric. The ellipses in these two potentials are resonance
orbits, since the ratio of the radial oscillation frequency,
$\Omega_1$, to that of the azimuthal oscillations, $\Omega_2$, is
2:1 for the quadratic potential, and 1:1 for the Keplerian one. As
the orbits are closed, the precession rate is zero.

Small deviation from these particular potentials results in a slow precession
with the frequency $\Omega_{\rm pr}$ much smaller than the typical frequencies of
orbital motion, $\Omega_1$ and $\Omega_2$. It occurs, for example, in the centers
of galaxies. In the absence of a central point mass, the potential is almost
quadratic, so the ratio of radial and azimuthal frequencies is close to 2:1. If a
point mass is present, this ratio is close to 1:1 in the central region. The
deviation from the exact resonance (and thus slow precession) is caused here by
gravity of stars around the central point mass.

In smooth potentials, nearly radial orbits, with angular momenta
small compared, for example, to the angular momenta of circular
orbits with the same energies $E$), are almost resonant for any
amplitude of radial oscillations. This statement can be easily
proved. Indeed, the azimuthal frequency, $\Omega_2$, to radial
frequency, $\Omega_1$, ratio for a star in the gravitational
potential $\Phi(r)$ is, by definition:
$$
    \frac{\Omega_2(E,L)}{\Omega_1(E,L)}=\frac{\Delta\varphi}{\pi},
$$
where
\begin{align}
\label{eq:2.1}
    \Delta\varphi=\frac{1}{2}\oint\frac{L}{r^2}\frac{dr}{v_r(r;E,L)}
    =\int_{r_{\rm min}}^{r_{\rm
    max}}\frac{L}{r^2}\frac{dr}{\sqrt{2E-2\Phi(r)-{L^2}/{r^2}\phantom{\big|}}}
\end{align}
is the rotation angle of the stellar trajectory as its radius changes from
$r_{\rm max}$ to $r_{\rm min}$, $E$ is the energy, and $L$ is the modulus of
angular momentum.

Now let us calculate the asymptotic behavior of (\ref{eq:2.1}) for $L\to 0$. To
do this, we shall analyze the expression for radial velocity, $ v_r^2=
2E-2\Phi(r)-{L^2}/{r^2}$; its zeros define the turning points.

\medskip

In the case of a non-singular potential, $\Phi(0)$ is finite, and
one may assume $\Phi(0)=0$. Obviously, the left turning point is $
r_{\rm min}\approx{L}/{\sqrt{2E}}$. Since at low  $L$ the value of
$r_{\rm min}$ is small, the main contribution to  the integral
comes from a region near the lower limit.  So we have
 $$
    \int_{r_{\rm min}}^{r_{\rm
    max}}\frac{L}{r^2}\frac{dr}{\sqrt{2E-2\Phi(r)-{L^2}/{r^2}\phantom{\big|}}}
    \approx \int_{r_{\rm
    min}}^{\infty}\frac{L}{r}\frac{dr}{\sqrt{2Er^2-L^2\phantom{\big|}}}.
 $$
Substituting ${(\sqrt{2E}}/{L})\,r=x,$ we obtain
$$
    \int_{r_{\rm min}}^{\infty}\frac{L}{r}\frac{dr}{\sqrt{2Er^2-L^2\phantom{\big|}}}=
    \int_1^{\infty} \frac{dx}{x\sqrt{x^2-1\phantom{\big|}}}\
$$
and finally
 $
    \Delta\varphi=\pi/2.
 $
Thus, $2\Delta\varphi=\pi$, and the orbit is a straight line which
passes (almost) through the center. The absolute value of the
ratio $\Omega_2/\Omega_1$ is 1:2.

Since for the Keplerian potential this ratio is 1:1, the question arises: Is
there a continuous transition from the non-singular case, where the ratio is 1:2,
to the Keplerian case? If such a transition exists, i.e. if the angle
$\Delta\varphi$ can vary smoothly from $\pi/2$ to $\pi$, a full stellar
trajectory cannot be a straight line: a radial direction from the apogee to the
perigee should change as a star goes from the perigee to the
apogee.\footnote{Note that in the Keplerian limit radial orbits degenerate into a
``ray''. More precisely, the incoming and outgoing rays merge together.} The
angle of rotation is $\pi<2\,\Delta\varphi<2\pi$, and the trajectory looks like
spokes of a bicycle wheel. The number of spokes is finite if  $\Delta\varphi/\pi$
is rational, otherwise the spokes will fill up a circle.
\medskip

\smallskip Now let us consider, for the neighborhood of the center,
\begin{align}\label{eq:2.2}
    \Phi(r)=-\frac{\alpha}{r^{s}},\ \ 0<s<2.
\end{align}
The family of potentials (\ref{eq:2.2}) includes both the non-singular potential,
corresponding to $s=0$, and the Kelperian potential corresponding to $s=1$. Near
the center, the absolute value of the radial velocity is
 $$
    \sqrt{2E-2\Phi(r)-{L^2}/{r^2}\phantom{\big|}}\approx
    \sqrt{{2\alpha}{r^{-s}}-{L^2}/{r^2}\phantom{\big|}},
 $$
and thus $ r_{\rm min}= \left({L}/{2\alpha} \right)^{1/(2-s)}$. As before, the
main contribution to the integral comes from the lower limit, so
 $$
    \int_{r_{\rm min}}^{r_{\rm
    max}}\frac{L}{r^2}\frac{dr}{\sqrt{2E-2\Phi(r)-{L^2}/{r^2}\phantom{\big|}}}
    \approx \int_{r_{\rm
    min}}^{\infty}\frac{L}{r}\frac{dr}{\sqrt{2\alpha\,
    r^{2-s}-L^2\phantom{\big|}}}.
 $$
Substitution $\bigl({2\alpha}/{L^2}\bigr)\,r^{2-s}=x^2 $ gives
$\Delta\varphi={\pi}/{(2-s)}$,  which leads to the frequency ratio
\begin{align}\label{eq:2.3}
    {\Omega_2}/{\Omega_1}={1}/{(2-s)}.
\end{align}
The relation (\ref{eq:2.3}) connects the non-singular and Keplerian potentials.
We would like to stress again that in this case the stellar trajectory has a
sharp turn almost in the center.

\begin{figure}
 \begin{center}
 \includegraphics[bb=141 290 460 610, width=55mm, clip=true, draft=false]{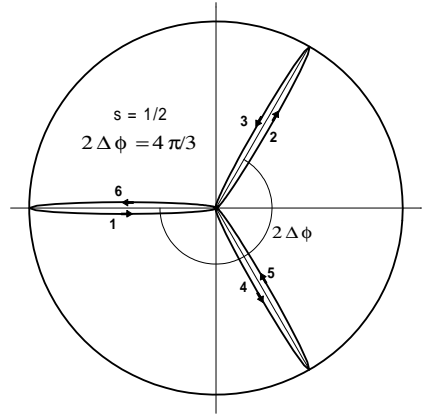}
 \end{center}
\caption{\small Trajectory of a star with a small angular positive
momentum $L$. Parameter $s=0.5$
 }
 \label{fig3}
\end{figure}

Fig.\,\ref{fig3} shows a schematic trajectory in the potential
$\Phi(r)\sim -r^{-1/2}$ ($s=1/2)$. One can see the sharp turn of
the orbit with the rotation angle
$2\Delta\varphi=2\pi/(2-s)=4\pi/3= 240^{\circ}$. Numbers
$1,\ldots,6$ trace a star in the trajectory. The star moves
counter-clockwise, in accordance with the positive sign of the
angular momentum, $L>0$. The trajectory is closed since $1/(2-s)$
is a rational number (2/3).

Further we focus our attention on spherical systems with
near-radial orbits in two special gravitational potentials: 1) for
a singular near-Keplerian potential, and 2) for an arbitrary
non-singular potential. We shall refer to 1:1-orbits in the former case,
2:1-orbits in the latter case. Recall that the most
obvious difference between these orbits is revealed in the
degenerate case of radial motion: a 1:1-orbit turns into a ray travelling
from the center, while a 2:1-orbit turns into a line segment,
symmetric to the center.

It turns out (see Introduction) that the gravitational loss-cone instability
occurs if the orbit precession is retrograde. Thus it is useful to have
expressions for the precession rate, $\Omega_{\rm pr}$, for both types of the
orbits at hand.

\begin{description}
    \item[\textbf{1:1-orbits.}] In the case of
    a low-mass spherical cluster around the central mass
    $M_c$, one can write
 $$
    (\Omega_1)^{-1}=\frac{1}{\pi}
    {\displaystyle{\int\limits_{r_{\rm min}}^{r_{\rm max}}}
    \dfrac{dr}{\sqrt{2E+{2\,GM_c}/{r}-2\Phi_G(r)-{L^2}/{r^2}\phantom{\big|}}}}.
 $$
and
 $$
    \frac{\Omega_2}{\Omega_1}=\frac{L}{\pi}
    {\displaystyle{\int\limits_{r_{\rm min}}^{r_{\rm max}}}
    \dfrac{dr}{r^2\,\sqrt{2E+{2GM_c}/{r}-2\Phi_G(r)-{L^2}/{r^2}\phantom{\big|}}}}\equiv
    \frac{\Delta\varphi}{\pi},
 $$
where $\Delta\varphi$ is the rotation angle of a star (cf. (\ref{eq:2.1})) in the
trajectory between $r_{\rm min}$ and $r_{\rm max}$. If there is no precession,
i.e. the contribution of $\Phi_G(r)$ to the total potential is negligibly small,
the angle would be $\pi$. It is these small deviations of the angle that lead to
slow rotation of the elliptical orbit, or precession:
 \begin{align}
  \label{eq:2.4}
    \Omega_{\rm
    pr}=\Omega_2-\Omega_1=\Omega_1\,\left({\Delta\varphi}/{\pi}-1\right).
 \end{align}

The expression for the precession velocity has obvious sense. After rewriting it
in the form
 $$
    \Omega_{\rm pr}=\frac{\Delta\varphi-\pi}{T/2},\ \ \
    T=\frac{2\pi}{\Omega_1},
 $$
one can see, that if during one half oscillation, from $r_{\rm min}$ to $r_{\rm
max}$ (for the time $T/2$), a star travels an angle exceeding $\pi$, it would
imply that the apogee drifts with angular velocity (\ref{eq:2.4}).

Our goal now is to derive an expression for precession velocity  of order ${\cal
O}(\Phi_G)$. Since ${\Delta\varphi}/{\pi}-1={\cal O}(\Phi_G)$, it is sufficient
to retain the ${\cal O}(1)$ order only in calculating the $\Omega_1$ factor in
(\ref{eq:2.4}). Then we have
 \begin{align}\label{eq:2.5}
    &\Delta\varphi
    \approx L\,\frac{\p}{\p E}
    \Bigl[(2|E|)^{1/2}\int\limits_{r_{\rm min}}^{r_{\rm max}}\frac{dr}{r^3}\,
    \sqrt{(b-r)(r-a)} - \Big. \nonumber\\
    &-\Big.(2|E|)^{-1/2}
    \int\limits_{r_{\rm min}}^{r_{\rm max}}\frac{\Phi_G(r)\,dr}
    {r\,\sqrt{(b-r)(r-a){\phantom{\big|}}}}\Bigr].
 \end{align}
Here we introduced new variables $a(E,L)=r_{\rm min}(E,L)$ and
$b(E,L)=r_{\rm max}(E,L)$, instead of $E$ and $L$, where
 $
    a,b={GM_c}/{(2|E|)}\,\mp\,\bigl\{\bigl[{GM_c}/(2|E|)\bigr]^2-
    {L^2}/(2|E|)\bigr\}^{1/2}.
    \ \
 $
For later purposes, we need some useful relations valid in the
Keplerian potential:
 \begin{align}\label{eq:2.6}
    &a\!+\!b=\frac{GM}{|E|},\ a\,b=\frac{L^2}{2|E|},\ 
     L=\sqrt{\frac{2GM_c\,ab}{a+b}}.
\end{align}
One can show that the first item in r.h.s. of (\ref{eq:2.5}) is
zero, so one obtains for the precession velocity
 $$
    \Omega_{\rm pr}=-\frac{\Omega(E)}{\pi}\,\frac{\p}{\p E}
    \Bigl[{L}\,(2|E|)^{-1/2}\,
    \int\limits_a^b\frac{\Phi_G(r)\,dr}
    {r\,\sqrt{(b-r)(r-a){\phantom{\big|}}}}\Bigr],
 $$
where $\Omega(E)=\Omega_1(E)=\Omega_2(E)\approx {(2|E|)^{3/2}}/{GM_c}.$ Using
(\ref{eq:2.6}), it is easy to obtain an expression for precession velocity in
variables $(a,b)$:
 \begin{multline}\label{eq:2.7}
    \Omega_{\rm pr}=-\dfrac{\Omega}{\pi\,G M_c}\,\dfrac{b+a}{b-a}\,
    \Bigl(b^2\,\dfrac{\p}{\p b}-a^2\,\dfrac{\p}{\p a}\Bigr) \times\\
    \times\Bigl[\sqrt{a\,b}
    \int\limits_a^b\dfrac{\Phi_G(r)\,dr}
    {r\,\sqrt{(b-r)(r-a){\phantom{\big|}}}}\Bigr],
\end{multline}
where $ \Omega=(GM_c)^{1/2}\,\bigl[{2}/(a+b)\bigr]^{3/2}.$ The expression for
$\Omega_{\rm pr}$ can be simplified, assuming $a$ to be small (nevertheless, we
retain $a$ in the denominator of the argument of the square root):
 \[
    \Omega_{\rm pr}\approx-\dfrac{\Omega(b)}{\pi\,G M_c}\,
    b^2\,\dfrac{\p}{\p b}
    \Bigl[\sqrt{a\,b}
    \int\limits_a^b\dfrac{\Phi_G(r)\,dr}
    {r\,\sqrt{(b-r)(r-a){\phantom{\big|}}}}\Bigr],
     \]
where $a^{1/2}={L}/{\sqrt{2GM_c\phantom{\big|}}}$. It is possible
to show that integral $J(a,b) \equiv{\displaystyle \int
\limits_a^b} \dfrac{\Phi_G(r)\,dr}
{r\,\sqrt{(b-r)(r-a){\phantom{\big|}}}}$ may be written as
 $$
    J(a,b)=\int\limits_a^b\Phi_G(r)\,d\Bigl[\frac{1}{\sqrt{ab}}\,\arcsin
    \frac{(b+a)\,r-2ab}{r\,(b-a)}\Bigr].
 $$
Integrating by parts and then differentiating over $b$ yields:
 \begin{multline*}
    \frac{\p}{\p b}\,\left[\sqrt{ab}\,J(a,b)\right]=
    \int\limits_a^b\frac{d\Phi_G}{dr}\,\frac{1}{b-a}\,
    \sqrt{\frac{a}{b}}\,\sqrt{\frac{r-a}{b-r}}\,dr\approx\\
    \approx\frac{a^{1/2}}{b^{3/2}}
    \int\limits_0^b\frac{d\Phi_G}{dr}\,\sqrt{\frac{r}{b-r}}\,dr.
\end{multline*}
Finally, we obtain for near-radial orbits:
 \begin{align}\label{eq:2.8}
    &\Omega_{\rm pr}(E,L) =\varpi(E)\,L,\nonumber\\
    &\varpi(E)=-\frac{2}{\pi G M_c\,b}
    \int\limits_0^b\frac{d\Phi_G}{dr}\,\sqrt{\frac{r}{b-r}}\,dr
\end{align}
with $b={GM_c}/{|E|}$. From Eq. (\ref{eq:2.8}), it is clear that
for such orbits and potentials the precession is \textit{always}
retrograde since $\Phi_G'(r)=G{\cal M}_G(r)/r^2>0$, where ${\cal
M}_G(r)$ is the mass within a sphere of radius $r$. (Moreover, it
is easy to show also from (\ref{eq:2.7}) that for near-Keplerian
orbits precession is retrograde even for the case of arbitrary
eccentricity \citep[see also][]{T05}).

Thus, for spherical near-Keplerian systems, the retrograde precession is common.
As we will see later, for 2:1-orbits, the situation is vice versa: in real
potentials, the precession is prograde, whereas the retrograde precession occurs
in systems with exotic density distributions.

\medskip

 \item[\textbf{2:1-orbits.}] There are several different expressions
for the precession velocity of near-radial orbits of this type. The expression
given in \cite{P92} is somewhat inconvenient for practical use, because it
contains a procedure of passage to limit. We shall give a more convenient
formula.

For the precession velocity, one has:
 $$
    \Omega_{\rm
    pr}(E,L)
    =\Omega_1(E,L)\,\Bigl[\frac{\Omega_2(E,L)}{\Omega_1(E,L)}
    -{\ccase{1}{2}}\Bigr].
 $$
The goal is to calculate the derivative of the precession velocity at constant
$E$ and $L=0$, i.e. $\bigl[{\p\Omega_{\rm pr}(E,L)}/{\p L}\bigr]_{L=0}$. Taking
into account that $\Omega_1(E,L)=\Omega(E)+{\cal O}(L^2)$, one obtains
 $$
    \Omega_{\rm pr}(E,L)\approx
    \Omega(E)\left({\Delta\varphi}/{\pi}-{\ccase{1}{2}}\right),
 $$
where
 $$
    \Delta\varphi=
    \!\!\!\int\limits_{r_{\rm min}(E,\,L)}^{r_{\rm max}(E,\,L)}\frac{L}{r^2}
    \frac{dr}{\sqrt{2E-2\Phi_0(r)-{L^2}/{r^2}}},
 $$
is the rotation angle of a star in the trajectory between $r_{\rm min}$ and
$r_{\rm max}$, which is equal to  $\frac{1}{2}\,\pi$ for any non-singular
potential, in the leading order in $L$. The next (linear) term in expansion of
$\Delta\varphi$ gives a value of the precession velocity and defines the
direction of precession (prograde or retrograde), for a highly elongated orbit,
with a small angular momentum. It is clear that if $\Delta\varphi>\pi/2$ (or
$2\Delta\varphi>\pi$), the precession direction coincides with the rotation of a
star (prograde precession), and vice versa. Thus, the sign of the derivative $
{\p/\p L} \left(\Delta\varphi-\pi/{2}\right) $ determines the precession
direction. Let us denote $\phi(E,L)=\Delta\varphi-{\pi}/{2}$, so that
 \begin{align}\label{eq:2.19}
    \varpi(E)\equiv \bigl[{\p\Omega_{\rm pr}(E,L)}/{\p
    L}\bigr]_{L=0}=\frac{\Omega(E)}{\pi}\,\bigl[{\p\phi(E,L)}/{\p
    L}\bigr]_{L=0}.
 \end{align}
We have
 \begin{multline*}
    \Delta\varphi=
    \!\!\!\int\limits_{r_{\rm min}(E,\,L)}^{r_{\rm max}(E,\,L)}\frac{L}{r^2}
    \frac{dr}{\sqrt{2E-2\Phi_0(r)-{L^2}/{r^2}}}=\\
    \frac{\p}{\p E}
    \int\limits_{r_{\rm min}(E,\,L)}^{r_{\rm
    max}(E,\,L)}\frac{L}{r^2}\,\,
    \sqrt{2E-2\Phi_0(r)- {L^2}/{r^2}}\,dr.
 \end{multline*}
Integrating by parts yields
 \begin{multline*}
    \Delta\varphi=
    \frac{\p}{\p E}
    \int\limits_{r_{\rm min}}^{r_{\rm
    max}}\frac{L^3}{r^4}\,\,
    \frac{dr}{\sqrt{2E-2\Phi_0(r)- {L^2}/{r^2}}}-\\
    -L\,\frac{\p}{\p E}
    \int\limits_{r_{\rm min}}^{r_{\rm
    max}}\frac{\Phi_0'(r)}{r}\,\,
    \frac{dr}{\sqrt{2E-2\Phi_0(r)-{L^2}/{r^2}}}.
 \end{multline*}
One can show that the main contribution to the first integral
comes from the lower limit, thus one can extend the integration to
infinity. Since, for non-singular potentials, one can neglect
$\Phi_0(r) \approx \Phi(r_{\rm min})$ (as compared to $E$) in the
root argument, we obtain
 \begin{multline*}
    \int\limits_{r_{\rm min}(E,\,L)}^{r_{\rm
    max}(E,\,L)}\frac{L^3}{r^4}\,\,
    \frac{dr}{\sqrt{2E-2\Phi_0(r)- {L^2}/{r^2}}}  =\\
    =2E
    \int_1^{\infty}\frac{dx}{x^3\sqrt{x^2-1}}=
    {\ccase{1}{2}}\,\pi E,
 \end{multline*}
so for $\phi(E,L)$ we get:
 $$
    \phi(E,L)\approx - L\,\frac{d}{d E}
    \int\limits_0^{r_{\rm
    max}(E,\,0)}\frac{\Phi_0'(r)}{r}\,\,
    \frac{dr}{\sqrt{2E-2\Phi_0(r){\phantom{\big|}}}}.
 $$

Substituting this into (\ref{eq:2.19}), we find
 \begin{align}\label{eq:2.21}
    \varpi(E)=-\dfrac{\Omega(E)}{\pi}\,\dfrac{d}{d E}
    \int\limits_0^{r_{\rm
    max}(E)}\dfrac{\Phi_0'(r)}{r}\,\,
    \dfrac{dr}{\sqrt{2E-2\Phi_0(r){\phantom{\big|}}}}.{\phantom{\Bigg|}}
 \end{align}
There is also an alternative expression, without differentiation
with respect to $E$. Omitting details, we give the final
expression:
 \begin{align}\label{eq:2.22}
    \varpi(E)&=\dfrac{\Omega(E)}{\pi}\,\Bigl[ \int\limits_{0}^{r_{\rm
    max}(E)}\dfrac{dr}{r^2}\,\Bigl(\dfrac{1}{\sqrt{2E-2\Phi_0(r){\phantom{\big|}}}}
    -\dfrac{1}{\sqrt{2E}}\Bigr) \Bigr. - \nonumber\\
    \Bigl. &-\dfrac{1}{r_{\rm
    max}(E)\,\sqrt{2E}}\Bigr].
 \end{align}
From the derived expressions it follows, for example, that for the expansion of a
potential, $\Phi_0(r) = \Phi_0(0) + \Omega_0^2 \left(\ccase{1}{2}
\,r^2+\beta\,r^4 + \cdots\right)$, at small $r$, the precession direction is
determined by the sign of $\beta$, since $\varpi=-\beta$. Thus, stars with small
radial amplitude have retrograde precession if $\beta$ is positive. From
Poisson's equation one immediately has
 $$
 \rho(r) = \frac1{4\pi G r^2} \frac {d}{dr}\,r^2 \frac{d}{dr}\,\Phi_0(r) =
 \frac{3\Omega_0^2}{4\pi G}\,\Bigl(1+\frac{20}{3}\beta r^2\Bigr),
 $$
i. e. density increases with radius. Such behavior is unrealistic, and we
conclude that the gravitational loss-cone instability is impossible in spherical
clusters with non-singular potentials. Note that in disk systems this instability
is possible
\citep{P91b}.

\end{description}

\section{Gravitational Loss-Cone Instability in Disks}

\subsection{Derivation of an equation for eigenmode spectra}

The simplest relations for describing the instability under
study can be obtained in a model with an active stellar
subsystem in the form of a disk with nearly-radial orbits. The aim
of this Section is to derive characteristic equations for
small perturbations in such a model. The derivation involves a
series of successive simplifications of the initial linearized
Vlasov equations.

Omitting a number of those steps, we take, as the starting point, the following
integral equation\footnote{Actually, (\ref{eq:3.1}) is a set of connected
integral equations, but for brevity we shall call it simply 'equation' hereafter.
(\ref{eq:3.1}) was earlier obtained in a paper by one of the authors \citep{EP05}
using the actions -- angles formalism suitable for problems of this type.}:
\begin{multline}\label{eq:3.1}
    \Phi_l(E,L)=\frac{G}{2\pi}\int\int
    \frac{dE'\,dL'}{\Omega_1(E',L')}\\ \times
    \sum\limits_{l'=-\infty}^{\infty}\frac{\bigl[m\,\Omega_2(E',L')
    {\phantom{\Bigg|}}\!\!+l'\,\Omega_1(E',L')\bigr]\,{\p F}/{\p
    E'}+m\,{\p F}/{\p L'}}{\omega - m\,\Omega_2(E',L') -
    l'\,\Omega_1(E',L') {\phantom{\Bigg|}}} \times
    \\
    \times \Pi_{l,\,l'}(E,L;E'L')\,\Phi_{l'}(E',L').
\end{multline}
Here $G$ is the gravitational constant, the frequencies $\Omega_i = \p H_0 / \p
I_i$, or more briefly, ${\bf \Omega}({\bf I}) = \p H_0/\p {\bf I}$, $H_0({\bf
I})$ is the Hamiltonian of a star in unperturbed state, ${\bf I}=(I_1,I_2)$ are
the actions, $\omega$ is the frequency of a perturbation, $E$ and $L$ are the
energy and the angular momentum, respectively, $F=F(E,L)$ is the unperturbed DF,
and the Fourier components $\Phi_l(E,L)$ and the kernel $\Pi_{l,\,l'}(E,L;E',L')$
are defined as follows:
 \begin{align*}
    \Phi_l(E,L)\!=\!\frac{1}{2\pi}\int_{-\pi}^{\pi}\!\!\!dw_1\,\Phi[\,r(E,L;w_1)]\,
    \cos[\,lw_1\!+\!m\,\chi(E,L;w_1)],
 \end{align*}
where $\Phi(r)$ is the radial part of the perturbed potential
$\Psi(r,\varphi;t)=\Phi(r)\,e^{-i\omega\,t+ i\,m\varphi}$, $t$ is the time,
$\varphi$ is the polar angle, $m$ is the azimuthal index, $w_1$ is the angle
conjugate to the radial action $I_1$, $l$ is the integer index,
\begin{multline*}
    \Pi_{l,\,l'}(E,L;E',L')=\int_{-\pi}^{\pi} dw_1
    \cos\,[\,lw_1+m\,\chi(E,L;w_1)]\\ \times
        \int_{-\pi}^{\pi}\!\!\!dw_1'
    \cos\,[\,l'w_1'+m\,\chi(E',L';w_1')]\,\\ \times \psi[r(E,L;w_1),r'(E',L';w_1')],
    \\
    \psi(r,r')=\int_0^{2\pi}d\theta\,\frac{\cos
    m\theta}{\sqrt{r^2+r'^2-2rr'\cos\theta{\phantom{\big|}}}}.
\end{multline*}
The function $\chi$ is defined by
 \begin{multline*}
    \chi(E,L;w_1)=\Omega_2(E,L)\!\!\!\!\!\!\!\!\!\int\limits_{r_{\rm
    min}(E,\,L)}^r\!\!\!\!\!\!\!\frac{dx}{v_r(E,L;x)}-
    \\
    -L\!\!\!\!\!\!\!\!\!\int\limits_{r_{\rm min}(E,\,L)}^r\!\!\!\!\!\!\!\!\!\frac{dx}{x^2 v_r(E,L;x)}
 \end{multline*}
(where $v_r$ is the radial velocity of a star), or
  $  \chi(E,L;w_1)=(\Omega_2/\Omega_1)\,w_1-\delta\varphi,$\  \
where
 \begin{align*}
    \delta\varphi=
    L\!\!\!\int\limits_{r_{\rm min}(E,\,L)}^r\frac{dx}{x^2
    v_r(E,L;x)}.
 \end{align*}

For definiteness, we consider systems consisting exclusively of
2:1-orbits in this Section. However, the final form of the desired
equation for the case of 1:1-orbits (i.e., for near-Keplerian
systems) is practically identical to that for 2:1-orbits. It is
evident that slow modes, with angular rates of the order of the
typical precession velocity, in the systems with 2:1-orbits are
possible only when the azimuthal index $m$ is even. In case of
1:1-orbits the azimuthal number $m$ is arbitrary.

Although the equation (\ref{eq:3.1}) is an exact linear integral
equation that allows one to determine the spectrum of eigenmodes for an
arbitrary distribution of stars in the disk, for a stellar system with
low angular momenta we are interested in, this equation is inconvenient
as the system can include stars with both direct ($L>0$) and inverse
($L<0$) orbital rotation. The point is that the orbital frequency
$\Omega_2(E,L)$ is discontinuous at $L=0$:
   $ \Omega_2(E,L=\pm\, 0)={\ccase{1}{2}}\,{s_L}\,\Omega_1(E,L=0),  \ \
    s_L\equiv{\rm sign}\,(L)$.
So all items in the sum over $\ l'$ are also discontinuous at
$L'=0$. The function $\Phi_l(L)$ itself and the kernel of the
integral equation $\Pi_{l,l'}(L,L')$ are discontinuous too.
(Hereafter we shall omit the arguments $E$ and $E'$, provided this
creates no difficulties.) The discontinuity is very inconvenient.
In fact it arises from a poor choice of the angle variables in the
set of actions -- angles, for the problems of interest involving
stars with $L<0$ (note that discontinuity is quite inessential in
problems with nearly-circular orbits (see, e.g., \citealt{EP05}).
However, this difficulty is fictitious, and actually the equation
can be transformed into a continuous form by means of a proper
procedure that involves shifting of indices and transformation of
the functions. The procedure is simple but cumbersome, so we give
the final form of the integral equation without going into detail:
 \begin{multline}\label{eq:3.7}
    \phi_{n}(E,L)=\frac{G}{2\pi}\int\int
    \frac{dE'\,dL'}{\Omega_1(E',L')}\\ \times
    \sum\limits_{n'=-\infty}^{\infty}\frac{\left({\p F}/{\p
    L'}\right)_{LB}
    {\phantom{\Bigg|}}\!\!+({n'}/{m})\,\Omega_1(E',L')\,{\p
    F}/{\p E'}}{\Omega_p-\Omega_{\rm
    pr}(E',L')-({n'}/{m})\,\Omega_1(E',L'){\phantom{\Big|}}}\,\,
    \\
    \times
    R_{n,\,n'}(E,L;E',L')\,\phi_{n'}(E',L').
 \end{multline}
Here $\Omega_{\rm pr}\, {\p F}/{\p E}+ {\p F}/{\p L}= \left({\p
F}/{\p L}\right)_{LB}$
 is the so called Lynden-Bell derivative of
the DF which is by definition a derivative with respect to the
angular momentum at fixed Lynden-Bell's adiabatic invariant
(\citeyear{LB79}), $J_f=I_1+\frac{1}{2}\,|I_2|$:
 \begin{align*}
    \left({\p F}/{\p L}\right)_{LB} \equiv \left({\p F}/{\p L}\right)_{J_f}
    =-{\ccase{1}{2}}\,s_L\left({\p F}/{\p I_1}\right)_{I_2}+\left({\p F}/{\p
    I_2}\right)_{I_1},
 \end{align*}
and the precession velocity is
  $$  \Omega_{\rm pr}(L')=\Omega_2(L')
  -{\ccase{1}{2}}\,s_{L'}\,\Omega_1(L').$$
Note that the function $\Omega_{\rm pr}(L)$ is continuous at $L=0$
(passes through zero). In the equation (\ref{eq:3.7}), the new
function (continuous at $L=0$)
\begin{multline}\label{eq:3.11}
    \phi_n(E,L)=\dfrac{1}{2\pi}\displaystyle\int_{-\pi}^{\pi}dw_1\,\Phi\,[\,r(E,L;w_1)]\,
    \\ \times \cos\,[\,n\,w_1+m\,{\tilde\chi}(E,L;w_1)],{\phantom{\Bigg|}}
\end{multline}
and the new kernel {continuous at $L=0$ and $L'=0$}
\begin{multline}\label{eq:3.12}
    R_{\,n,\,n'}(E,L;E',L')=\displaystyle\int_{-\pi}^{\pi} dw_1
    \cos\,[\,n\,w_1+m\,{\tilde\chi}(E,L;w_1)]
    \\
    \times
    \displaystyle\int_{-\pi}^{\pi} dw_1'
    \cos\,[\,n'\,w_1'+m\,{\tilde\chi}(E',L';w_1')]\,
    \\ \times \psi\,[\,r(E,L;w_1),r'(E',L';w_1')]
\end{multline}
appear. In (\ref{eq:3.11}) and  (\ref{eq:3.12})
\begin{multline*}
    {\tilde\chi}(E,L;w_1)=\Omega_{\rm pr}(E,L)\!\!\!\!\!\!\!\!\int\limits_{r_{\rm
    min}(E,\,L)}^r\!\!\!\!\!\!\!\!\frac{dx}{v_r(E,L;x)}-
    L\!\!\!\!\!\!\!\!\int\limits_{r_{\rm
    min}(E,\,L)}^r\!\!\!\!\!\!\!\!\frac{dx}{x^2 v_r(E,L;x)},
\end{multline*}
that is $ {\tilde\chi}(E,L;w_1) = \chi(E,L;w_1)
-{\ccase{1}{2}}\,s_L\,w_1$.

Then the equation (\ref{eq:3.7}) can be treated as the initial
exact integral equation. Since below we shall concentrate
on studying distributions localized in the vicinity of
 $L=0$, it is important to keep in mind that the functions $\phi_n(E,L)$
 and $R_{n,\,n'}(E,L;E',L')$,  and also the frequency $\Omega_1(E,L)$
 are not only continuous, but {\it smooth} at $L=0$ è $L'=0$ as well:
 $$
 \phi_n(E,L)=\Phi_n(E,0)+\alpha_n(E)\,L+{\ldots},
 $$
 $$
 \Omega_1(E,L)=\Omega_1(E,0)+{\cal O}(L^2),\ \
 R_{n,\,n'}(E,L;E',L')=
 $$
 $$=R_{n,\,n'}(E,0;E',0)+\beta_n(E)\,L+
 \beta_{n'}(E')\,L'+\ldots.
 $$

Although proof of this statement is rather non-trivial, we leave it beyond the
scope of the article. Here it is particularly important for us that the
coefficients $\alpha_n(E)$ and $\beta_n(E)$ for $n=0$ and $n'=0$ tend to zero,
hence
\begin{multline}\label{eq:3.15}
    \phi_0(E,L)=\phi_0(E,0)+{\cal O}(L^2),\\
    R_{0,\,0}(E,L;E',L')=R_{0,\,0}(E,0;E',0)+{\cal O}(L^2)+{\cal
    O}(L'^2).
\end{multline}
It is precisely this fact that allows us to consider the kernel $R_{0,\,0}$
and the function $\phi_0$ to be constant at the DF localization scales
(with respect to $L$), when studying slow modes.

Our following step is  to transform (\ref{eq:3.7}) into an equation describing
slow disk modes with frequencies of the order of precession velocities. The
latter are always less than the orbital frequencies $\Omega_1$ and $\Omega_2$,
and for nearly-radial orbits $\Omega_{\rm pr} \ll \Omega_1$, $\Omega_2$. As it
was explained in considerable detail in a paper by \cite{EP04}, for  slow modes,
only the items with $n'=n=0$ dominate in the sum of (\ref{eq:3.7}), since they
have minimal denominators. If these items are only taken into account, we obtain
the equation
 \begin{multline}\label{eq:3.16}
    \phi_{0}(E,L)=\frac{G}{2\pi}\int\int
    \frac{dE'\,dL'}{\Omega_1(E',L')}\\ \times \frac{\left({\p F}/{\p
    L'}\right)_{LB} {\phantom{\big|}}\!\!}{\Omega_p-\Omega_{\rm
    pr}(E',L'){\phantom{\Big|}}}\,\,
    R_{\,0,\,0}(E,L;E',L')\,\phi_{0}(E',L').
 \end{multline}
This is the desired equation for slow modes.

The next step consists in transforming (\ref{eq:3.16}) into an equation
convenient for a disk model with nearly-radial orbits. Let us assume that in
the domain of small angular momenta, the DF is
   $ F(E,L)=f^{(E)}(E)\,f^{(L)}(L)$.
The scale of localization domain, $\delta L$, for the function
$f^{(L)}(L)$ near $L=0$ is assumed to be small. In general, the
exact meaning of this smallness needs to be refined, but in any
case the scale must be smaller than the characteristic length of
variation in momentum, $\Delta L$, for all the functions appearing
in the equation (\ref{eq:3.16}), i.e. $\phi_0(E,L)$,
$R_{0,0}(E,L;E',L')$, $\Omega_1(E,L)$, $\Omega_{\rm pr}(E,L)$. The
characteristic scale of variation for these functions is
determined exclusively by the behavior of unperturbed potential.
As for the latter, we do not suggest any peculiar behavior and
consider this potential to be non-singular at $r=0$. So we can
assume that $\delta L\ll \Delta L$. In this case, due to the
relations (\ref{eq:3.15}), in the localization domain of the
function $f^{(L)}(L)$, we can take
 $
    \phi_0(E,L)\approx\Phi_0(E,0)\equiv \Phi(E),\ \ R_{0,0}(E,L;E',L')\approx
    R_{0,0}(E,0;E',0)\equiv {P}(E,E'),
 $
 $
    \Omega_1(E,L)\approx \Omega_1(E,0) \equiv\Omega(E).
 $
For the precession velocity in a domain of small values of L, we have
  $ \ \Omega_{\rm pr}(E,L)\approx\varpi(E)\,L,\ \
    \varpi\equiv\bigl[{\p\Omega_{\rm pr}(E,L)}/{\p
    L}\bigr]_{L=0}.$

As a result, the two-dimensional integral equation (\ref{eq:3.16})
reduces to a one-dimensional equation with the kernel depending
on $E$ and $E'$ only:
 \begin{multline}\label{eq:3.19}
    \Phi(E)=\frac{G}{2\pi}\int\frac{dE'}{\Omega(E')}\,P(E,E')\,f^{(E)}(E')\,\Phi(E')
    \\ \times
    \int dL'\,
    \frac{{df^{(L)}(L')}/{dL'}}{\Omega_p-\varpi(E')\,L'}.
 \end{multline}
To find the spectrum of eigenmodes, a numerical solution is
required. However, one can predict immediately some qualitative
consequences. For ${P}(E,E')$, we obtain
 \begin{multline*}
    {P}(E,E')= 4\,\Omega(E)\,\Omega(E')\!\!\!\!{\displaystyle\int\limits_0^{r_{\rm
    max}(E)}}\!\!\!\!\!\!\!\dfrac{dr}{v_r(E,r)}\!\!\!\!\!
    \int\limits_0^{r_{\rm
    max}(E')}\!\!\!\!\!\!\!\dfrac{dr'}{v_r(E',r')}\,\,\psi(r,r')
 \end{multline*}
with $v_r(E,r)=\sqrt{2E-2\Phi_0(r)\phantom{\big|}}$. It can easily be shown from
(\ref{eq:3.19}) that in the ``cold'' case of purely-radial orbits, i.e. for
$f^{(L)}(L) =\delta(L)$,
 \begin{align*}
    \Phi(E)=-\frac{G}{2\pi\Omega_p^2}\int\frac{dE'}{\Omega(E')}\,\,
    f^{(E)}(E')\,\varpi(E')\,{P}(E,E')\,\Phi(E').
 \end{align*}

It is easy to verify (see, e.g., \citealt{P92}) that ${P}(E,E')$
is a positive quantity. Then it is evident that instability or
stability depends exclusively on the sign of $\varpi(E)$. Namely,
instability occurs when it is positive, $\varpi>0$. This is the
radial orbit instability. Recall that near the center, where the
potential $\Phi(r)=\Phi(0)+\Omega_0^2\bigl(\frac{1}{2}\,r^2
+\beta\,r^4+\ldots)$, the quantity $\varpi$ is positive when
$\beta<0$, so that instability must occur. Note that such behavior
of the potential is typical for most surface density distributions
decreasing with radius.

To investigate the spectrum of eigen oscillations in more
detail, let us add another simplifying assumption. Namely, let us
consider a model with monoenergetic distribution over energy,
i.e. $F(E,L)=\delta(E-E_0)\,f(L)$. In this case the integral
equation (\ref{eq:3.19}) reduces to a simple characteristic
equation for the complex (generally speaking) velocity $\Omega_p$:
 \begin{align*}
    1=\frac{G}{2\pi\Omega(E_0)}\,{P}(E_0,E_0)\int\limits_{-\infty}^{\infty}
    \frac{d\,f(L)/dL} {\Omega_p-\varpi(E_0)\,L}\,dL.
 \end{align*}
One can turn from distribution over the angular momentum $L$ to
that over the precession velocities: $ \Omega_{\rm
pr}=\varpi(E_0)L\,\equiv \nu. $ After denoting
$f(L)=f\left({\nu}/{\varpi}\right)\equiv f_0(\nu), $ we obtain
 \begin{align*}
    1=\frac{G}{2\pi\,\Omega(E_0)}\,{P}(E_0,E_0)\,{\rm
    sign}\,(\varpi)\int\limits_{-\infty}^{\infty}
    \frac{d\,f_0(\nu)/d\nu} {\Omega_p-\nu}\,d\nu.
 \end{align*}
Hereafter we are primarily interested in the case of retrograde
precession, $\varpi<0$, therefore let us write the characteristic equation
in the following final form:
 \begin{align}\label{eq:3.25}
    1=-{\cal A}\int
    \frac{d\,f_0(\nu)/d\nu} {\Omega_p-\nu}\,d\nu, \ \
    {\cal A}=\frac{G}{2\pi\Omega(E_0)}\,{P}(E_0,E_0)>0,
 \end{align}
or equivalently,
 \begin{align}\label{eq:3.26}
    1={\cal A}\int
    \frac{f_0(\nu)}{(\Omega_p-\nu)^2}\,d\nu.
 \end{align}
It is easy to check that the equation (\ref{eq:3.26})
coincides\footnote{Excepting an inessential slip in the
coefficient ${\cal A}$ in \cite{P91b}.} with the characteristic
equation obtained \cite{P91b} in the so-called
``spoke''-approximation. The derivation above is in effect a
formal justification for this equation obtained earlier by V.
Polyachenko (\citeyear{P91b}) using a semi-intuitive approach. In
this approach, a set of stars moving along the same elongated
orbit is  regarded as a new elementary object replacing individual
stars, and the dynamics of stars reduces to the dynamics of spokes
(for slow processes); for an extended discussion see the relevant
papers by V. Polyachenko (\citeyear{P91a, P91b}). The advantage of
the spoke approach is that it is much simpler than the general
methods commonly used. It is this approach that is appropriate for
studying low-frequency oscillations and instabilities. However,
its rigorous justification required the above, rather cumbersome
calculations. This procedure was however necessary in order to
make sure that the spoke approximation is reliable. These
calculations were also useful in that they illustrated suggestions
and assumptions required for the approach.

Now we can proceed to the study of the resulting characteristic equation.

\subsection{Loss-cone instability of a disk in the spoke
approximation}

Let us represent  Eq. (3.8) in the form
\begin{align}\label{eq:3.27}
    Q=\int\limits_{-\infty}^{\infty}
    d\nu\,\frac{{d\,f(\nu)}/{d\nu}}{\nu-\Omega_p}, \ \ Q={\cal
    A}^{-1}>0,
 \end{align}
where for a distribution with a loss cone, i.e. with a deficiency of stars with
low angular momenta, the DF $f(\nu)$ has a zero minimum at $\nu=0$:
\begin{align}\label{eq:3.28}
    f(0)=0,\quad f'(0)=0,\quad f''(0)>0.
 \end{align}
We assume that this minimum is unique, and the DF looks like what
is shown in Figs. 4 or 5. An important point is that the quantity
$Q$ in Eq. (\ref{eq:3.27}) is positive only when the orbit
precession is retrograde. It is this circumstance that allows us
to lean upon the analogy with plasma and use the formalism of the
plasma theory when studying the instability. Recall that for the
case of direct precession, when $Q<0$, Eq. (\ref{eq:3.27})
describes only the radial orbit instability (the modification of
the Jeans instability for very elongated orbits) leading to the
spokes merging together \citep{PS72,Ant}. Based on the
Penrose\,--\,Nyquist criterion (see, e.g., \citealt{Pen}, or
\citealt{Mih}), \cite{P91b} showed that a new instability must
occur in systems of orbits with retrograde precession. Here we
reproduce this proof using a somewhat different language which is
more physically evident and even more constructive: e.g., it
allows us to determine, in a relatively simple manner, the
instability boundary in the parameter $Q$. This approach is based
on considering neutral modes.

\begin{figure}
\includegraphics[bb=69 304 545 620, width=78mm, clip=true, draft=false]{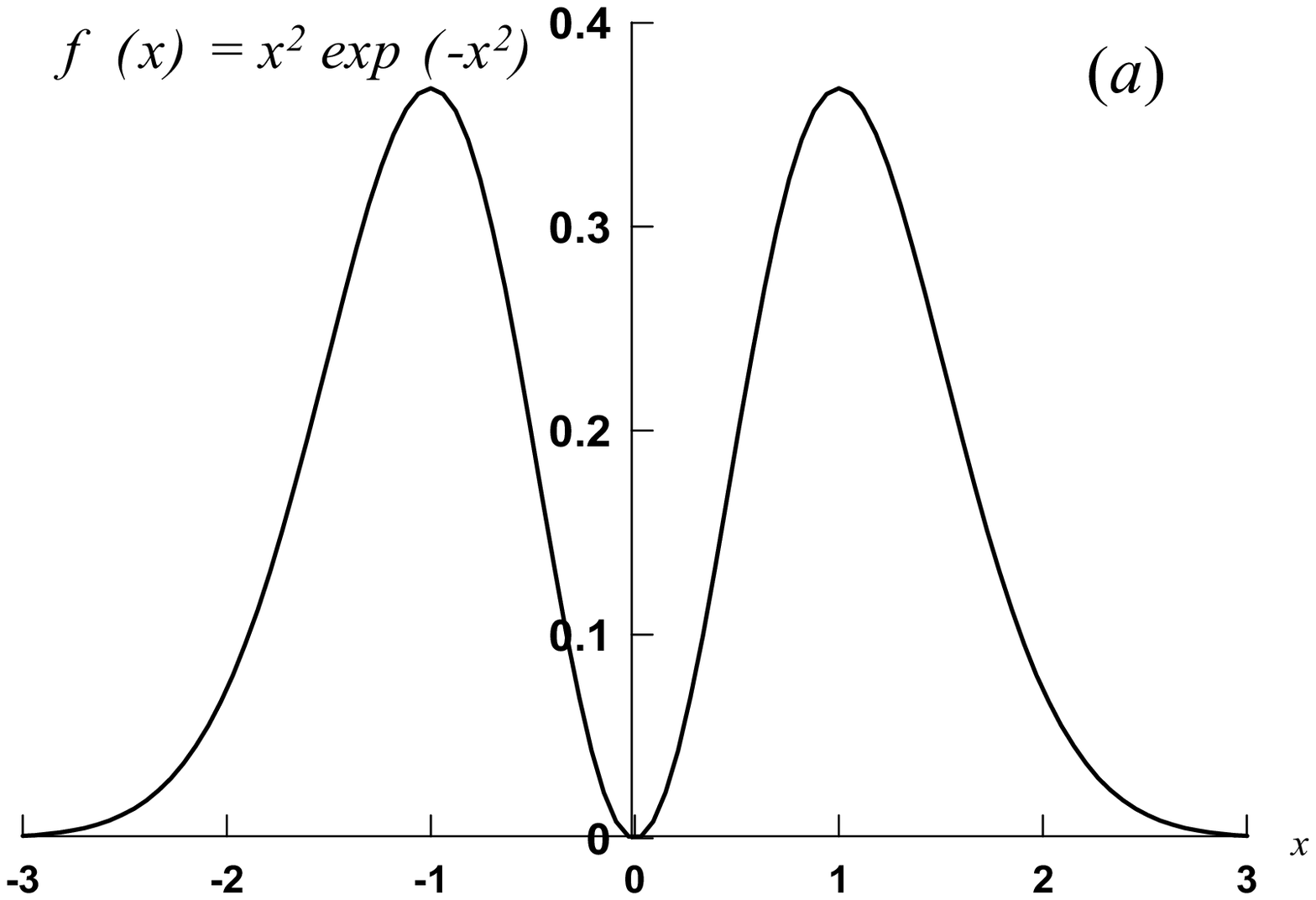}
\includegraphics[bb=45 245 537 725, width=78mm, clip=true, draft=false]{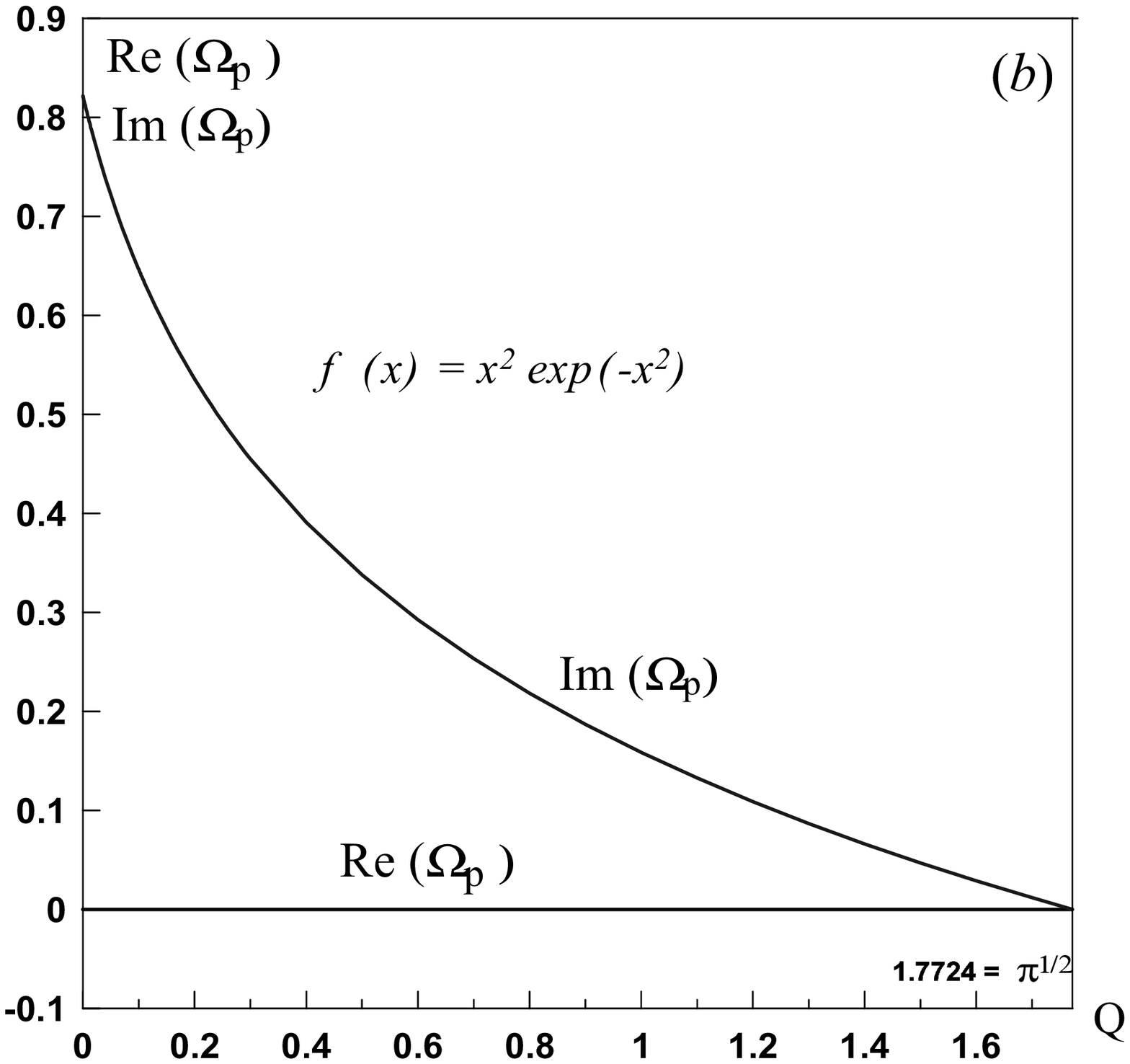}
 \caption{\small Loss-cone instability in disk.
\textit{a} -- symmetric distribution over precession velocity
(\ref{eq:3.13}) with $a=0$; \textit{b} -- dependence of ${\rm
Re}(\Omega_p)$ and ${\rm Im}(\Omega_p)$ on parameter $Q$.}
 \label{fig4}
 \end{figure}

 \begin{figure}
 \includegraphics[bb=81 291 543 609, width=78mm, clip=true, draft=false]{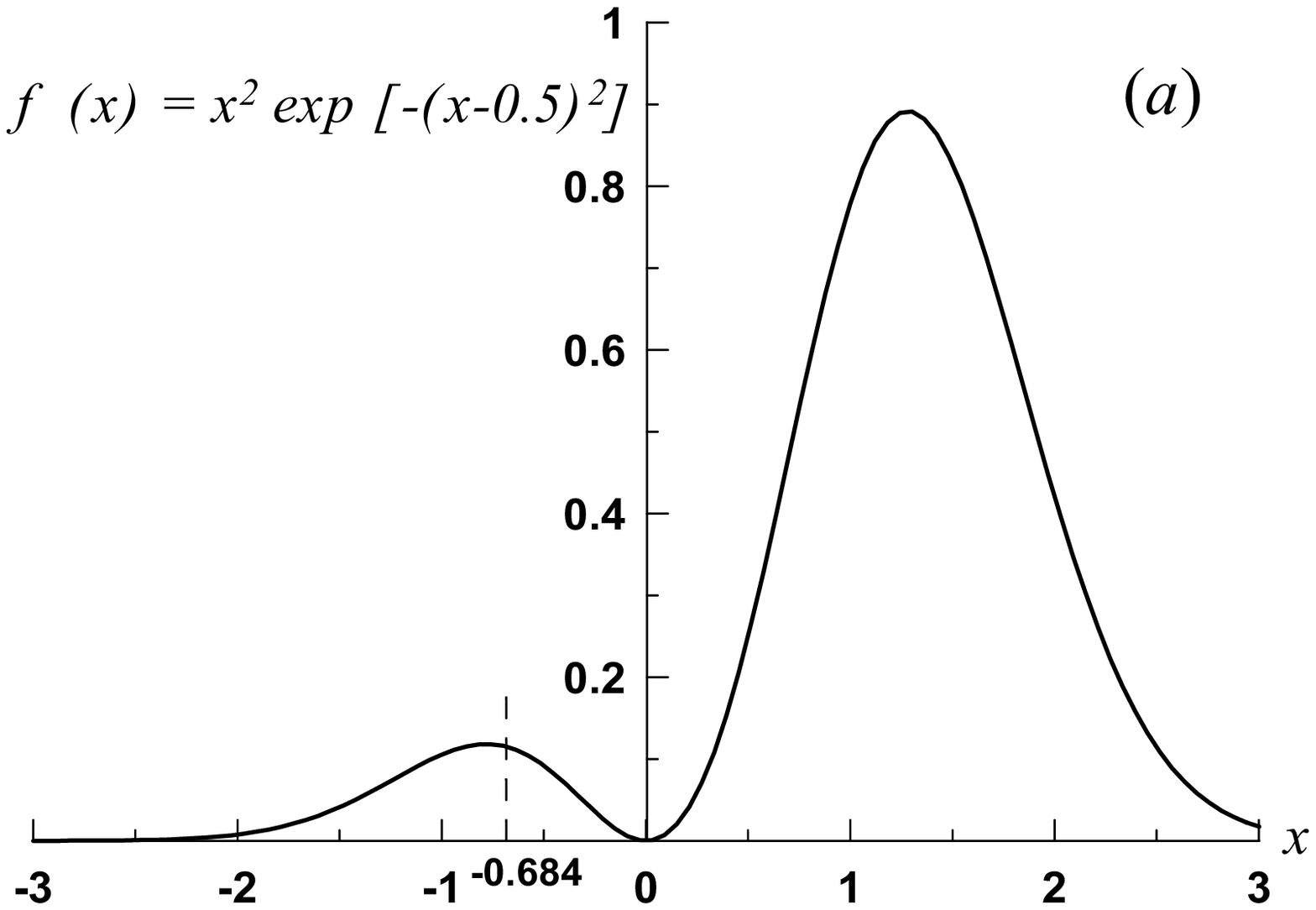}
 \includegraphics[bb=21 207 568 698, width=78mm, clip=true, draft=false]{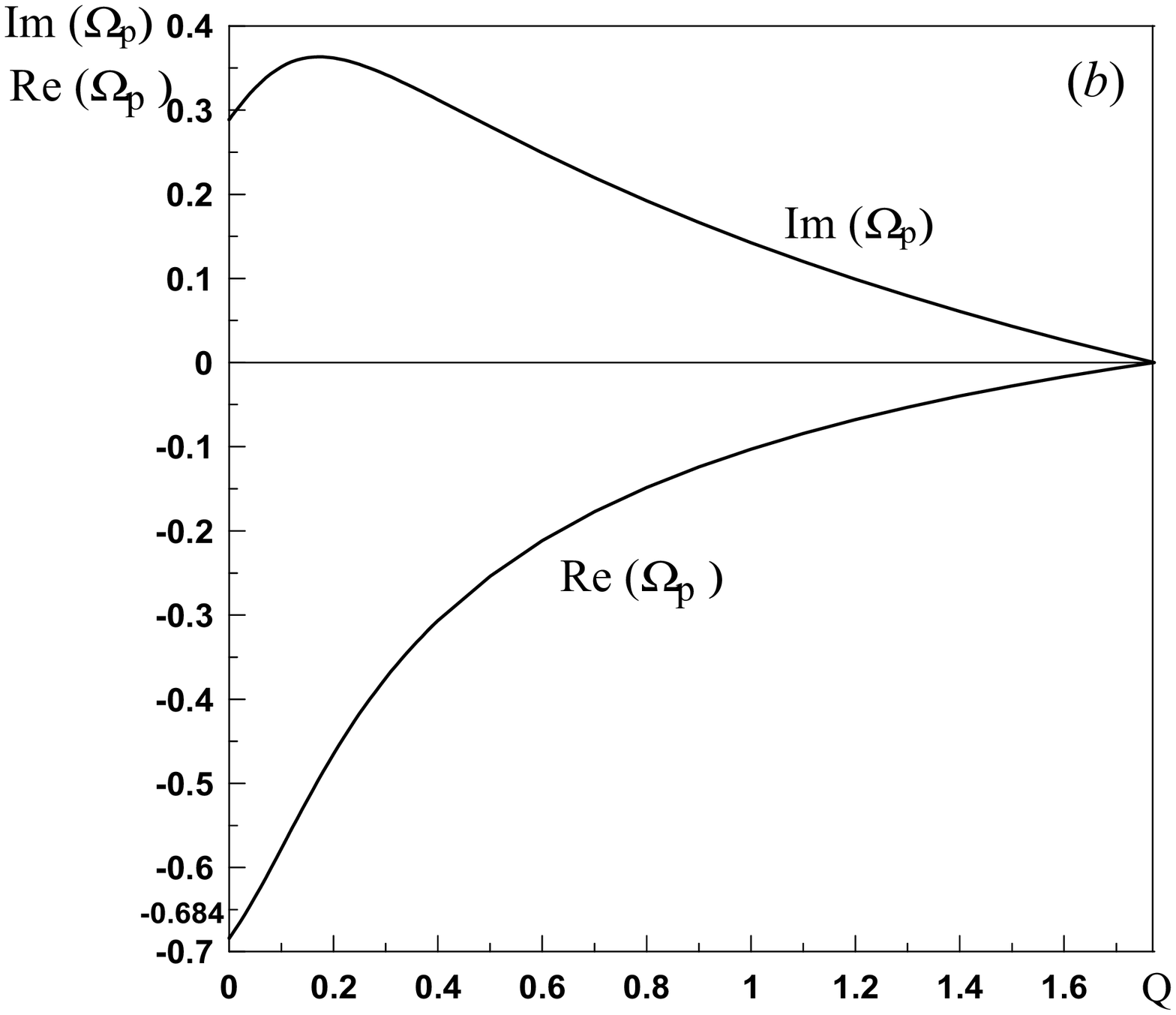}
 \caption{\small The same as in Fig. 4 for asymmetric
 distribution (\ref{eq:3.13}) with $a=0.5$.}
  \label{fig5}
 \end{figure}

The essence of the approach is as follows. Suppose that, when the
parameter $Q$ changes, the initially stable system becomes
unstable. This means that some value of $Q=Q_c$ exists for which a
neutral mode appears. For such a neutral mode, the location of the
resonance should coincide with an extremum of the DF $f(\nu)$ --
otherwise Eq. (\ref{eq:3.27}) will have a pole on the real axis,
bypassing which would necessarily result in an imaginary
contribution into the right side of (\ref{eq:3.27}); so that Eq.
(\ref{eq:3.27}) could not be fulfilled. For distribution with one
minimum and two maxima, one can show that (i) a neutral mode with
a resonance at the location of a higher maximum cannot exist; (ii)
when one maximum is much higher than the other, and two maxima are
sufficiently separated from each other, the neutral mode
corresponding to the lower maximum is possible, this mode
belonging to the same oscillation branch as the neutral mode
related to the minimum of the DF (see Sec. 3.3). Consequently, for
the neutral mode at the minimum, $\Omega_p=0$, while for the other
neutral mode, $\Omega_p=u_1$, where $\nu=u_1$ is the location of
the lower maximum.

Putting $\Omega_p=0$ in (\ref{eq:3.27}), we find the value of
$Q=Q_c^{({\rm min})}$, for which the neutral mode connected with
the minimum can exist:
\begin{align}\label{eq:3.29}
    Q_c=\int_{-\infty}^{\infty} d\nu\,\nu^{-1}\,{f'(\nu)}
    =\int_{-\infty}^{\infty} d\nu\,\nu^{-2}\,{f(\nu)}.
 \end{align}
Here we have no need to elucidate the meaning of these integrals.
Due to the condition (\ref{eq:3.28}), they converge at $\nu=0$ in
the ordinary sense. Obviously, the right side of (\ref{eq:3.29})
is positive. So $Q_c^{({\rm min})}$ always exists, for arbitrary
locations and heights of maxima. As to the sign of $Q_c^{({\rm
max})}$, it can be either (as mentioned above). If it is positive,
the second neutral mode exists. Note in addition that one more
neutral mode, with $Q_c=0$ always exists; formally, it corresponds
to the resonance at infinity.

With a knowledge of the values of $Q_c$ for neutral modes,  one
can determine domains of instability in the parameter $Q$ ($Q>0$)
as $Q_c$ are the margin values. For this purpose, let us apply the
perturbation theory.

First we consider the region near the boundary $Q=Q_c^{({\rm
min})}$. Let us deflect from critical value: $Q=Q_c^{({\rm
min})}+\delta Q$. The frequency $\Omega_p$ acquires an addition
$\delta\Omega_p$, which is simply equal to $\Omega_p$. We would
like to find out what direction should be taken to reach
instability. Varying the quantity $Q$ relative to $Q_c^{({\rm
min})}$ and $\Omega_p$ relative to zero in (\ref{eq:3.27}):
 $\delta Q=\delta\Omega_p\int_{-\infty}^{\infty}d\nu\,
    \nu^{-2}\,{f'(\nu)}$.
Here the pole in the integral must be bypassed below. This gives
 $$
    \delta Q=\delta\Omega_p\,\bigl[\int\limits_{-\infty}^{\infty}\hspace{-11.7pt}-\
    d\nu\, \nu^{-2}\,{f'(\nu)}+i\pi f''(0)\bigr].
 $$
Denoting
 $
    A=\displaystyle{\int\limits_{-\infty}^{\infty}}\hspace{-11.7pt}-\  d\nu\,
    \nu^{-2}\,{f'(\nu)},\ \ B=\pi f''(0)>0,
    $
we obtain
   $$  \Omega_p=\delta\Omega_p={\delta
    Q}/(A+i\,B)=\bigl[(A-i\,B)/(A^2+B^2)\bigr]\,\delta Q.$$
As $B>0$, the instability appears when $Q$ decreases below the
critical value $Q_c^{({\rm min})}$.

For the second neutral mode (with the resonance at the location of
lower maximum), we use the same procedure of the perturbation
theory to find that the instability is possible when $Q$ is
above the critical value $Q_c^{({\rm max})}$, as
$B=\pi\,f''(u_1)<0$ in the maximum.

As a result, we conclude that in the absence of the
neutral mode associated with the lower maximum, i.e. if
$Q_c^{({\rm max})}<0$, the unstable domain lies in the range $0<Q<Q_c^{({\rm
min})}$. If such a neutral mode exists, the range of instability becomes
$Q^{({\rm max})}<Q<Q_c^{({\rm min})}$. (It can be shown that $Q_c^{({\rm
max})}<Q_c^{({\rm min})}$).

As for the question of where the resonance shifts when the
parameter $Q$ is deflected
from the corresponding margin value into the unstable domain,
it depends on the sign of $A$, defined by the integrals in the
sense of principal value. In principle, these integrals can have any sign. In
the particular case of symmetric DF, showed in Fig.\,\ref{fig4}, the quantity
$A$ is zero (for $Q$ deviating into the domain $Q < Q_c^{(\min)}$).
Stability (or instability) is governed only by the sign of $B$,
and this does not depend on the sign of $A$, which can be
associated with the angular momentum of the wave.

We would like to say in this connection that here it is impossible to consider
our instability in terms of exchanged angular momentum between the wave and the
resonance stars, as is done in plasma physics or in the theory of galactic
structures which has undergone appreciable development since the well-known paper
by Lynden-Bell and Kalnajs (\citeyear{LBK72}). The language for explaining the
instability uses considerations operating with the momentum exchange between the
wave and resonance stars. For instance, if the wave momentum is positive and the
resonance stars lose their momentum transferring it to the wave, the momentum of
the latter increases. This is the instability. However, such a language does not
work in the case under consideration. The reason is that our systems are not
weakly-dissipative, as is usual in plasma. In the latter we have a well-defined
wave. All stars contribute into the wave dispersion properties, while the
dissipation is determined by a small portion of resonant stars. Consequently, the
dissipation is only a small correction. But now we have a completely different
situation: the dissipation and dispersion parts of the wave (i.e., roughly
speaking, the imaginary and real parts of dielectric permittivity) are of the
same order. So our instability is not kinetic, in the ordinary sense, and such
considerations do not work. However, in the case when the maxima are sufficiently
separated, and their heights strongly differ from each other, we return to the
usual, weakly-dissipative situation. Then the sign of phase velocity (i.e., the
sign of $A$) must correlate with the inclination of the DF. In the following
Sec.\,3.3, we consider this case as well.

\subsection{Neutral modes with resonance at DF maxima.
 Investigation of stability in a model two-humped distribution}
To illustrate the above reasoning, we shall study a model example with DF
 \begin{multline}\label{eq:3.13}
    f_a(\nu)=(\nu^2/\nu_T)\,\exp\,[-(\nu-a\nu_T)^2/\nu_T^2]\\ \equiv
    \nu_T\,x^2\,\exp[-(x-a)^2],
     \end{multline}
where we can control the maximum locations and heights.

But first we shall prove the statement formulated in the preceding subsection
that $Q_c$ for the neutral mode related to the higher maximum is always
negative, while $Q_c$ for the lower maximum can have any sign.
Suppose that the distribution has one maximum, similar to that showed in
Fig.\,4. Let the left maximum be at $\nu=u_1$, and the right maximum at
$\nu=u_2$, the right maximum being larger than the left one: $f(u_1)<f(u_2)$.
We shall consider these two variants separately.

(i) First let us assume that the neutral mode has frequency
$\Omega_p=u_2$. Let us rewrite (\ref{eq:3.27}) in the form
 \begin{multline}\label{eq:3.14}
    Q_c^{({\rm max})}=\int\limits_{-\infty}^{\infty} d\nu\,\frac{d\,f(\nu)/{d\nu}}
    {\nu-u_2}= \int\limits_{-\infty}^{\infty}
    d\nu\,\frac{[f(\nu)-f(u_2)]'} {\nu-u_2}=\\ =
    \int\limits_{-\infty}^{\infty} d\nu\,\frac{[f(\nu)-f(u_2)]}
    {(\nu-u_2)^2}.
 \end{multline}
Note that all integrals here can be considered in the usual sense since no
problems arise concerning their convergence at $\nu=u_2$. Since for the higher
maximum, $\nu=u_2$, $f(\nu)<f(u_2)$ everywhere, then the right side of
(\ref{eq:3.14}) is obviously negative. Thus, we proved that there cannot be a
neutral mode related to the higher maximum.

(ii) Let us now suppose that $\Omega_p=u_1$. Let us split the integral
into two parts:
 \begin{multline*}
    Q_c^{({\rm max})}=\int\limits_{-\infty}^0 d\nu\,\frac{d\,f(\nu)/{d\nu}}
    {\nu-u_1}+ \int\limits_{0}^{\infty} d\nu\,\frac{d\,f(\nu)/{d\nu}}
    {\nu-u_1} =
    \\
    =\int\limits_{-\infty}^{0} d\nu\,\frac{[f(\nu)-f(u_1)]}
    {(\nu-u_1)^2}+ \int\limits_{0}^{\infty}
    d\nu\,\frac{f(\nu)} {(\nu-u_1)^2}.
 \end{multline*}
Obviously, the integrals in the right side have opposite signs, so that the
resulting sign can be either. In the case of the model (\ref{eq:3.13}), the
quantity $Q_c^{({\rm max})}$ can be readily calculated:
$$Q_c^{({\rm max})}(a) =
2\sqrt{\pi}\,\,[\,a\sqrt{\ccase{1}{4}\,a^2+1}-\ccase{1}{2}\,(a^2+1)],
$$
so that at $a>a_c=2^{-1/2}\approx0.71$, the quantity $Q_c^{({\rm
max})}(a)$ becomes positive. Note that in the given model,
$Q_c^{({\rm min})}$ does not depend on $a$ and equals
$\sqrt{\pi}$.

Fig.\,\ref{fig4} shows the dimensionless complex phase velocity of the wave,
$\Omega_p$, as a function of $Q$ (in units of $\nu_T$, for the model
(\ref{eq:3.13}) with $a=0$. It is evident that owing to symmetry the real part of
the frequency $\Omega_p$ (and the quantity $A$ as well) is equal to zero.

Fig.\,\ref{fig5} shows $\Omega_p(Q)$ for the model (\ref{eq:3.13}) with $a=0.5$.
Since $a<a_c$, then this model (as is the case with the model with $a=0$) sees
only a neutral mode related to the minimum. As the model is asymmetric, the real
part of the frequency is not zero. It is negative for $|\delta Q|\ll 1$ as $A>0$.
Calculation shows that it remains negative when $Q$ is far from the instability
boundary.

For sufficiently large values of $a$ ($a>a_c=1/\sqrt{2}$), the DF maxima
(\ref{eq:3.13}) will be highly separated, and one of them becomes much higher
than the other. Then $Q_c^{({\rm max})}$, for the neutral mode corresponding to
the lower maximum, becomes positive, and a second neutral mode appears. As a
result, the instability domain looks like a horizontal band (converging with
increasing $a$), between these two neutral modes (see Fig.\,6). Fig.\,7 shows the
tracks of a complex eigenvalue $x_0$ in a complex $x_0$-plane for various values
of $a$. When the parameter $Q$ decreases from $Q^{({\rm min})}=\pi^{1/2}$ to
$Q^{({\rm max})}(a)$ the position of the point on the corresponding curve changes
so that ${\rm Re}(x_0)$ moves from 0 to $x_{\rm max}(a)$, where $x_{\rm max}(a)$
is the position of the lower maximum, while ${\rm Im}(x_0)>0$ and tends to zero
on both ends of the curve.

 \begin{figure}
 \vspace{1cm}
 \begin{center}
 \includegraphics[width=78mm]{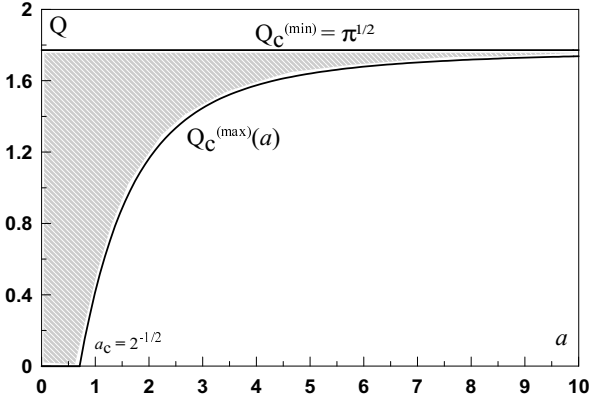}
 \end{center}
\caption{\small Unstable region on $Q$\,--\,$a$ - plane for the
model (\ref{eq:3.13}) is bounded from above by the straight line
$Q=Q_c^{({\rm min})}=\sqrt{\pi}$, from below by the curve
$Q_c^{({\rm max})}(a) = 2\sqrt{\pi}\,\, [\,a \sqrt{\frac14\,a^2+1}
-\frac12\,(a^2+1)]$ (shaded).}
 \label{fig6}
 \end{figure}

Based on the plasma analogy, there is no difficulty in understanding the physical
essence  of the instability under sufficiently large values of $a$. When $a\gg
1$, the DF becomes identical to the DF of plasma particles with a weak beam
moving at a rate significantly higher than the thermal velocity of particles in
the main plasma (a beam at the tail). Then the instability degenerates into the
well-known beam instability. It occurs when the wave phase velocity is on the
slope of the beam DF oriented towards the main plasma. For our model
(\ref{eq:3.13}) with $a>0$ the phase velocities of unstable modes must be
negative, which is the case in our calculations (see Fig.\,7).

\begin{figure}
\begin{center}
\includegraphics[width=78mm]{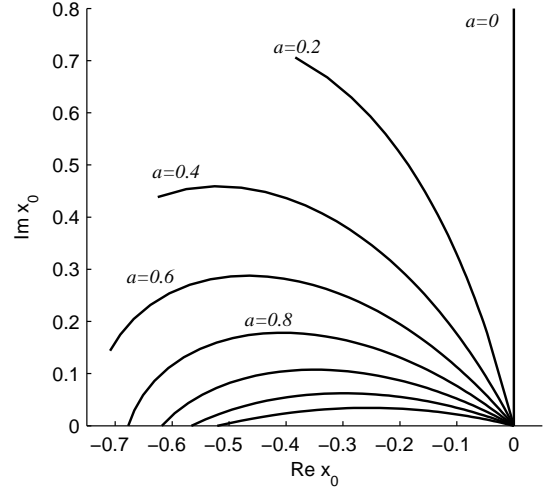}
\end{center}
\caption{\small The tracks of eigenvalue $x_0$ in complex $x_0$-plane for various values $a$,
indicated near corresponding curves.}
  \label{Fig7}
 \end{figure}

\section{Loss-cone instability in spherically-symmetric systems}

\subsection{Basic set of integral equations}

As we did in the case of disks, here we start with an exact equation governing
perturbations in a spherical system with  DF $F=F(E,L)$, where
  $  E=\frac{1}{2}\,v_r^2+\frac{1}{2}\,{L^2}/{r^2}+\Phi_0(r)$
is the star energy,
  $  L=rv_{\bot}\equiv r\,(v_{\theta}^2+v_{\varphi}^2)^{1/2}$
is the absolute value of angular momentum, $\Phi_0(r)$ is the
unperturbed gravitational potential. We assume that the DF does
not depend on the third integral $L_z=rv_{\varphi}\sin\theta$. As
is well-known, the spectrum of eigenvalues $\omega$ in this case
is independent of the azimuthal number $m$. Thus, instead of a
general representation of the potential and density in the form of
the sectorial harmonic
$\Phi(t;r,\theta,\varphi)=\chi(r)\,Y_l^m(\theta, \phi)\,
e^{-i\omega t}$ and
$\rho(t;r,\theta,\varphi)={\hat\rho}(r)\,Y_l^m(\theta,\phi)
\,e^{-i\omega t}$, we can restrict our consideration to a simpler
variant
   $ \Phi(t;r,\theta,\varphi)=\chi(r)\,P_l(\cos\theta)\,e^{-i\omega t},
    \ \ \rho(t;r,\theta,\varphi)={\hat
    \rho}(r)\,P_l(\cos\theta)\,e^{-i\omega t},$
where $P_l(x)$ is the Legendre polynomial. The derivation of the
basic integral equation (a set of integral equations, to be
precise) is also based on the action--angle formalism (as in the
disk case) and presented in Appendix A. Note that this formalism
was first used in the paper by Polyachenko \& Shukhman
(\citeyear{PS81}) as applied to spherical gravitating systems.

Thus, the initial exact set of integral equations has the form
($l_1,l'_1=-\infty,\dots,\infty$; $l_2,l'_2 = -l,\dots,l$):
 \begin{multline}\label{eq:4.8}
    \chi_{\,l_1,\,l_2}(E,L)
    =\frac{4\pi
    G}{2l+1}\sum\limits_{l_1'=-\infty}^{\infty}\sum\limits_{l_2'=-l}^l
    D_l^{l_2'} \int\frac{dE'\,
    LdL'}{\Omega_1(E',L')}\\
    \times\chi_{\,l_1'\,l_2'}(E',L')
    \,\Pi_{l_1,\,l_2;\,l_1',\,l_2'}(E,L;E',L')\times
    \\
    \times
    \,\frac{\Bigl[\,l_1'\,\Omega_1(E',L')+l_2'\,\Omega_2(E',L')\Bigr]\,{\p
    F}/{\p E'}+l_2'\,{\p F}/{\p
    L'}}{\omega-l_1'\,\Omega_1(E',L')-l_2'\,\Omega_2(E',L')}
 \end{multline}
with the kernel
 \begin{multline}\label{eq:4.9}
    \Pi_{l_1,\,l_2;\,l_1',\,l_2'}(E,L;E',L')\\=\int\limits_0^{2\pi}dw_1
    \int\limits_0^{2\pi}dw_1'
    \,{\cal F}_l\left[\,r\,(E,L;w_1),r'(E',L';w_1')\right]\times
    \\
    \times
    \exp{\left\{\,i\left[\Bigl(\,l_1'\,w_1'+l_2'\,{\p S_1}/{\p
    I_2'}\Bigr)-\Bigl(l_1\,w_1+l_2\,{\p S_1}/{\p
    I_2}\Bigr)\right]\right\}}.
 \end{multline}
 This is a two-dimension set of integral equations relative to unknown functions
 $\chi_{l_1,\,l_2}(E,L)$, which are related to the radial part of perturbed potential
 by
 \begin{align}\label{eq:4.10}
    \chi_{l_1,\,l_2}(E,L)=\int\limits_0^{2\pi}e^{-(l_1 w_1+l_2\p
    S_1/\p I_2)}\chi\bigl[r(E,L,w_1)\bigr]\,dw_1.
 \end{align}
In Eq. (\ref{eq:4.8}), we denote
 \begin{align}\label{eq:4.11}
    D_l^k=\left\{
    \begin{array}{cc}
    \dfrac{1}{2^{\,2\,l}}
    \,\dfrac{(l+k)!\,\,(l-k)!}{\left[\Bigl(\frac{1}{2}\,(l-k)\Bigr)!\,
    \Bigl(\frac{1}{2}\,(l+k)\Bigr)!{\phantom{\bigg|}}\right]^2},& |l-k| \  {\rm even},\\
    \\
    0& |l-k|\ \ {\rm odd}.
    \end{array}
    \right.
 \end{align}
In  Eq. (\ref{eq:4.9})
$${\cal F}_l(r,r')=\frac{(r_<)^l}{(r_{>})^{l+1}},\ \  r_<={\rm
min}(r,r'),\ \  r_>={\rm max}(r,r'),
$$ while the function of  radial action $S_1$ in
(\ref{eq:4.9}) and (\ref{eq:4.10}) is
 \begin{align*}
    S_1=\int\limits_{r_{\rm min}}^r dr'
    \sqrt{2E\,(\bI)-2\Phi_0(r')-{(I_2+|I_3|)^2}/{r'^2}\phantom{\big|}},\ \
     \end{align*}
where $\bI=(I_1,I_2,I_3)$ and the actions $I_2$ and $I_3$ are related to the
integrals of motion $L$ and $L_z$ by
  $ \ L=I_2+|I_3|,\ L_z=I_3$.
The dependence $r\,(E,L;w_1)$ is determined from
 \begin{align*}
    w_1={\p S_1}/{\p I_1}=\Omega_1\int\limits_{r_{\rm min}}^r
    \frac{dr'}{\sqrt{2E-2\Phi_0(r')-{L^2}/{r'^2}\phantom{\big|}}}.
 \end{align*}
We should also keep in mind that indices $l_1$ and $l_2$ correspond to
the spatial dependence of the perturbed potential expanded over
harmonics of the angular variables $w_1$ and $w_2$, conjugate to
the action variables
 \begin{align*}
    I_1=\frac{1}{2\pi}\oint p_r\, dr=\frac{1}{\pi}\int_{r_{\rm
    min}}^{r_{\rm max}}\sqrt{2E-2\Phi_0(r)-L^2/r^2\phantom{\big|}} dr,
 \end{align*}
 \begin{align*}
    I_2=\frac{1}{2\pi}\oint p_{\theta}\,
    d\theta=\frac{1}{\pi}\int_{\theta_0}^{\pi-\theta_0}
    \sqrt{L^2-L_z^2/\sin^2\theta}=L-|L_z|,
 \end{align*}
 respectively:
 \begin{align*}
    \Phi(\bI,w_1,w_2)=(2\pi)^{-2}
    \sum\Phi_{\,l_1,\,l_2}(\bI)\,e^{i\,(\,l_1\,w_1+l_2\,w_2)}.
 \end{align*}
In the case of $m=0$, the dependence on the angular variable $w_3$ is absent. Let
us also give an
alternative form for ${\cal F}_l(r,r')$ (sometimes it proves to be
more convenient):
\begin{align}\label{eq:4.7}
    {\cal F}_l(r,r')=(2l+1)\int\limits_0^{\infty} dk\,\frac{J_{\,l+{1}/{2}}(kr)}{\sqrt{k\,r}}\,
    \frac{J_{\,l+{1}/{2}}(kr')}{\sqrt{k\,r'}},
 \end{align} where $J_{\nu}(x)$ is a Bessel function (see \citealt{PS82}).

\subsection{Simplified equation for describing slow modes}

Let us assume that a massive nucleus or a black hole (with mass
$M_c$) is placed at the galactic center. Moreover, we assume that
the central mass dominates the unperturbed potential $\
\Phi_0(r)$, so that $\ \Phi_0(r)=\Phi_c(r)+\Phi_G(r),$ where $
\Phi_c(r)=-{G M_c}/{r}$, $\Phi_G(r)$ is the potential created by
the spherical subsystem of galaxy. Then the force acting on a star
from the central mass significantly exceeds the force from stars
in the galactic spherical component: $|\Phi_c'(r)|\gg
|\Phi_G'(r)|.$ In this case, stellar orbits are predominately
governed by the potential of the central massive point. This means
that we are dealing with 1:1-type orbits. Due to a small
additional potential $\Phi_G(r)$, the Keplerian ellipses precess,
the precession velocity is determined by the small difference
  $ \ \Omega_{\rm pr}(E,L)=\Omega_2(E,L)-\Omega_1(E,L),\ \Omega_{\rm
    pr}\ll \Omega_{1}\approx\Omega_2.$

An explicit expression for the precession velocity in terms of the
potential $\Phi_G(r)$ was found in the Section 2 (see formula
(2.8)). We shall be interested in slow modes, i.e., modes with
frequencies of the order of precession velocities, $\omega={\cal
O}(\Omega_{\rm pr})$. Then from all items with denominators
$\bigl[\omega-l'_1\, \Omega_1(E',L')-l'_2\,
\Omega_2(E',L')\bigr]$, we keep only those with $ l_1'=-l_2'$.
These denominators can be transformed into
  ${\omega-l_1'\,\Omega_1(E',L')-l_2'\,\Omega_2(E',L')}=
    {\omega-l_2'\,\Omega_{\rm pr}(E',L')}$.
The frequency constructions in numerators of these contributions
are equal to
  \begin{multline}\label{eq:4.20}
    \Bigl[\,l_1\,\Omega_1(E',L')+l_2'\,\Omega_2(E',L')\Bigr]{\p F}/{\p E'}+
    {\p F}/{\p L'}\\= l_2'\,\left[\Omega_{\rm pr}(E',L')\,{\p F}/{\p
    E'}+{\p F}/{\p L'}\right].
 \end{multline}
The expression in square brackets is a so-called Lynden-Bell
derivative \citep{LB79} of the DF:
 \begin{align}\label{eq:4.21}
    \left({\p F}/{\p E}\right)_{L}\Omega_{\rm pr}+\left({\p
    F}/{\p L}\right)_{E}=\left({\p F}/{\p L}\right)_{LB}.
 \end{align}
Recall that it is defined as the derivative with respect to the
absolute value of angular momentum, $L=I_2+|I_3|$, with the
adiabatic invariant, $J_f=I_1+I_2+|I_3|$, and the projection of
angular momentum, $L_z$, being constant.

Denoting
 \begin{multline*}
  \chi(E,L)_{l_1=-l_2,\,l_2}=\phi_{l_2}(E,L),\\
  \Pi_{l_1=-l_2,\,l_2;\,l_1'=-l_2',\,l_2'}(E,L;E',L')=P_{l_2,\,l_2'}(E,L;E',L')
 \end{multline*}
and keeping only items with $l_1'=-l_2'$ in the sum (\ref{eq:4.8}) over $l_1'$,
we find a ``slow'' equation for quantities
 $\phi_n(E,L)$:
 \begin{multline}\label{eq:4.24}
    \phi_{n}(E,L) =\dfrac{4\pi G}{2l+1}\sum\limits_{n'=-l}^l D_l^{n'}
    \int\dfrac{dE'\, L' d L'}{\Omega_1(E',L')}\\
   \times \,P_{\,n;\,n'}(E,L;E',L')\,\dfrac{n'\,\left({\p F}/{\p
    L'}\right)_{LB}}{\omega-n'\,\Omega_{\rm pr}(E',L')}\,\phi_{\,n'}(E',L'),
 \end{multline}
with the kernel
\begin{multline*}
    P_{\,n,\,n'}(E,L;E',L')=\\= \int\limits_0^{2\pi}dw_1 \int\limits_0^{2\pi}dw_1'
    \,{\cal F}_{\,l}\left[\,r\,(E,L;w_1),r'(E',L';w_1')\right]\times
    \\
    \times
    \exp{\left\{\,i\left[n'\,\Bigl({\p S_1}/{\p
    I_2'}-w_1'\Bigr)-n\,\Bigl({\p S_1}/{\p
    I_2}-w_1\Bigr)\right]\right\}}.
\end{multline*}
The kernel $P_{\,n,\,n'}$ can be rewritten in an
explicitly real form if one changes
the limits of integration over $w_1$ and $w_1'$ from  $[0,2\pi]$ to $[-\pi,\pi]$.
Then it becomes evident that $r(w_1,E,L)$ is the symmetric function of $w_1$, and
\begin{multline*}
    P_{\,n,\,n'}(E,L;E',L')= \\
    4{\displaystyle\int\limits_0^{\pi}}dw_1
    {\displaystyle\int\limits_0^{\pi}}dw_1'
    \,{\cal F}_{\,l}\left[\,r\,(E,L;w_1),r'(E',L';w_1')\right]
    \\
    \times
    \cos\left[n'\,\Bigl({\p S_1}/{\p
    I_2'}-w_1'\Bigr)\right]\,\cos\left[n\,\Bigl({\p S_1}/{\p
    I_2}-w_1\Bigr)\right].
\end{multline*}

The quantity $\phi_{\,n}(E,L)$ can also be written in a simpler
form
 \begin{align*}
    \phi_{\,n}(E,L)=2{\displaystyle\int\limits_0^{\pi}} \cos
    \left[\,n\,\Bigl ({\p S_1}/{\p I_2}-w_1\Bigr) \right] \,
    \chi\, \bigl[r(E,L,w_1)\bigr]\,dw_1.
 \end{align*}

\subsection{Simplified equation for the case of nearly radial orbits}

In the case of orbits with low angular momenta $L$ we are interested in the
set of equations (\ref{eq:4.24}) for slow modes allows further significant
simplifications. The requirements imposed on the width, $\delta L$, of
localization domain of the DF in angular momentum were
discussed in Sec. 3.

Based on the fact that the kernel $P_{n\,n'}(E,L;E',L')$, for nearly radial
orbits, depends only on energies $E$ and $E'$ (accurate to terms
quadratic in $L$ and $L'$)\footnote {The proof of this statement (and other
statements concerning analytical properties of the functions involved, near
$L=0$) is omitted here.} and acquires the form
  $ \ P_{n\,n'}(E,L;E',L')\approx
  P_{n\,n'}(E,0;E',0')=(-1)^{n+n'}\Pi(E,E'),\ $
where
 \begin{multline}\label{eq:4.29}
    \Pi(E,E')=4\int\limits_0^{\pi}dw_1\int\limits_0^{\pi}dw_1'\,{\cal F}_l
    [\,r(E,w_1), r'(E',w_1')]=
    \\
    = 4\Omega_1(E)\,\Omega_1(E')\!\!\int\limits_0^{b(E)} \dfrac{dr}{v_r(E,r)}
    \int\limits_0^{b(E')}
    \dfrac{dr}{v_r(E',r')}\,\,
    {\cal F}_{\,l}(r,r'),
 \end{multline}
  $\Omega_1(E)={(2|E|)^{3/2}}/(GM_c)$,
    $b(E)\equiv r_{\rm max}(E)={GM_c}/{|E|}$,
    $v_r(E,r)=\sqrt{2|E|}\,\sqrt{(b-r)/{r}}.
$
The unknown function $\phi_n(E,L)$ also depends only on $E$ (accurate to
${\cal O}(L^2)$):
  $\  \phi_n(E,L)\approx (-1)^n\Phi(E).$
Moreover, according to (\ref{eq:4.21}), the Lynden-Bell derivative
in $L$ coincides (accurate to ${\cal
O}\bigl[(M_G/M_c)\,L^2\bigr]$) with the derivative in $L$, with
the energy $E$ constant. As a result the equation (\ref{eq:4.24})
transforms into the one-dimension integral equation
 \begin{multline}\label{eq:4.32}
    \Phi(E) =\dfrac{4\pi G}{2l+1} \int\int\dfrac{dE'\,
    L'dL'}{\Omega_1(E')}\,\Pi(E,E')\,\Phi(E')\\
    \times\sum\limits_{s\,=-\,l}^l
    s\,D_l^{s}\,\dfrac{\p F(E',L')/\p L'}{\omega-s\,\varpi(E')\,L'}.
 \end{multline}

\subsection{Model case of monoenergetic distribution}

Let us consider again the model distribution
$F(E,L)=f(L)\,\delta(E-E_0)$, as in the disk case.
 Integrating (\ref{eq:4.32}) over $E'$ and putting $E=E_0$ in the resulting
 equation, we obtain the characteristic equation in the form
  \begin{align*}
    1=\dfrac{4\pi G}{2l+1} \dfrac{\Pi(E_0,E_0)}
    {\Omega_1(E_0)}\,\,\int L\,dL\sum\limits_{s\,=-\,l}^l
    s\,D_l^{s}\,\dfrac{df(L)/dL}{\omega-s\,\varpi(E_0)\,L}.
 \end{align*}
Recall (see Sec. 2) that for the case of near-Keplerian orbits
(i.e., orbits of the 1:1 type), the orbit precession is retrograde
for arbitrary distributions of the potential $\Phi_G(r)$, i.e.,
$\varpi<0$.

For convenience, we can turn from the variable $L$ to the variable
$\nu=|\Omega_{\rm pr}|=|\varpi(E_0)|L=-\varpi L>0$. Denoting $
f(L)=f\bigl({\nu}/{|\varpi|}\bigr)\equiv f_0(\nu),$ we write
 \begin{align}\label{eq:4.34}
    1=-\dfrac{4\pi G}{2l+1} \dfrac{\Pi(E_0,E_0)} {|\varpi(E_0)|\Omega_1(E_0)}\,\,\int
    \nu\,d\nu\sum\limits_{s\,=-\,l}^l
    s\,D_l^{s}\,\dfrac{df_0(\nu)/d\nu}{\omega-s\,\nu}.
 \end{align}
The equation (\ref{eq:4.34}) coincides with the Eq. (2) of \cite{P91a},
derived immediately in the spoke approximation\footnote{Excluding the unessential
factor
 $\Omega_1/\pi$ lost in r.h.s. of Eq. (2) of \cite{P91a}.} The rather cumbersome
derivation above provides the basis for the spoke approach
(together with the one for disks, in the previous Section).

To conclude this subsection, let us note that the monoenergetic model under
consideration corresponds to the specific density distribution of spherical
cluster $\rho_0(r)$ (and  the potential $\Phi_G(r)$) and has the finite radius
$R={GM_c}/{|E_0|}$. For this distribution, the quantity $\varpi(E_0)$ can be
explicitly calculated:
 \begin{align}\label{eq:4.37}
    \varpi(E_0)=-\frac{M_G}{M_c}\,\frac{8}{\pi^2}\,\frac{1}{R^2}.
 \end{align}
Here $M_G$ is the total mass of the spherical cluster (recall that
it is assumed that $M_G\ll M_c$). The kernel $\Pi(E_0,E_0)$ can
also be calculated in the explicit form. Using the relations
(\ref{eq:4.7}) and (\ref{eq:4.29}), we obtain
 \begin{multline}\label{eq:4.38}
    {\Pi}(E_0,E_0)=\frac{8\pi^2(2\,l+1)}{R}\,C_l,\ \
    \\ C_l=\int\limits_0^{\infty}
    \frac{dz}{z}\,\left[J_{\,(\,l+1)/2}(z)J_{\,l/2}(z)\right]^2.
 \end{multline}
The first seven coefficients $C_l$ calculated from (\ref{eq:4.38})
are presented in Table 1. After substituting these coefficients
into (\ref{eq:4.34}), we find
 \begin{multline}\label{eq:4.39}
    1=-{\cal A}_l\int \nu\,d\nu\sum\limits_{s\,=-\,l}^l
    s\,D_l^{s}\,\dfrac{d\,f_0(\nu)/d\nu}{\omega-s\,\nu}, \\ {\cal
    A}_l=\frac{16\pi^3\,G\,C_l}{|\varpi|}\,\left(\frac{R}{2GM_c}\right)^{1/2}.
 \end{multline}
 This equation can also be written in the form
 \begin{align}\label{eq:4.40}
    1=-2{\cal A}_l\sum\limits_{s\,=1}^l s^2\,D_l^{s}\,\int
    d\nu\,\dfrac{\nu^2 d\,f_0(\nu)/d\nu}{\omega^2-s^2\,\nu^2}.
 \end{align}
The DF $f_0(\nu)$ is normalized by the condition that
the total mass of spherical
cluster is equal to $M_G$, i.e., $\int F d\,\Gamma=M_G$. This gives
 \begin{align}\label{eq:4.41}
    \int f_0(\nu)\,\nu\,d\nu={\ccase{1}{2}}\,(2\pi)^{-3}\,[\varpi(E_0)]^2\Omega_1(E_0)M_G.
 \end{align}

\begin{table}
\begin{center}
 \begin{tabular}{|c|c|c|c|c|c|c|c|}
\hline
 $l$&1&2&3&4&5&6&7\\
 \hline
 $C_l$&0.135&
 0.063&
 0.037&
 0.025&
 0.018&
 0.014&
 0.011\\
  \hline
 \end{tabular}
\end{center}
 \caption{The values of coefficients  $C_l$.}
 \end{table}

\subsection{Stability of the $l=1$ and $l=2$ modes}

 We begin with studying the stability of modes $l=1$ and $l=2$. As
 we shall show, these modes are stable.  For definiteness, let us
 take the mode $l=2$. From the considerations below, it will be
 immediately obvious that they are valid for the mode $l=1$ as
 well.

 We start with the equation (\ref{eq:4.40}) for the mode $l=2$. An
important point is
 that (\ref{eq:4.40}) for this mode (as with $m=1$) includes
only one item in the
 sum over $s$ (with $s=2$) (correspondingly, two items $s=\pm 2$ in
(\ref{eq:4.39})). This follows from definition (\ref{eq:4.11}) of
the quantity $D_l^s$ and the equation (\ref{eq:4.40}). For $l=2$,
$D_2^2={\ccase{3}{8}}$, so that
 \begin{align}\label{eq:4.42}
    Q=-4\int\limits_0^{\infty}\dfrac{\nu^2\,\bigl[d\,f_0(\nu)/d\nu\,\bigr]}
    {\omega^2-4\,\nu^2}\,d\nu,\ \ \ Q={\ccase{4}{3}}\,{\cal A}_2^{-1}>0.
 \end{align}
Let the DF has a beam-like form due to deficiency of stars with
low angular momenta or, which is the same, with small precession
velocities $\nu$. We assume that the distribution has only one
maximum, located at $\nu=u$ on the semi-axis $0\le\nu<\infty$. If
there is a neutral mode, at some value of $Q=Q_c$, the
corresponding resonance must coincide with the maximum of
$f_0(\nu)$,\footnote{The absence of the neutral mode with
resonance at $\nu=0$ is established trivially.} i.e.,
$\omega^2=4u^2$. This means that for $Q=Q_c$
 \begin{align}\label{eq:4.43}
    Q_c=-\int\limits_0^{\infty}\dfrac{\nu^2\,
    \bigl[d\,f_0(\nu)/d\nu\,\bigr]}{u^2-\nu^2}\,d\nu.
 \end{align}
Evidently, for any one-hump distribution, the right side of
(\ref{eq:4.43}) is negative, as the integrand is free of
singularities and positive everywhere. Since $Q>0$, we conclude
that the neutral mode is impossible. The absence of neutral mode
means that the marginal value of $Q$, $Q=Q_c$, which separates
stable and unstable distributions, is also absent: each
distribution is either stable for all values of $Q$ or unstable
everywhere. Since stable one-hump distributions obviously exist,
we conclude that the mode $l=2$ is always stable.

The above considerations make it also clear that the conclusion
about the stability of the mode $l=2$ is valid only for the case
of retrograde precession, when the quantity $Q$ in Eq.
(\ref{eq:4.42}) is positive. In the case of prograde precession,
$Q<0$, so that the neutral mode (as well as the instability)
exists. This is the well-known radial orbit instability. True,
here it develops in a non-monotonic distribution with an empty
loss cone, instead of the usual distributions when most stars are
concentrated at near-radial orbits.

However, the conclusion that the instability is utterly impossible
in the case of retrograde precession would be premature.
The matter is that we obtain the above result concerning the mode $l=2$ due to a
formal reason: for this mode there is only one summand in the sum over $s$. So,
indeed, the instability is absent for $l=2$ (and $l=1$ as well). However, for
modes with $l\ge 3$, when there are at least two summands in that sum, the
instability becomes possible under suitable conditions. In the following
subsection, we study the mode $l=3$ in detail and demonstrate that the
instability can occur here.

\subsection{The mode $l=3$}

Considering the case of retrograde precession and restricting
ourselves only to the mode $l=2$, we showed (Sec.\,4.5) that
neutral modes (and consequently the instability) are absent for
one-hump distributions. However, this is valid only in the special
case that the mode has one resonance, as with $l=2$ or $l=1$.
Recall that the proof is based on the fact that the resonance must
then be located at the DF maximum. In such a situation, the
characteristic relation cannot be satisfied as the signs of right
and left sides of (\ref{eq:4.43}) are necessarily opposite, when
$\omega$ equals the frequency of neutral mode.

This proof, however, fails if a neutral mode has two (or more)
resonances. Indeed, the resonances can then be located so
that the resulting growth at one group of resonances is totally
cancelled by an equal damping at another group. In these conditions,
a neutral mode can exist. This means that the former group of
resonances must be located right of the maximum while the latter group
to the left. In the simplest case when there are only two
resonances, we must conclude that these resonances necessarily lie
on different sides of the maximum. Then it is hard to make a certain
conclusion about the sign of the integrand in the characteristic
relation that involves the principal value integrals, with a
singularity at each resonance. One may hope therefore that we can
find neutral modes (and consequently the instability) for $l\ge 3$
when there is at least a couple of resonances. Now we study the
possibility of neutral mode in the simplest suitable case of
$l=3$.

For $l=3$, we have $D_3^1=\frac{3}{16},\ \ \ D_3^3=\frac{5}{16}$.
The characteristic equation (\ref{eq:4.40}) gives
 \begin{align}\label{eq:4.44}
    Q=-\int\!d\nu\,\nu^2\,\frac{d\,f_0(\nu)}{d\nu}\left(\frac{1}
    {\omega^2-\nu^2}+\frac{15}{\omega^2-9\nu^2}\right)\!, \ Q={\ccase{8}{3}}\,
    {\cal A}_3^{-1}\!.
 \end{align}

Let us suggest that the neutral mode with the frequency $\omega=\omega_0$ occurs
at some value of $Q=Q_c$. For definiteness, we assume that
$\omega_0>0$.\footnote{It is apparent that with a given neutral mode of the
frequency $\omega_0$, a neutral mode of the frequency $-\omega_0$ also exists
(with the same $Q=Q_c$). Thus we can seek a frequency $\omega_0$ squared for
neutral mode.} For this frequency, there are two resonances:
 \begin{align}\label{eq:4.45}
    \nu=\nu_1=\omega_0, \ \ \ \nu=\nu_2={\ccase{1}{3}}\,\omega_0.
 \end{align}
Obviously, the resonance corresponding to smaller $\nu$ (i.e.,
$\nu=\omega_0/3$), must lie to the left of the maximum of the function
$f_0(\nu)$ (we denote its position $u$), while the resonance
corresponding to larger $\nu$ (i.e., $\nu=\omega_0$), must lie to
the right of the maximum: ${\ccase{1}{3}}\, \omega_0 < u < \omega_0.$
Bypassing the singularity in the complex plane $\nu^2$ from below
and equating the imaginary part of the full integral
(\ref{eq:4.44}) to zero, we find
\begin{align}\label{eq:4.46}
    f_0'(\omega_0)+{\ccase{5}{9}}\,f_0'\left({\ccase{1}{3}}\,\omega_0\right)=0.
 \end{align}
 Here we should explain that direction of bypassing in the complex
 plane $\nu^2$ coincides with that in the complex plane
 $\nu$ (i.e., it is from below) because of $\omega_0>0$. The condition
 (\ref{eq:4.46}) expresses the balance between growth at one of
 resonances and damping at the other. Eq. (\ref{eq:4.46}) determines
 the frequency of the neutral mode which exists when $Q=Q_c$,
 the latter found from the condition
 \begin{align}\label{eq:4.47}
    Q_c=-\int\hspace{-10.1pt}-\ d\nu\,\nu^2\,\frac{d\,f_0(\nu)}{d\nu}\left(\frac{1}
    {\omega_0^2-\nu^2}+\frac{15}{\omega_0^2-9\nu^2}\right).
 \end{align}

Eq. (\ref{eq:4.47}) involves the principal value integrals. The
pair of equations, (\ref{eq:4.46}) and (\ref{eq:4.47}), determines
the frequency of the neutral mode and the critical value of
parameter $Q$, $Q=Q_c$ such that the system has a neutral mode.
This is in fact the condition on some DF parameter (say,
dispersion of precession velocities, $\nu_T$), or on mass, $M_G$,
of spherical component (i.e., on value of its self-gravitation).
Under this condition, the system is at the stability boundary. If
the parameter deviates from its critical value in a certain
direction, the system becomes unstable. This direction is yet to
be determined.

It is not evident beforehand that the right side of Eq. (\ref{eq:4.47}) will be
positive for potential neutral modes. (Their frequencies are determined from Eq.
(\ref{eq:4.46}). It is easy to understand that for one-hump distributions, a
suitable solution of this equation always exists.)

Moreover, we showed above for the mode $l=2$ (in fact, for the
mode $l=1$ as well) that in principle the neutral modes are absent
as the resulting value of $Q$ turns out to be negative.
\medskip

In order to clarify the possibility of neutral modes $l\ge 3$ in
more detail, we consider the series of specific models in the form
of one-hump distributions
 $ \
 f_0^{(n)}(\nu)=N_n(\nu^2)^n\,\exp\left(-{\nu^2}/{\nu_T^2}\right),$
where $N_n$ is the normalized coefficient, $n=1,2,...$ . On the
assumption that the DF is normalized to some total mass ${\bar M}$
by the condition
 \begin{align}\label{eq:4.49}
    \int_0^{\infty} \nu\,d\nu f_0^{(n)}={\bar M},
 \end{align}
then $ N_n=\bigl[2{\bar M}/(\nu_T^2)^{n+1}\bigr](n!)^{-1}$.
Dimensionless variables are introduced for convenience: $
x={\nu^2}/{\nu_T^2},\ \ x_0={\omega_0^2}/{\nu_T^2}$. Besides, we
shall use the function $f_n(x)=f_0^{(n)} \left(\nu_T \sqrt{x}
\right) = N_n\,(\nu_T^2)^n\,\, x^n\,e^{-x}$, instead of the
function $f_0^{(n)}(\nu)$. In new designations, the set of
equations (\ref{eq:4.46}) and (\ref{eq:4.47}) takes the form
 \begin{align}\label{eq:4.50}
    f_n'(x_0)+{\ccase{5}{27}}\,f_n'\left({\ccase{1}{9}}\,x_0\right)=0,
 \end{align}
 \begin{align}\label{eq:4.51}
    Q_c=-\int\limits_0^{\infty}\hspace{-10.1pt}-\ dx\,\ x\,\frac{d\,f_n(x)}{dx}\left(\frac{1}
    {x_0-x}+\frac{15}{x_0-9x}\right),
 \end{align}
where
 \begin{align}\label{eq:4.52}
    f_n'(x)= (N_n\,\nu_T^{2n})\, x^{n-1}(n-x)\,e^{-x}.
 \end{align}
One can see from (\ref{eq:4.52}) that the maximum of the function
$f_n(x)$ is located at $x=n$, so that (\ref{eq:4.45}) gives the
condition on the value of dimensionless frequency squared: $
n<x_0<9n$. The value $x_0$ is found from Eq. (\ref{eq:4.50}), that
takes the form of a transcendental equation
 \begin{align}\label{eq:4.53}
    (9n-x_0)-{\ccase{1}{5}}\,3^{2n+3}\,\exp\left(-{\ccase{8}{9}}\,x_0\right)\,(x_0-n)=0.
 \end{align}
It is evident that at least one root always exists as the left
side of Eq. (\ref{eq:4.53}) has opposite signs at the ends of the
interval under consideration.  More detailed calculations (or
plotting the left side) show that there are actually three roots
satisfying the condition $ n<x_0<9n$, for each $n$. All roots are
candidates for a possible neutral mode. The numerical solution of
Eq. (\ref{eq:4.53}), for $n=1,...,7$, gives squared dimensionless
frequencies  of neutral modes listed in Table 2.

\begin{table}
\begin{center}
 \begin{tabular}{|c|c|c|c|}
\hline
 $n$&$x_0^{(1)}$&$x_0^{(2)}$&$x_0^{(3)}$\\
 \hline
 1&1.66&3.44&8.85\\
 2&2.27&5.343339&17.99\\
 3&3.09&7.733736&26.99\\
 4&4.03&10.17&35.99\\
 5&5.01&12.63&44.99\\
 6&--&15.09&--\\
 7&--&17.55&--\\
  \hline
 \end{tabular}
\end{center}
 \caption{\small The values of squared dimensionless frequencies --
 candidates to neutral mode.}
 \label{Tab 2}
 \end{table}

We next substitute the obtained values of $x_0$ into the relation
(\ref{eq:4.51}) that determines $Q_c$. Let us represent it in the
form convenient for numerical calculations. To do this, we
introduce the function $g_n(x)$:
 \begin{align*}
    g_n(x)=e^{-x}\int\limits_{-\infty}^x\hspace{-11.8pt}-\
    \frac{dt}{t}\,\,e^{\,t}\,\,
    [(x-t)^n\,(x-t-n)],\ \ \ \ x>0.
 \end{align*}
Using this function, Eq. (\ref{eq:4.51}) can be written as
$$\ {\bar
Q}_c^{(n)}=g_n(x_0)+{\ccase{5}{3}}\,\,g_n\left({\ccase{1}{9}}\,{x_0}\right),
$$
where we denoted
  $\  {\bar Q}^{(n)}\equiv {Q}/[{\nu_T^{2n}\,N_n}]=n!\,({Q}/{2{\bar
    M}})\,\nu_T^2,$
and the quantity ${\bar M}$ (see (\ref{eq:4.49})) is determined by
the normalization requirement for the monoenergetic model,
(\ref{eq:4.41}), i.e.,
  $$\  {\bar
  M}={\ccase{1}{2}}(2\pi)^{-3}\,\varpi(E_0)^2\Omega_1(E_0)\,M_G.$$
Substituting all three roots $x_0^{(1)}, x_0^{(2)}, x_0^{(3)}$
into (\ref{eq:4.51}), we obtain the following results.
\medskip

(i). For the DF with $n=1$, the right side of (\ref{eq:4.51}) turns out to be
negative for each potential neutral mode. This means that the mode
$l=3$ has no neutral modes (consequently, the mode is stable) for the model
$n=1$.

\medskip
(ii). For the models with $n\ge 2$, there is one neutral mode. It corresponds to
the middle root $x_0$ (in Table 2, it is the root $x_0^{(2)}$). Only for this
root, the right side of (\ref{eq:4.51}) turns out to be positive. This means that
these models can be unstable if the parameter $Q$ differs from the critical value
we found. Recall that the deviation direction of $Q$ is yet to be determined. The
critical values $Q_c^{(n)}$ are presented in Table 2.

\begin{figure}
 \begin{center}
 \includegraphics[bb=81 181 550 666, width=60mm, clip=true, draft=false]{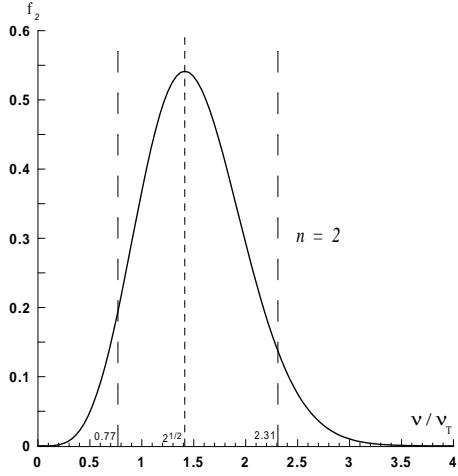}
 \end{center}
 \caption{\small Positions of resonances for the mode
$l=3$ for DF $f_{n=2}(\nu)\propto\,\nu^4\,\exp(-\nu^2/\nu_T^2)$.
The frequency of a neutral mode is
$\omega_0=\sqrt{5.34}\,\nu_T=2.31\nu_T$. The DF is maximum in the
point $\nu=u=\sqrt{2}\,\nu_T$.}
 \label{fig8}
\end{figure}

Fig.\,\ref{fig8} shows the resonances for the neutral mode in the model with
$n=2$, for illustration. As is seen in Table 1, the dimensionless frequency
$\omega_0/\nu_T$ is equal to $\sqrt{5.34}\approx 2.31$. So the resonances lie at
 $\nu/\nu_T=\omega_0/(3\nu_T)=0.77$ and $\nu/\nu_T=\omega_0/\nu_T=2.31$.

We are thus convinced that the neutral modes $l\ge 3$ (and consequently
the loss-cone instability) indeed exist for one-hump distributions for
suitable parameters ($n\ge 2$).

\subsection{The perturbation theory near $Q={\bar Q}_c^{(n)}$.
The instability criterion.}

Let us impose an increment $\delta{\bar Q}$ on the parameter
${\bar Q}$ and calculate a correction to the squared dimensionless
frequency, $\delta x_0$, using the perturbation theory. Then
the instability corresponds to the positive imaginary part of $\delta
x_0$. Indeed, $\delta
x_0=\delta(\omega^2)/\nu_T^2=2\omega_0\delta\omega/\nu_T^2$, and
the signs of imaginary parts of $\delta x_0$ and $\delta\omega$
coincide as $\omega_0>0$ by agreement.

Let us write the characteristic equation in the form
(\ref{eq:4.44}), using the new variables $x=\nu^2/\nu_T^2$,
$x_0=\omega_0^2/\nu_T^2$ and a new function ${\bar
f}_n(x)=x^ne^{-x}$.
 \begin{align}\label{eq:4.57}
    {\bar Q}^{(n)}=\int\limits_0^{\infty}\ dx\,\ x\,
    \frac{d\,{\bar f}_n(x)}{dx}\left(\frac{1}
    {x-x_0}+\frac{5/3}{x-\frac{1}{9}\,x_0}\right),
 \end{align}
where $ {\bar f}_n^{\,'}(x)= x^{n-1}(n-x)\,e^{-x}$. We find
 \begin{align*}
    \delta{\bar Q}^{(n)}=\delta x_0\int\limits_0^{\infty}\
    dx\,\ x\,\frac{d\,{\bar f}_n(x)}{dx}\Biggl[\frac{1}
    {(x-x_0)^2}+\frac{5/27}{\left(x-\frac{1}{9}\,x_0\right)^2}\Biggr],
 \end{align*}
where the integration involves bypassing the singularities from
below. Now we apply the relation useful for integrating the expressions with
peculiarities of the type $(x-x_0)^{-2}$ (with indentation due to
the singularity in the complex plane):
 \begin{align}\label{eq:4.59}
    \int\limits_a^b\frac{F(x)}{(x-x_0)^2}\,dx ={\rm FP}\int\limits_a^b
    \frac{F(x)}{(x-x_0)^2}\,dx +i\pi F'(x_0),
 \end{align}
where $a<x_0<b$ and FP means ``Finite Part'' of the integral. It is easy to
reproduce the derivation of this formula when we recall the definition:
$$
{\rm FP}\!\!\int\limits_a^b \!\!\frac{f(x)\,dx}{(x-x_0)^2}\equiv
\lim\limits_{\eps\to 0} \Bigl[\!\int\limits_a^{x_0-\eps}\!\!\!\!
\frac{f(x)\,dx}{(x-x_0)^2}\,+\!\!\int\limits_{x_0+\eps}^b\!\!\!\!
\frac{f(x)\,dx}{(x-x_0)^2}-\frac{2\,f(x_0)}{\eps}\Bigr],
$$
where $a<x_0<b$, and the function $f(x)$ is regular at $x_0$. Evidently, the
direction of bypassing (either from below or above) has no influence on the real
part of the result (i.e., the value of the FP integral). Changing the direction
of indentation, we change only the sign of the imaginary part in (\ref{eq:4.60}).
[Note that the concept of FP integral is well-known in hydrodynamics.
 It is widely used in problems
 related to the so-called critical layer, i.e.,
 a narrow domain near the resonance of the wave and
 the shear flow of fluid (see, e.g., Hickernell 1984, or Shukhman 1991).]
We obtain
 \begin{multline}\label{eq:4.60}
    \delta{\bar Q}^{(n)}=\delta x_0\Biggl\{i\pi\Bigl[\Bigl(x\frac{d{\bar
    f}_n(x)}{dx}\Bigr)_{x=x_0}'+{\ccase{5}{27}}\,\bigl(x\frac{d{\bar
    f}_n(x)}{dx}\bigr)_{x=x_0/9}'\,\Bigr]+
    \\
    +{\rm FP}
    \int_0^{\infty}\
    dx\,\ x\,\frac{d\,{\bar f}_n(x)}{dx}\Bigl[\frac{1}
    {(x-x_0)^2}+\frac{5/27}{\bigl(x-\frac{1}{9}\,x_0\bigr)^2}\Bigr]\Biggr\}.
 \end{multline}
The imaginary part (\ref{eq:4.59}) can be simplified using the
relation (\ref{eq:4.50}), reflecting the balance between growth
and damping for a neutral mode. Then we obtain
 \begin{multline}\label{eq:4.61}
    \delta{\bar Q}^{(n)}=\delta x_0\Biggl\{i\pi x_0\left[{\bar
    f}_n''(x_0)+{\ccase{5}{243}}\,{\bar
    f}_n''\left({\ccase{1}{9}}\,x_0\right)\right]\\
    +{\rm FP}
    \int_0^{\infty}\!\!\!
    dx\,\ x\,{\bar f}_n'(x)\Bigl[\frac{1}
    {(x-x_0)^2}+\frac{5/27}{\left(x-\frac{1}{9}\,x_0\right)^2}\Bigr]\Biggr\}.
 \end{multline}
Writing (\ref{eq:4.61}) in the form $\ \delta{\bar Q}^{(n)}=(A_n+i
B_n)\,\delta x_0, $ where
 $$
    A_n={\rm FP}
    \int\limits_0^{\infty}
    dx\,\ x\,{\bar f}_n^{\,'}(x)\Biggl[\frac{1}
    {(x-x_0)^2}+\frac{5/27}{\left(x-\frac{1}{9}\,x_0\right)^2}
    \Biggr],$$
    $$
    B_n=\pi x_0\left[{\bar
    f}_n^{\,''}(x_0)+{\ccase{5}{243}}\,{\bar
    f}_n^{\,''}\left({\ccase{1}{9}}\,x_0\right)\right],
 $$
we find for real and imaginary parts of $\delta x_0$:
 \begin{align}\label{eq:4.62}
  {\rm Re}\,(\delta x_0)=\frac{A_n}{A_n^2+B_n^2}\,\delta{\bar Q}^{(n)}, \ \
  {\rm Im}\,(\delta x_0)=-\frac{B_n}{A_n^2+B_n^2}\,\delta{\bar Q}^{(n)}.
 \end{align}
We see that the stability criterion is determined only by the sign of $B_n$.
Calculating the growth rate itself requires
the values of both quantities, $A_n$ and $B_n$. Using the functions
\begin{align*} u_n(x)=x\,{\bar
f}_n{''}(x)=x^{n-1}\,\Bigl[(n-1-x)\,(n-x)-x\Bigr]\,e^{-x}, \\
 h_n(x)=e^{-x}\,{\rm FP} \int\limits_{-\infty}^x\frac{dt}{t^2}\,e^t\,
(x-t)^n(n-x+t),\ \ \ \ x>0,
\end{align*}
the result can be written in a compact
form
$$
 A_n=h_n(x_0)+{\ccase{5}{27}}\,h_n\left({\ccase{1}{9}}\,x_0\right),\
 B_n=\pi\,\Bigl[u_n(x_0)+{\ccase{5}{27}}\,u_n\left({\ccase{1}{9}}\,x_0\right)\Bigr].
$$
The values of $A_n$ and $B_n$ are calculated numerically, for the values of $x_0$
found above. They are presented in Table 3. It is seen from this Table that $B_n$
is positive for all values of $n$. Consequently, the instability occurs when
$\delta {\bar Q}^{(n)}<0$, or $\ {\bar Q}^{(n)}<{\bar Q}^{(n)}_c,\,$ where the
critical values of ${\bar Q}^{(n)}_c$ are presented in Table 3. If we recall the
definition of ${\bar Q}^{(n)}=n!({Q}/{2{\bar M}})\,\nu_T^2$, the instability
condition of the mode $l=3$  for the monoenergetic model can be reformulated in a
form of criterion on the ratio of the angular momentum dispersion,
$L_T=\nu_T/\varpi$,  to the angular momentum of a star with the same energy $E_0$
in a circular orbit, $L_{\rm circ}=\sqrt{\frac{1}{2}\,GM_cR}$:
\begin{align}\label{eq:4.63}
    \frac{L_T}{L_{\rm circ}}<\frac{\pi}{4}\sqrt{\frac{6C_3\,{\bar
    Q}_c^{(n)}}{n!}},\ \ \ \ C_3\approx 0.0373
 \end{align}

Particularly, for the model with $n=2$, when ${\bar Q}_c^{(2)}=0.5426$, we obtain
from (\ref{eq:4.63})
 ${L_T}/L_{\rm circ}<0.193$, and for the model with $n=3$, when ${\bar
 Q}_c^{(3)}=3.4866$ we obtain ${L_T}/L_{\rm circ}<0.283$.
Note that the criterion in such a form does not involve mass of the spherical
component, $M_G$.\footnote {Since these critical values $(L_T/L_{\rm circ})_{\rm
crit}$ proved to be not too small, it means that in accepted ``spoke
approximation'' (where it is supposed that $L/L_{\rm circ}\ll 1$) instability
criterion is knowingly satisfied. However the rigorous calculation of the
instability boundary in terms of parameter $L_T/L_{\rm circ}$ requires an exit
from a framework of this approximation. This problem is under consideration now.}

\begin{table}
\begin{center}
 \begin{tabular}{|c|c|c|c|c|c|}
\hline
 $n$&$x_0$&${\bar Q}_c^{(n)}$&${\bar Q}_c^{(n)}/n!$&$A_n$&$B_n$\\
 \hline
  2&5.343339&0.542582&0.2713&0.188022&0.732157\\
 3&7.733736&3.486635&0.5811&0.218900&1.884489\\
 4&10.17&15.6856&0.6536&0.08&5.46\\
 5&12.63&74.7372&0.6228&-1.40&18.32\\
 6&15.09&399.7582&0.5552&-11.86&70.95\\
 7&17.55&2425.3916&0.4812&-\,83.65&312.89\\
  \hline
 \end{tabular}
\end{center}

 \caption{\small Dimensionless frequency squared $x_0$,
 the value of critical parameter ${\bar Q}_c^{(n)}$
 and the values of $A_n$ and $B_n$
 in the expression for complex frequency squared
 $\delta x_0\equiv\delta(\omega^2/\nu_T^2)$.
The unperturbed DF is $f_n(x)=(N_n\nu_T^{2n})\,{\bar f}_{n}(x)$,\
\ ${\bar f}_{n}(x)=x^n\exp\,(-x)$. (For $n=1$ there are no neutral
modes.)}
 \label{Tab 3}
 \end{table}

\begin{figure}
\includegraphics[bb=62 315 579 581, clip=true, width=78mm]{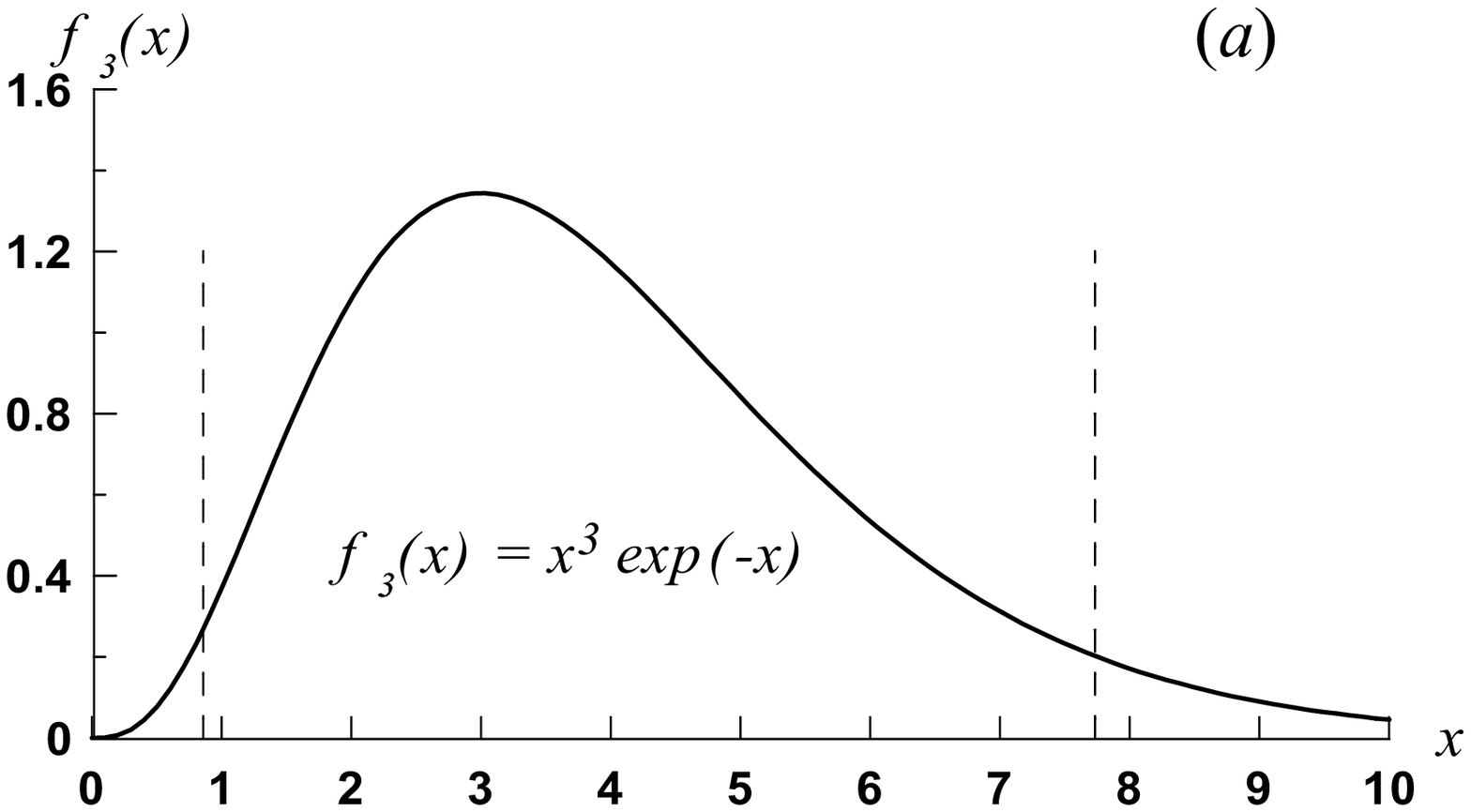}
\includegraphics[bb=54 202 537 709, clip=true, width=78mm]{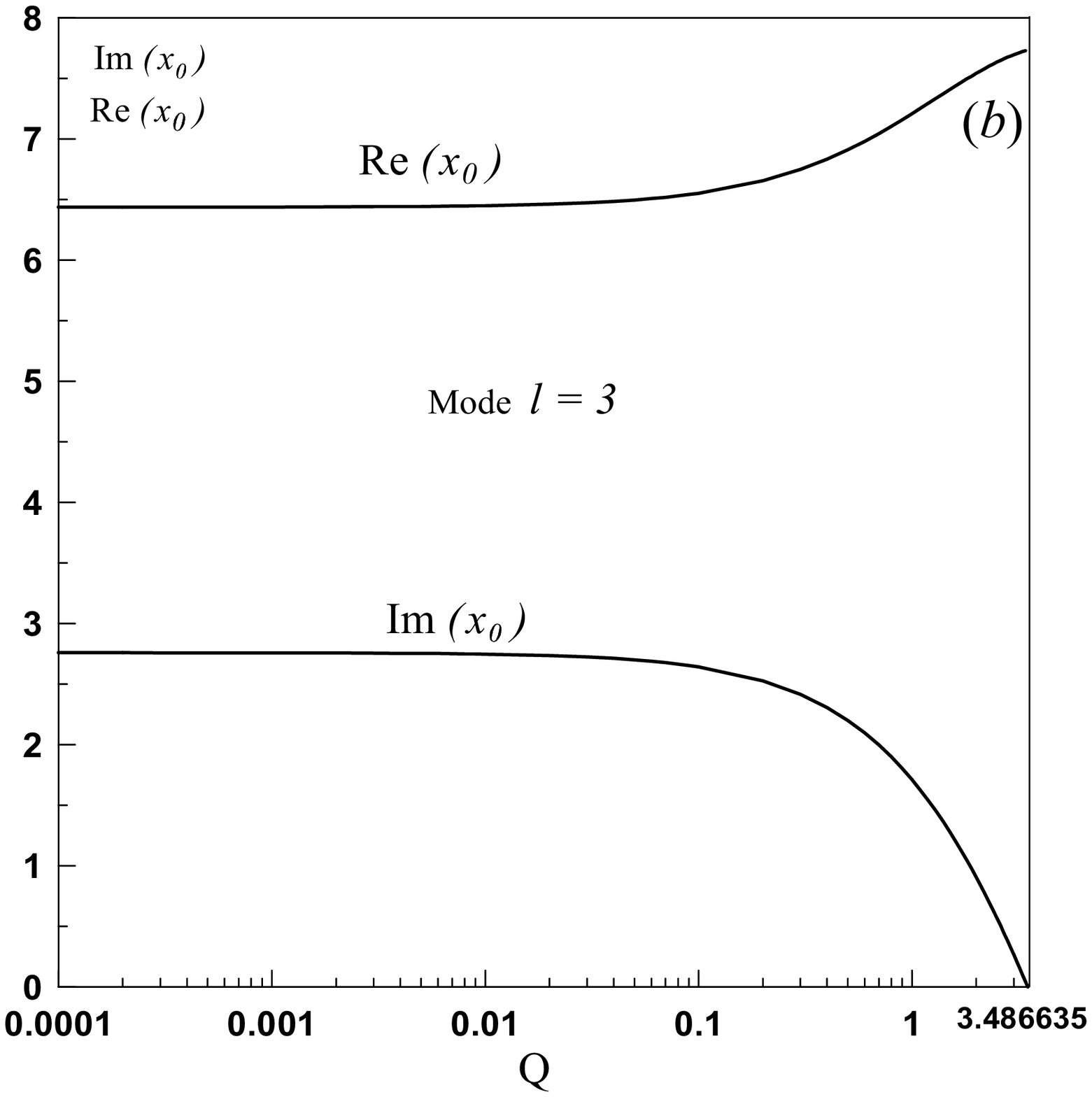}
\caption{\small Loss-cone instability in spherical system.
\textit{a} -- DF over precession angular velocity
 ${\bar f}_{n=3}=x^3\,\exp(-x)$.
 The dash lines indicate positions of resonances for neutral mode.
(Squared dimensionless frequency  is  $x_0=7.733736$.);
 \textit{b} -- Behavior of squared dimensionless frequency $x_0$
 of unstable mode.  The values of ${\rm Re}\,(x_0)$ and ${\rm Im}\,(x_0)$ are shown
against parameter ${\bar Q}$.}
 \label{fig9}
\end{figure}

When the supercriticality of $\delta Q$ is not small,
the perturbation theory above does not
allow us to calculate the complex eigenfrequency. In this
case, the characteristic
equation (\ref{eq:4.57}) was solved numerically, for values of $n=2,\ldots, 5$.
The qualitative behavior of real and imaginary parts of the frequency is similar
for all calculated cases. So we restrict our illustrations only to the model
with
$n=3$ (see Fig.\,\ref{fig9}). Note also that the results of computations, for
small deviations from the stability boundary, coincide with the asymptotic
results obtained using the perturbation theory (\ref{eq:4.62}).

\subsection{General instability criterion and study of specific distributions}

The above results, based on the neutral mode approach, can also be
obtained using a suitable analogue by means of the well-known
Penrose\,--\,Nyquist theorem (Penrose 1960, Mikhailovsky 1974).
Recall that this theorem is widely used in plasma physics.
Employing the theorem helped to establish numerous general results
in the theory of plasma instabilities.

First we represent our characteristic equation in the form
\begin{align}\label{eq:59}
{Q}=C_l\int\limits_0^{\infty}
dx\,x\,\frac{d\,{f}(x)}{dx}\sum\limits_{s\,=\,s_{\rm min}}^{l}\frac{D_l^s}
 {x-z/s^2},
\end{align}
where the quantity $Q$ is independent of $l$. In the case of retrograde
precession we are interested in, $Q>0$. Recall
that $s_{\rm min}$ is equal to 1 or 2
depending on evenness of $l$, $x=\nu^2$ is the precession angular velocity
squared, in dimensionless units (the units of $\nu_T$, where $\nu_T$ is the
precession velocity dispersion, is common), $f(x)$ is the unperturbed DF,
$z=\omega^2$ is the frequency squared, in the same dimensionless units.

We are also reminded that the singularity of the integral in the
right side of (\ref{eq:59}), for $z$ on the real axis, must be
bypassed from below if ${\rm Re}\,(\omega)>0$, and above if ${\rm
Re}\,(\omega)<0$. From this indentation rule and the form of Eq.
(\ref{eq:59}), it immediately follows that complex unstable roots,
$z_0$, form pairs: if $z=z_0=a+i\,b,\ \ a\ne 0,\ \ b>0$ -- the
root of Eq. (\ref{eq:59}), and the corresponding eigenfrequency is
$\omega=\omega_0=\alpha+i\,\beta$, ($\alpha > 0,\ \ \beta>0$), the
complex conjugate root, $z={\bar z}_0=a-i\,b$, also satisfies Eq.
(\ref{eq:59}) and describes the mode with the same growth rate,
but opposite sign of frequency, $\omega=-{\bar
\omega}_0=-\alpha+i\,\beta$.

Note that the form of Eq. (\ref{eq:59}) shows that the aperiodic
instability, for which ${\rm Re}\,(z_0)<0, \ {\rm Im}\,(z_0)=0$,
i.e., $a=0$, is absent here. Indeed, putting $z_0=-|z_0|<0$ in
(\ref{eq:59}) and integrating by parts, it is easy to verify that
the right side of (\ref{eq:59}) is negative. By the way, this is a
further distinction of Eq. (\ref{eq:59}) from the disk
characteristic equation that allows aperiodic instabilities for
symmetric distributions $f(\nu)$.

Eq. (\ref{eq:59}) involving $[\frac{1}{2}\,(l+1)]$ items can be reduced to a
one-term equation. Indeed, the  substitution
\begin{align}\label{eq:61}
F^{(l)}(x)=C_l\sum\limits_{s\,=\,s_{\rm min}}^{l}{D_l^s}\,f\bigl({x}/{s^2}\bigr).
 \end{align}
transforms (\ref{eq:59}) into the equation
\begin{align}\label{eq:60}
{Q}=\int\limits_0^{\infty}dx\,x\,\frac{d\,{F}^{(l)}(x)/dx}{x-z}.
\end{align}
Thus, for arbitrary values of $l$, we obtain a one-item equation
(\ref{eq:60}) similar to that for the mode $l=2$ (or $l=1$).
However, now the integral involves the function $F^{(l)}(x)$,
instead of the initial function $f(x)$ (with one maximum and
tending to zero at $x=0$ and infinity). Starting with modes $l=3$,
the new function, $F^{(l)}(x)$, can have minima. It is easy to
understand that the frequencies (candidates for a neutral mode)
calculated in subsection 4.6 from the condition of balance between
growth and damping at resonances on different sides of the maximum
of the initial function $f(x)$, are precisely those coinciding
with the extrema of the new function, $F^{(l)}(x)$, i.e., $z=x_j$,
$F'(x_j)=0$. Correlating this with earlier results for disks, and
also with the sphere DF $f_n(x)=x^n \exp\,(-x)$, we become
convinced that the largest (more often, the only) positive
critical value of $Q_c$ for the neutral mode should necessarily be
related to a {\it minimum} of the DF $F^{(l)}(x)$. As an
illustration of this statement, Fig. 10 shows three functions
$F^{(l)}_n(x)$ of this series, for the mode $l=3$. From the figure
(and also Table 2), it is seen that only the central of these
three extrema, i.e., the minimum, gives rise to the neutral mode
with positive values of $Q_c$ (for $n\ge1$).

\begin{figure}
\begin{center}
\includegraphics[bb=64 181 534 644, clip=true, width=78mm] {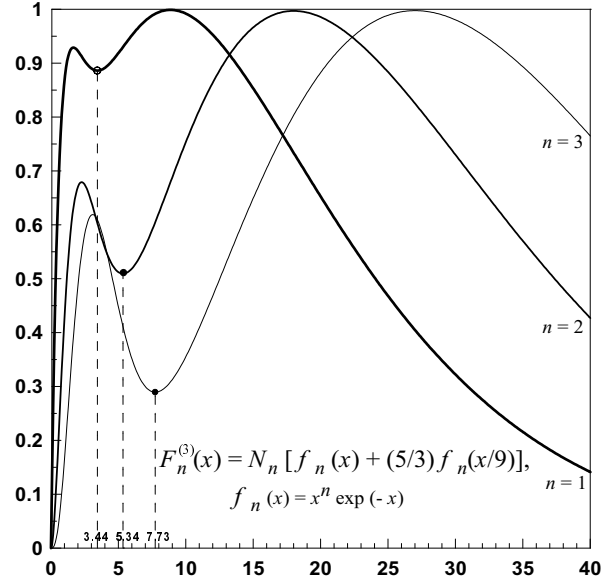}
\end{center}
\label{Fig.10} \caption{\protect\footnotesize -- Function $F_n^{(l)}(x)$ for a
model $f(x)=x^n e^{-x}$ with $l=3$ ($n=1,\,2$ and 3).}
\end{figure}

Thus, we see that the availability of minima is of fundamental
importance for the existence of neutral modes with positive $Q_c$
(and consequently for instability). We have already seen that for
$l=1$ and $l=2$, when the DF $F^{(l)}(x)$ coincides with the
initial DF $f(x)$, so that the former has no minima, the
corresponding neutral modes (and instability) are absent.

Recall that the possible-in-principle neutral mode with $z=0$
corresponding to the resonance at the minimum $x=0$, has $Q_c=0$,
so that it cannot be assumed to be a candidate for a neutral mode
with the property required for instability ($Q_c>0$). Therein lies
a fundamental difference from disks where any two-hump
distribution with a zero minimum at $\nu=0$ always has the neutral
mode with $Q_c>0$ at the minimum. So such a distribution is always
unstable (when $Q<Q_c$) independently of other DF details.

We have a right to expect a neutral mode (and instability) related
to the minimum, for $l\ge3$ only. In fact, it has already been
demonstrated above using a somewhat more cumbersome method.
Besides, a question remains unsolved in the approach we apply: why
are not all distributions with one maximum that vanishes at the
ends of the positive semi-axis $0\le x<\infty$ generally unstable,
even though $ l\ge 3$. Empirically, by considering various series
of distribution functions, we found the qualitative instability
condition. Its rough formulation is: the instability is possible
if the DF function is well-localized around its maximum.

Now a possibility appears for a more rigorous (in fact,
quantitative) formulation of the instability condition. Though Eq.
(\ref{eq:60}) differs from the equation
\begin{align}\label{eq:62}
 {Q}=\int\limits_{-\infty}^{\infty}\frac{f'(v)\,dv}{v-c},
\end{align}
(where $c=\omega/k$ is the complex phase velocity,
$Q=k^2/\omega_0^2>0$, $k$ is the wave number, $\omega_0^2=4\pi n_0
e^2/m$ is the plasma frequency squared), for which  \cite{Pen}
obtained his well-known criterion, here we also can obtain an
analogous criterion -- i.e. a counterpart of the
Penrose\,--\,Nyquist criterion, for our equation (\ref{eq:60}).
First we formulate it in terms of neutral modes.

{\bf Theorem}. The distribution $F^{(l)}(x)$ is stable if neutral modes
corresponding to minima of $F^{(l)}(x)$ are absent. Alternatively, if at least
one neutral mode corresponding to a minimum occurs, then a sufficiently small $Q$
always exists, for which the system will be unstable relative to perturbations
with a given $l$.

The instability condition for any $l$ follows immediately from the
theorem. Indeed, if for at least one of $l$ ($l=1,2,\ldots$), a
neutral mode exists for the corresponding distribution
$F^{(l)}(x)$, then such a sufficiently small $Q$ occurs, for which
the system is unstable.

It is useful to give another formulation of the theorem with a
maximally possible similarity to that of \cite{Pen} for Eq.
(\ref{eq:62}).

{\bf Theorem} (another formulation). The distribution $F^{(l)}(x)$
is stable if and only if for all points $x_j$, at which the
modified DF $F^{(l)}(x)$ has a minimum (i.e., $F'(x_j)=0,\ \
F''(x_j>0$), the condition
\begin{align}\label{eq:63}
{Q_c}\equiv\int\limits_0^{\infty}dx\,x\,\frac{d\,{F}^{(l)}(x)/dx}
 {x-x_j}<0.
\end{align}
is met. Conversely, if at least for one minimum the opposite
inequality is satisfied, then such a sufficiently small $Q$
exists, for which the system is unstable for perturbations with a
given $l$.

The proof of this theorem can be found in Appendix B, but now we
discuss a correlation between the instability condition following from this
criterion and our qualitative condition formulated above.

From the results obtained for disks, we know that under
sufficiently deep minimum, the corresponding $Q_c$ can become
positive. Thus, it can rigorously be shown that $Q_c>0$ (with a
finite margin) in the limit when the minimum is exactly equal to
zero.

Indeed, let us assume that $x_j$ is the position of minimum, at
which $F(x_j)=0$, $F'(x_j)=0$. Then we obtain by integrating in
(\ref{eq:63}) by parts
$$ {Q_c}\equiv\int\limits_0^{\infty}dx\,x\,\frac{d\,{F}^{(l)}(x)/dx}
 {x-x_j}=x_{j}\int\limits_0^{\infty}dx\,\frac{d\,{F}^{(l)}(x)/dx}
 {x-x_j}=$$
 \vspace{-14pt}
 $$
 =x_j\int\limits_0^{\infty}dx\,\frac{{F}^{(l)}(x)}
 {(x-x_j)^2}>0.
   $$

It becomes impossible to prove in a similar manner that the
integral is also positive for a non-zero minimum. Actually, it can
have any sign. However, positive contributions into the integral
increase the closer the minimum is to zero. So the integral should
eventually become positive with increasing depth of minimum.

In light of this fact, it becomes clear that our qualitative
instability condition means that a minimum of $F^{(l)}(x)$ is
sufficiently deep, so that the related neutral mode occurs. This
becomes evident when we consider how the function $F^{(l)}(x)$ is
built from the initial DF $f(x)$. The instability is unavailable
if a minimum is absent or is not sufficiently deep. In Fig.\,10,
the functions $F_n^{(l)}(x)$ constructed from the functions $
f_n(x)=x^n\,e^{-x}$, for the mode $l=3$, are for convenience
calibrated so that the value of the highest maximum is equal to
unity for all values of $n$. We see that the only minimum becomes
deeper with increasing $n$. This is in complete agreement with the
results of Sec. 4.7 where we found that the mode $l=3$ is stable
for $n=1$ and unstable for $n \ge 2$. (For completeness we
additionally checked that instability emerges when $n>1.55$).
Moreover, we checked that the instability in the model with $n=1$
is also absent for any $l$.

The formulated criterion allows a purposeful search for such DF
$f(x)$ which gives  a new function $F(x)$ with a minimum capable
of `generating'' a neutral mode. In other words, the integral
(\ref{eq:63}),
\begin{align}\label{eq:64}
Q_c=\int\limits_0^{\infty}dx\,x\,\frac{d\,{F}^{(l)}(x)/dx}
 {x-x_j}\equiv x_j\int\limits_0^{\infty}dx\,\frac{F^{(l)}(x)-F^{(l)}(x_j)}
 {(x-x_j)^2},
 \end{align}
is positive, so that the instability occurs when $Q<Q_c$. It turns
out that suitable distributions are known in plasma physics. In
particular, \cite{Pen} has pointed to a case of such a
distribution. Namely, he demonstrated that a plasma distribution
becomes unstable provided this distribution has a sufficiently
sharp minimum (then the Penrose integral similar to (\ref{eq:64})
becomes positive). For instance, such a minimum appears at the
electron DF when  a sufficiently cool electron beam is injected
into the Maxwellian plasma, provided the beam velocity is larger
than the electron thermal velocity of main plasma.

It is interesting that distributions with sharp minima appear in
our problem quite naturally. Indeed, let us assume that the star
distribution with respect to angular momenta (or, which is the
same, to precession angular velocities) is Gaussian. In terms of
the variable $x=\nu^2/\nu_T^2$, it has the form $f(x)= N\,e^{-x}$,
$0<x<\infty$, $N$ is the normalized constant. Now we suggest that
the stars enclosed by the loss cone elude the distribution so that
the resulting distribution arises:
 \begin{align}\label{eq:65}
 f(x)=N\,H(x-a)\,e^{-x},
  \end{align}
where $H(t)$ is the Heaviside step function. Any physically
admissible distribution should of course be smooth. So, instead of
discontinuous function (\ref{eq:65}), we assume a nearly identical
(but continuous and smooth) distribution
\begin{multline}\label{eq:66}
f(x)=N\,{\cal R}_a(x)\, e^{-x},\\
 {\cal R}_a(x)={\case{1}{2}}\left[\tanh\,
\Bigl(\frac{x-a}{\delta}\Bigr)+\tanh\,\Bigl(\frac{a}{\delta}\Bigr)\right],
\end{multline}
where  $\delta \ll a$ in the ``cutting factor'' ${\cal R}_a(x)$.
 The function
 \[
 F(x,a)=
\frac{1}{16}\,C_3\,\bigl[3\,f(x)+{5}\,f(\case{1}{9}\,x)\bigr]
 \]
for the mode $l=3$, corresponding to the DF (\ref{eq:66}) with
$a=0.1$, $a=0.01$ and $\delta=\frac{1}{20}\,a$, is shown in Fig.
11\,({\it a}). It is seen that the distribution $F(x)$ has only
one minimum, this minimum being sufficiently sharp. Direct
calculations show that the neutral mode ($Q_c>0$) corresponding to
the minimum occurs under an arbitrarily small size of a ``slot''
(i.e., a value of $a$). Fig. 11\,({\it b}) shows the marginal
curve on the plane $(Q-a)$, where the modes with arbitrary values
of $l$ are taken into account. It is seen that for not too large
values of $a$  the boundary of instability is nevertheless
actually  determined by the first unstable mode $l=3$ only.

\begin{figure}
\begin{center}
 \includegraphics[bb=56 90 552 796, clip=true, width=78mm]{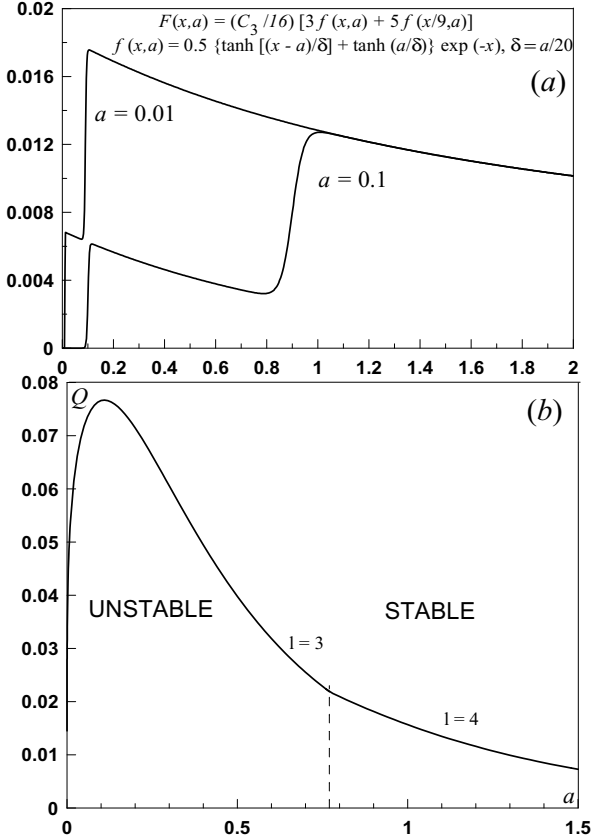}
\end{center}
 \caption{\protect\footnotesize ({\it a}) -- Function
$F(x,a)=\frac{1}{16}\,C_3\,[3\,f(x)+5\,f(\frac{1}{9}\,x)]$  for
the model (8) with $l=3$ and $a=0.01$, $a=0.1$ ($\delta =a/20$).
 ({\it b})-- the marginal
curve on the $(Q-a)$ -- plane, calculated for all $l$.}
 \label{fig11}
\end{figure}

Thus, we see that an empty loss cone, even if it is very
narrow, inevitably leads to the instability, for suitable
distributions (i.e., the dispersion $\nu_T$ is less than the
critical value $(\nu_T)_c$ determined by the parameter $Q_c$.)

\section{Discussion}

First we list the results of the paper.

1. The paper presents a systematic derivation, from general
linearized Vlasov equations (written in the action-angle
variables), of simple characteristic equations for small
perturbations in disk and spherical stellar systems with
near-radial orbits.

2. On the one hand, our analysis of these characteristic equations
confirms the presence (already discussed earlier, Polyachenko
1991b, Tremaine \citeyear{T05}) of gravitational loss-cone
instability in disks. On the other hand, we succeeded to prove, in
this paper, a possibility of this instability in spherical
clusters.

3. It is shown that the physical reason of the instability under
consideration is escape of stars through the loss cone due to
destruction of stars with sufficiently low angular momenta. As a
result, the DF in angular momenta will assume an unstable
(``beam-like'') form. This is very similar to the situation in
plasma traps (mirror machines) where plasma particles with low
transversal (relative to the axis of the trap) velocities escape
those systems. For this reason, distribution over these
transversal velocities also becomes ``beam-like'', so that the
classical loss-cone instability develops.

4. We highlight retrograde precession of orbits as a necessary
condition for the gravitational loss-cone instability. Expressions
for precession velocity both in non-singular and near-Keplerian
potentials are derived. In particular, they helped to obtain
conditions for the the precession to become retrograde.

5. While deriving the characteristic equations, we justify the
obvious (and very convenient for practical use) rotating-spokes
approximation.

6. For analyzing the characteristic equations, a specific method
is developed. It is based on preliminary search of neutral modes.

7. We also developed a method based on generalization of the
plasma (and, in fact, gravitating disk) Penrose\,--\,Nyquist
theorem that establishes the criterion of stability and
instability. First, from an initial DF of stars in a spherical
cluster, $f$, we turn to another, effective DF, $F$. The latter is
constructed from $f$ according to a simple recipe (see the
beginning of Subsec. 4.8). Using this new function, the many-term
characteristic equation describing perturbations in spherical
systems reduces to the simplest one-term disk-like equation. In
turn this allows us to formulate and prove the above-mentioned
generalization of the Penrose\,--\,Nyquist criterion.

8. It is shown that this criterion allows us to justify the
following qualitative criterion of the gravitational loss-cone
instability: the instability is possible if the DF is
well-localized about its maximum. Using the criterion, we can
perform a purposeful search of unstable distributions. In
particular, we succeeded in proving an empty loss cone, even if
very narrow, to be able to lead to instability.

\medskip

Let us remind that for spherical near-Keplerian systems, Tremaine
does not find the instability, although his integral equation (55)
is equivalent to our ``slow'' equation (4.8), and the problem
definition is very similar to ours: To study the stability of
spherical models against comparatively slow perturbations (with
typical times of the order of inverse precession frequency)
provided that DF possesses positive derivative $\p F/\p L>0$ at
low angular momentum due to the loss-cone.

Still, there are two considerable differences. The first one is
that \cite{T05} has used Goodman's (1988) criterion, which is
fundamentally restricted to the modes $l \le 2$. The second one is
that Tremaine studied only monotone DF (from radial orbits $L=0$
up to circular orbits $L=L_{\rm circ}$). In that case the {\it
variational principle} takes place, which claims that squares of
eigen frequencies should be real (${\rm Im}(\omega^2)=0$). So the
unstable modes are aperiodic, if any. \cite{T05} has shown that
for the models considered ($\log$-models, see (98) in his paper),
stabilizing and destabilizing contributions cancel each other for
$l=1$ leading to neutrally stable lopsided mode
$\omega^2=0$,\footnote{The lopsided zero mode cannot be found with
our spoke-orbit approximation (4.15). To obtain the mode, one
should hold terms of the order of ${\cal O}(L^2)$ in the expansion
of functions in (4.8).} while for $l=2$ the stabilizing
contribution dominates.

In our case {\it non-monotone} DFs violate the variational
principle, so that the squares of eigen frequencies should not be
real. Moreover, we have shown (Sect. 4.8) that the aperiodic
instability is impossible, i.e. the eigenmodes are oscillating,
${\rm Re} (\omega)\ne 0$. We have found the instability at $l \ge
3$ for spherical models, if DFs have maximum somewhere in the
region $0<L<L_{\rm circ}$.\footnote{Strictly speaking, in the
spoke-orbit approach we use here, DFs describe almost radial
orbits, $L \ll L_{\rm circ}$. However, we believe that presence of
maximum, rather than high degree of orbit elongation is of
fundamental importance for the instability.} It is plausible that
we deal with instabilities of somewhat different nature.

By considering two-humped models in Sect. 3, we demonstrated that
the gravitational loss-cone instability can arise in disks with
nearly radial orbits. One can show that two-humped DFs is crucial
for the instability. Such distributions arise naturally from
originally Maxwell-like DFs after consumption of stars with low
angular momenta by a black hole. The instability takes place for
arbitrary azimuthal number $m$. Meanwhile \cite{T05} has
considered distributions in the interval $-L_{\rm circ}<L< L_{\rm
circ}$, which is symmetric around $L=0$. On the example of
lopsided mode $l=1$, he has shown that disk is generally unstable.
Note that condition (3.11) is implicitly suggested in his
derivation of instability criterion (Eq. (114) in \cite{T05}).

For the final conclusion on which of two differences is decisive
for the instability, i.e. monotonity or mode number $l \ge 3$,
some additional work is needed. We hope to attack this problem in
future using our ``slow'' equation, derived in Sect. 4. The
problem seems to be of importance since in some numerical models
(e.g., Cohn and Kulsrud, 1978) collisions result in the
establishment of distributions growing monotonically (from the
loss-cone boundary up to circular orbits). It is likely that such
distributions will prove to be stable (see recent N-body
experiments by \cite{B05}). On the other hand, we also emphasize
the fact that such a ``near-isotropy'' observed in some numerical
simulations is not an unambiguously established fact at present.
Not to mention that distributions with prevalent elongated orbits
can be quite natural in some circumstances (e.g., periods between
their formations due to, say, collisionless collapse, and the
moment of relaxation).

In the paper by \cite{B05} N-body simulations have been used to
model the loss cone in the vicinity of the (binary) black hole.
The model allowed them to construct the lose cone which was
maintained in nearly-empty state during simulations. However the
black hole feeding rate was at the level consistent with standard
collisionally-repopulated loss-cone theory. The discrepancy
between their and our results is probably explained by difference
in character of distribution over angular momentum. \cite{B05}
have assumed initially homogeneous system, which leads to the
near-isotropic one in the course of collision evolution (more
exactly, to the distribution slightly growing with $L$ if to
ignore a small region of empty loss-cone), whereas we have proved
instability for systems with nearly radial orbits. Yet, this
N-body modelling can be a weighty argument in favour of stability
of monotone distribution functions.

\medskip

In conclusion we present preliminary estimations of efficiency of
the collective mechanism under consideration. For the most
interesting, near-Keplerian, case, such estimates were made by
Tremaine (\citeyear{T05}), and we use these estimates below.

There are several characteristic time scales. The first is the
dynamical time, $t_{\rm dyn} \sim\Omega^{-1}\sim
(R^3/GM_c)^{1/2}$, where $R$ is the typical orbital radius, $M_c$
is the central point mass. The orbit precession determines, using
Tremaine's (\citeyear{T05}) terminology, the secular time scale,
$$
 t_{\rm sec} \sim t_{\rm dyn} \frac {M_c}{M_G} \sim t_{\rm dyn}
 \frac {M_c}{N_G m},
$$
where $ M_G$, $N_G$ and $m$ is the cluster mass, the number of
stars and the mass of one star, respectively. The gravitational
loss-cone instability develops precisely on this time scale (cf.
the formula (\ref{eq:4.37}) for the precession velocity in our
monoenergetic model, $\gamma={\rm Im}\,(\omega)\sim\Omega_{\rm
pr}= \varpi L\sim ({M_G}/{M_c})\,({L}/{R^2})\sim
\Omega\,({M_G}/{M_c})$.) The next important time scale defines a
period of collision relaxation,
$$
t_{\rm relax} \sim \frac {R^{3/2}M_c^{3/2}}{G^{1/2}m\,M_G} \sim t_{\rm dyn} \frac
{M_c^2 }{m\,M_G} \sim t_{\rm dyn} \frac {M_c^2}{m^2\,N_G}.
$$
These three time scales are well-known. \cite{T05} introduces another
(less known) time scale, citing his paper \citep{RT96} -- the time of resonance
relaxation of angular momenta,
$$
t_{\rm res} \sim \frac {R^{3/2}M_c^{1/2}}{G^{1/2}m} \sim t_{\rm dyn} \frac {M_c
}{m}.
$$

For near-Keplerian systems (when $M_G\ll M_c$, $N_G\gg 1$), these
four time scales are highly separated:
$$
t_{\rm dyn}\ll t_{\rm sec}\ll t_{\rm res}\ll t_{\rm relax}.
$$
Thus, according to these estimates by Tremaine, the instability
should grow faster than collisional (and resonant) relaxation,
whether or not a cluster mass is small.

Note, however, that the estimates of time scales presented here
are insufficient to claim that the collective mechanisms under
consideration should dominate. There is a need to calculate and
compare the star fluxes onto the black hole. In this connection,
we should remind the reader that so far we only attempted to prove
that the instability is possible in principle.

\section*{Acknowledgments}
We are grateful to A. M. Fridman who drew our attention to this
interesting problem. We thank Scott Tremaine for interesting
comments concerning the lopsided zero mode. We also express our
gratitude to S. M. Churilov for helpful discussion of issues
concerning the proof of the possibility of instability in
spherical systems, and to V. A. Mazur for stimulating interest in
our work.  The work was supported in part by Russian Science
Support Foundation, RFBR grants No. 05-02-17874 and No.
07-02-00931, ``Leading Scientific Schools'' Grant No. 7629.2006.2
and ``Young doctorate'' Grant No. 2010.2007.2 provided by the
Ministry of Industry, Science, and Technology of Russian
Federation, the ``Extensive objects in the Universe'' Grant
provided by the Russian Academy of Sciences, and also by Programs
of presidium of Russian Academy of Sciences No 16 and OFN RAS No.
16.

\subsection*{\large APPENDIX A. Derivation of the integral equation for
perturbations in a
spherically-symmetric gravitating system in terms of the action-angle formalism}
\medskip

\begin {center}
{\bf 1. The action-angle variables in a spherically-symmetric potential}
\end {center}
\medskip

Let us recall the action-angle variables  in a spherically-symmetric potential $
\Phi_0 (r) $.

{\it  The action variables:}
$$
I_1=\frac{1}{2\pi}\oint p_r dr=\frac{1}{\pi}\int\limits_{r_{\rm min}}^{r_{\rm
max}} \sqrt{2E-2\Phi_0(r)-L^2/r^2} dr,
 \eqno {(A1)}
$$
$$
I_2=\frac{1}{2\pi}\oint p_{\theta} d\theta =
\frac{1}{\pi}\int\limits_{\theta_0}^{\pi-\theta_0}d\theta\,\sqrt{L^2-L_z^2/\sin^2\theta}=
L-|L_z|,
 \eqno {(A2)}
 $$
 $$
 I_3=\frac{1}{2\pi}\oint p_{\varphi}\,d\varphi=L_z.
 \eqno {(A3)}
 $$
Here $E=\frac{1}{2}\,{v_r^2}+\frac{1}{2}\,{L^2}/{r^2}+\Phi_0(r) $
is the particle energy, $ L=r\sqrt{v_{\theta}^2+v_{\varphi}^2}=
\sqrt{p_{\theta}^2+{p_{\varphi}^2}/{\sin^2 \theta}\phantom{\big|}}
$ is  its angular momentum magnitude, and $
L_z=r\,\sin\theta\,v_{\varphi} $ is a projection of the angular
momentum on the axis $z$. The angle $\theta_0$ is defined as $
\sin^2\theta_0={L_z^2}/{L^2}, $ and the generalized impulses are
defined as follows $ p_r=\dot{r}=v_r, \ \
p_{\theta}=r^2\,\dot{\theta}=r\,v_{\theta}, \ \
p_{\varphi}=r^2\sin^2\theta\,\dot{\varphi}=r\sin\theta\,v_{\varphi}.
$ Note that it follows from (A2) and (A3) that $ L=I_2+|I_3 |,\ \
L_z=I_3. $
\medskip

    {\it The angular variables}.

    By definition angular variables $w_1,\ w_2,\ w_3$ are
    $ w_i={\p S}/{\p I_i}.
 $
The function of action $S$ in a spherically-symmetric potential is known to allow
for separating the variables and can be written as a sum
 $S (\bI;r,\theta,\varphi)=S_1+S_2+S_3,
$ where the components $S_1 $, $S_2 $ and $S_3 $ are
$$
S_1=\int\limits_{r_{\rm min}}^r dr'\sqrt{2E\,(\bI)-2 \Phi_0(r') -
{(I_2+|I_3|)^2}/{r'^{\,2}\phantom{\big|}}},
$$
\vspace{-15pt}
$$
S_2=\int_{\theta_0}^{\theta} p_{\theta}\,d\,\theta'=
\int\limits_{\theta_0}^{\theta}\sqrt {(I_2+| I_3
|)^2-{I_3^2}/{\sin^2\theta'}\phantom{\big|}}\, d\theta',$$ \vspace{-15pt}
$$
S_3 = \int\limits_0^{\varphi} I_3 \, d \varphi' =I_3 \, \varphi.
 $$

For convenience, we accepted a symbolical, ``vector'', designation $ \bI =
(I_1,I_2,I_3) $. We have
 $$
 w_1={\p S}/{\p I_1}={\p S_1}/{\p I_1}=\Omega_1\int\limits_{r_{\rm min}}^r
 \frac {dr'}{\sqrt{2E-2\Phi_0(r ')-{L^2}/{r'^{\,2}}\phantom {\big|}}},
 $$
 $$
 w_2={\p S_1}/{\p I_2}+\int \limits_{\theta_0}^{\theta} d\,\theta'
\frac{I_2+|I_3|}{\sqrt{(I_2+|I_3|)^2-{I_3^2}/{\sin^2 \theta'}}} $$
$$=
 {\p S_1}/{\p I_2}+\arccos\left(\frac{\cos\theta}{\cos\theta_0}\right),
 $$
 $$
 w_3={\p S_1}/{\p I_3}+{\p S_2}/{\p I_3}+\varphi.
 $$
Action variables are integrals of motion, and angular variables linearly depend
on time: $w_i(t)=w_i(0)+\Omega_i(\bI)\,t$,
 where frequencies
$\Omega_j(\bI)$ are $\Omega_j={\p E (I_1,I_2,I_3)}/{\p I_j}$.

  \begin {center}
{\bf 2. The solution of the kinetic equation, calculation of perturbed
density and derivation of the integral equation}
\end {center}
\medskip

The perturbation  of the DF  $f_1 $ is easily obtained from the
kinetic equation if
we write it down in terms of action-angle variables:
 $$
 \frac{df_1}{dt}\equiv\frac{\p f_1}{\p t}+\Omega_i\,\frac{\p f_1}{\p w_i}
 =\frac{\p F}{\p I_i}\,\frac{\p \Phi}{\p w_i}.
 $$
 We have
$$
f_1=-\frac{1}{(2\pi)^3}\sum\limits_{l_1,\,l_2,\,l_3}\Phi_{l_1,\,
l_2,\,l_3}(\bI)\,\frac{l_j\,{\p F}/{\p I_j}}{\omega-l_j\,\Omega_j}\,e^{i\,(l_j
\,w_j-\omega\,t)},
 \eqno {(A4)}
$$
This  involves summation over a repeating index $j=1,\,2,\,3$.
In what follows the
background DF  $F$ is supposed  to be dependent on $E$ and $L$ only. The
function
$\Phi_{l_1,\,l_2,\,l_3}(\bI)$  appearing in (A4) is
$$
\Phi_{l_1,\,l_2,\,l_3}(\bI) = \int \limits_0^{2 \pi} dw_1\int\limits_0^{2
\pi}dw_2\int\limits_0^{2 \pi} dw_3\,\,\Phi(\bI,\bw)\,\exp\,(i\,\bl\,\bw),
 \eqno {(A5)}
$$
where  another  symbolical ``vector''  designation are introduced for brevity: $
\bw= (w_1,w_2,w_3)$ and $\bl =(l_1,l_2,l_3)$. In Eq. (A5) the function
$\Phi(\bI,\bw)$ represents the perturbation of potential (without the factor
$e^{-i\omega t}$) expressed in variables $(\bI,\bw) $. We shall choose this
function in the form
$$
\Phi(r,\theta,\varphi)=\chi(r)\,P_{\,l}(\cos\theta),
 \eqno {(A6)}
$$
where $P_l(x)$ is Legendre  polynomial. In the main
text we have already given arguments why we
may confine ourselves to the case $m=0$, without
considering perturbations with
a more  general angular structure of the type $Y_l^m (\theta,\varphi) =P_l^m(\cos
\theta)\,e^{im\varphi}$, where $P_l^m(x)$ is the associated Legendre function.
This results in certain simplifications, in particular, we may only deal with
double, not triple, summation.
We have $\Phi_{l_1,\,l_2,\,l_3=0}\equiv
2\pi\,\Phi_{l_1,\,l_2},
 $
where, according to (A5),
  $$\Phi_{l_1,\,l_2} =
\int\limits_0^{2 \pi} dw_1 \int\limits_0^{2\pi}
dw_2\,\chi\,\bigl[r\,(I_1,I_2+|I_3|),w_1\bigr]$$
$$\times\,P_{\,l}\biggl[\cos\theta_0\,\cos \Bigl(w_2-{\p S_1}/{\p
I_2}\Bigr)\biggr]\,e^{-i\,(l_1\,w_1+l_2\,w_2)},
 \eqno {(A7)}
$$
and $\Phi(\bI,w_1,w_2)=(2
\pi)^{-2}\sum\Phi_{\,l_1,\,l_2}(\bI)\,e^{i\,(\,l_1\,w_1+l_2\,w_2)}.
$
 The expression for perturbed DF also retains a mere double
summation:
$$
f_1\!=\!-\frac{1}{(2\pi)^2}\!\!\sum\limits_{l_1,\,l_2}\!\Phi_{l_1,\,l_2}(\bI)\,\frac{l_1\,{\p
F}/{\p I_1}\!+\!l_2\,{\p F}/{\p
I_2}}{\omega-l_1\,\Omega_1-l_2\,\Omega_2}\,e^{i\,(l_1 \,w_1+l_2\,w_2)}.
$$

Expression for $\Phi_{l_1,\,l_2}(\bI) $ can be transformed into a
more compact form. For this purpose we shall first transform
$P_{\,l}\bigl[\cos \theta_0\,\cos\bigl(w_2-{\p S_1}/{\p
I_2}\bigr)\bigr]$, using the summation theorem  for Legendre
polynomials:
\begin{align*}
P_{\,l}(\cos\theta_1\cos\theta_2-\sin\theta_1\sin\theta_2\cos\varphi)\\
=\sum \limits_{k =-l}^{l} e^{-i\,k\,\varphi} P_{\,l}^{\,k}(\cos
\theta_1)\,P_{\,l}^{-k} (\cos \theta_2).
\end{align*}
 Substituting in this
 formula $\theta_2=\frac{1}{2}\,\pi,\ \ \ \theta_1=\frac{1}{2}\,\pi-\theta_0, \
\ \varphi=w_2-{\p S_1}/{\p I_2}+\pi,
$ let us write down
 $$
P_{\,l}\biggl[\cos\theta_0\,\cos\Bigl(w_2-{\p S_1}/{\p
I_2}\Bigr)\biggr]=$$
$$=\sum_{k =-l}^l P_{\, l}^{\, k} (\sin\theta_0)\,P_{\,
l}^{-k}(0)e^{-ik\,\left(\,w_2-{\p S_1}/{\p I_2}\right)} e^{-ik \pi}
 \eqno {(A8)}
$$
Substituting (A8) into (A7) and  integrating over $w_2 $, we obtain
$$
\Phi_{l_1,\,l_2}(E,L)=2\pi P_{\,l}^{\,l_2}(0)P_{\,l}^{-l_2}(\sin \theta_0)
 e^{i\,l_2 \pi} \chi_{l_1,\,l_2}(E, L),
 \eqno {(A9)}
 $$
  where
$$
\chi_{l_1,\,l_2}(E, L)=\int\limits_0^{2\pi} e^{-(l_1 w_1+l_2\p S_1/\p I_2)}\chi
\,\bigl[r(E,L,w_1)\bigr]\,dw_1.
 \eqno {(A10)}
$$

For perturbed density $\rho(r,\theta,t)= \rho(r,\theta)
\,e^{-i\omega t}$ we have
$$\rho(r,\theta)=\int f_1\,d\bv=$$
$$
=-\frac{1}{2\pi}\sum\limits_{l_1,\,l_2}\int d\bv\,
P_{\, l}^{\, l_2}(0)P_{\,l}^{-l_2}(\sin \theta_0)\,e^{i\,l_2\pi}\chi_{l_1,\,l_2}(E, L)\,
$$
$$\times
\dfrac{l_1\,{\p F}/{\p I_1}+l_2\,{\p F}/{\p I_2}}{\omega-l_1\Omega_1-l_2\Omega_2}\,
e^{i\,(\,l_1\,w_1+l_2\,w_2)}.
\eqno {(A11)}
$$
To close the system, we shall use the integral version of the
Poisson equation (which appears more convenient for our purposes
than the Poisson equation itself)
$$
\Phi
(r,\theta)\!=
-\,G \int \!\!\frac
{\rho\,(r',\theta')\,dV'}{\sqrt{r^2+r'^{\,2}-2\,rr'\cos\Theta{\phantom {\big
|}}}},
 \eqno {(A12)}
$$
where $dV'$ is a volume element and $\Theta $ is the angle between
vectors $\br$ and $\br'$: $\cos\Theta=\cos
\theta\cos\theta'+\sin\theta\sin\theta'\cos(\varphi-\varphi')$.
Using (A12) it is simple to obtain the integral connection between
radial parts of perturbed density ${\hat\rho}(r)$ and potential
$\chi(r)$. However, the simplest way of obtaining this connection
is to directly solve, relative to $\chi(r)$, the known ordinary
differential equation which follows from the Poisson equation
after separating the angular dependence :
$$
\frac{1}{r^2}\frac{d}{dr}\,\Bigl[r^2\,\frac{d\,\chi(r)}{dr}\Bigr]-\frac{l\,(l+1)}
{r^2}\,\chi(r)=4\pi G\,{\hat\rho}(r).
  \eqno {(A13)}
$$
Solving Eq. (A13) in terms of Green's function, we shall find the required
relation:
$$
\chi(r)=-\frac{4\pi G}{2l+1}\int r'^{\,2}dr'{\hat\rho}(r')\,{\cal F}_l(r, r ').
 \eqno {(A14)}
$$
where
$$
{\cal F}_l(r,r')=\frac{(r_<)^l}{(r_{>})^{l+1}},\ \  r_<={\rm min}(r,r'), \ \
r_>={\rm max}(r,r').
$$ For obtaining the integral equation in the desired
form it is necessary to write down  (A14) in action-angle
variables and to split it into harmonics $(l_1,l_2)$. For this
purpose it is necessary to select a radial component ${\hat
\rho}(r)$ from the general expression for density (A11) and then
to substitute it into the r.h.s. of  (A14). Further it is
necessary to use  relation (A10), connecting
$\chi_{\,l_1,\,l_2}(E, L)$ to $\chi(r)$, multiplying its both
parts by  $\exp{\bigl[-i\,(l_1\,w_1+l_2\p S_1/\p I_2 \bigr]}$ and
integrating over $w_1 $. Let us execute the above described
procedure. We have for ${\hat\rho}(r)$:
$$
{\hat
\rho}(r')=(l+{\case{1}{2}})\int\limits_0^{\pi}\rho(r',\theta')\,P_{\,l}(\cos
\theta')\,\sin\theta' d\theta'.
 \eqno {(A15)}
$$
Using an explicit form of expression for density (A11) and
substituting the function ${\hat\rho}(r')$ found with the help of
(A15) into r.h.s. of (A14), we obtain:
$$
\chi(r)=G\sum\limits_{l_1',\,l_2'}\int d\bv'\int r'^{\,2}dr'\int\limits_0^{\pi}
\sin \theta' d\theta'P_{\,l}(\cos\theta')\,{\cal F}_l(r,r')$$
$$\times \,P_{\,l}^{l_2'}(0)
 P_{\,l}^{-l_2'}(\sin\theta_0')\,e^{i\pi l_2'}\chi_{\,l_1'\,l_2'}
(E',L')\times
$$
$$ \times \frac {l_1'\,{\p F}/{\p I_1'}+l_2'\,{\p F}/{\p I_2 '}}
{\omega-l_1'\,\Omega_1(E',L')-l_2'\,\Omega_2(E',L')}\,e^{\,i(\,l_1' w_1'+l_2'
w_2')}.
 \eqno {(A16)}
$$
It is also possible to integrate explicitly over $w_2'$ in (A16). For this
purpose let us note that in r.h.s. of (A18) there is an integration over phase
volume $\int d\Gamma'=\int d\bv' dV'$, which, obviously, may be  represented as
${\displaystyle\int} d\Gamma'=
2\pi {\displaystyle\int} d\bI'dw_1'\,dw_2'$. Writing down again
 $P_{\,l}(\cos\theta')$,  similarly to (A8), as a series, we find
\begin{align*}
 \int\limits_0^{2\pi}dw_2' e^{\,il_2' w_2'}P_{\,l}(\cos \theta') \\ =2
\pi\,P_{\,l}^{\,l_2'}(\sin\theta')P_{\,l}^{-l_2'}(0)\,e^{-il_2'\pi}\,e^{\,i\,l_2\,\p
S_1/\p I_2'}.
\end{align*}
 As a result for $ \chi (r) $ we obtain an expression which already
retains only a single integration over angular variables, namely, over $w_1'$:
\begin{align*}
 \chi(r)=2\pi G\sum\limits_{l_1',\,l_2'}\int
dI_1'\,dI_2'\,dI_3'\Bigl[P_{\,l}^{l_2'}(0)\,P_{\,l}^{-l_2'}(\sin\theta_0')\Bigr]\,\\
\times \left[P_{\,
l}^{-l_2'}(0)\,P_{\,l}^{l_2'}(\sin\theta_0')\right]\chi_{\,l_1'\,l_2'}(E',L')\,
\\
 \times
 \frac{l_1'\,{\p F}/{\p I_1'}+l_2'\,{\p F}/{\p I_2'}}
 {\omega-l_1'\,\Omega_1(E',L')-l_2'\,\Omega_2(E',L')}\,\\
 \times\int\limits_0^{2 \pi}
dw_1'\,{\cal F}_l\left[r,r'(E',L';w_1')\right]\,e^{\,i(\,l_1'\,w_1'+l_2'\,\p S_1
/\p I_2')}.
 \end{align*}
Remembering that $F(\bI)=F(E, L)$ does not depend on  $L_z=I_3 $
we can perform explicitly another integration, namely, over
$\theta_0'$. Taking it into account we find
\begin{align*}
\int\limits_{-\pi/2}^{\pi/2}P_{\,l}^{\,l_2'}(\sin\theta_0')P_{\,l}^{-l_2'}(\sin
\theta_0')\,d\,(\sin\theta_0')\\=\int_{-1}^{1}
dz\,\bigl[P_{\,l}^{\,l_2'}(z)\bigr]^2 \frac{(l-l_2)!}{(l+l_2)!}= \frac {2}{2l+1}.
 \end{align*}
 Using a known relation for
$P_{\,l}^{k}(0)$ we obtain for $\chi(r)$:
\begin{align*} \chi(r)=\frac{4 \pi G}{2l+1}
\sum\limits_{l_1'=-\infty}^{\infty}\sum\limits_{l_2' =-l}^l D_l^{l_2
'}\int\frac{dE'\,L\,dL'}{\Omega_1(E',L')}\chi_{\,l_1'\,l_2'}(E',L')\\
 \times\,\frac{\Bigl[\,l_1'\,\Omega_1(E',L')+l_2'\,\Omega_2(E',L')\Bigr]\,{\p
F}/ {\p E'}+l_2'\,{\p F}/{\p
L'}}{\omega-l_1'\,\Omega_1(E',L')-l_2'\,\Omega_2(E',L')}\\
\times \int
\limits_0^{2 \pi} dw_1'
 \, {\cal F}_l\left[r,r'(E',L'; w_1')\right]\,
 e^{\,i\,(\,l_1'\,w_1'+l_2'\,\p S_1/\p I_2')}.
\end{align*}
The explicit expression for $D_l^k$ is presented in the main text (see
(\ref{eq:4.11})). Finally, we obtain the desired set of integral equations for
$\chi_{\,l_1,\,l_2}$ presented in the main text:
$$
\chi_{\,l_1,\,l_2}(E,L)=\frac{4\pi G}{2l+1}\sum\limits_{l_1'=-\infty}^{\infty}
\sum\limits_{l_2'=-l}^l D_l^{l_2'}\int\frac{dE'\,LdL'}{\Omega_1(E',L')}$$
$$\times\,\chi_{\, l_1'\,l_2'}(E',L')\,\Pi_{l_1,\,l_2;\,l_1',\,l_2'}(E,L;E',L')\times
 $$
 $$
\times\,\frac{\Bigl[\,l_1'\,\Omega_1(E',L')+l_2' \,\Omega_2(E',L')\Bigr]\,{\p F}
/{\p E'}+l_2'\,{\p F}/{\p L
'}}{\omega-l_1'\,\Omega_1(E',L')-l_2'\,\Omega_2(E',L')},
$$
where the kernel is
\begin{align*}
\Pi_{l_1,\,l_2;\,l_1',\,l_2'}(E,L;E',L')=\\=
\int\limits_0^{2\pi}dw_1\int\limits_0^{2\pi} dw_1'
 \,{\cal F}_l\left[\,r\,(E,L; w_1),r'(E',L'; w_1')\right]\times\\
 \times
  \exp{\left\{\,i\,\left[\Bigl(\,l_1'\,w_1'+l_2'\,{\p S_1}/{\p I_2'}
\Bigr)-\Bigl(l_1\,w_1+l_2\,{\p S_1}/{\p I_2}\Bigr)\right]\right\}}.
\end{align*}
\bigskip

\subsection*{\large APPENDIX B. Instability criterion for the characteristic equation
(\ref{eq:60}) -- counterpart of Penrose\,-\,Nyquist plasma criterion}

Let us write down the equation (\ref{eq:60}) in the form
$\varepsilon^{(l)}(z,Q)=0$, where
 $\varepsilon^{(l)}(z,Q)\equiv 1-{Q_c(z)}/{Q}$ and
complex function $Q_c(z)$ is defined as follows:
$$
Q_c(z)=\int\limits_0^{\infty}dx\,x\,\frac{d\,{F}(x)/dx}{x-z}.
 \eqno {(B1)}
$$
(Note that $Q_c(z)$ does not depend on $Q$.) In what follows  the
upper  index $l$ is omitted for brevity. It is easy to understand
that in the points of a real axis $z$, in which $z=z_j\equiv x_j$,
where $x_j$ is any of the extremum points of the function $F(x)$,
the function $Q_c(z)$ is real and coincides with the corresponding
value $Q_c$ of the ``neutral mode". Quotation marks  here are to
reflect the fact that the corresponding value $z_j$ is not
necessarily a squared frequency of a true neutral mode, since the
sign of $Q_c(z_j)$ can be any. Further we shall denote the real
numbers $Q_c(z_j)$ as $Q_c^j$ for brevity.

According to (B1) the function $Q_c(z)=Q_c(\omega^2)$, considered
as a function of $\omega$, is an analytic function in the
$\omega$-plane cut along the real axis ${\rm Im}\,(\omega)=0$.
From its definition (B1) it also follows that it is continuous and
bounded (it tends to zero when $|\omega|\to\infty$). We are
interested in unstable roots of the equations
$\varepsilon(\omega^2,Q)=0$, that is the roots lying in the upper
half plane $\omega$. The number of such roots, according to the
argument principle, coincides with the number of poles of the
function $\varepsilon^{-1}$ in the upper half plane and is equal
to $N=(2\pi i)^{-1}\int \limits_{C_{\omega}} d\omega\,\veps^{-1}\,
{d\varepsilon/d\omega} \,$ where the contour $C_{\omega}$ is shown
on 12\,({\it a}). Turning to a variable $z=\omega^2$, we find
$$
N=\frac{1}{2\pi i}\int\limits_{C_z}
\frac{d\varepsilon/dz}{\veps}\,dz,
$$
where the directed contour  $C_z$ is the image of a contour $C_{\omega}$ on a
complex plane $z$. It is shown in Fig. 12\,({\it b}).

\begin{figure}
 \begin{center}
 \includegraphics[bb=40 80 564 795, width=78mm, clip=true, draft=false]{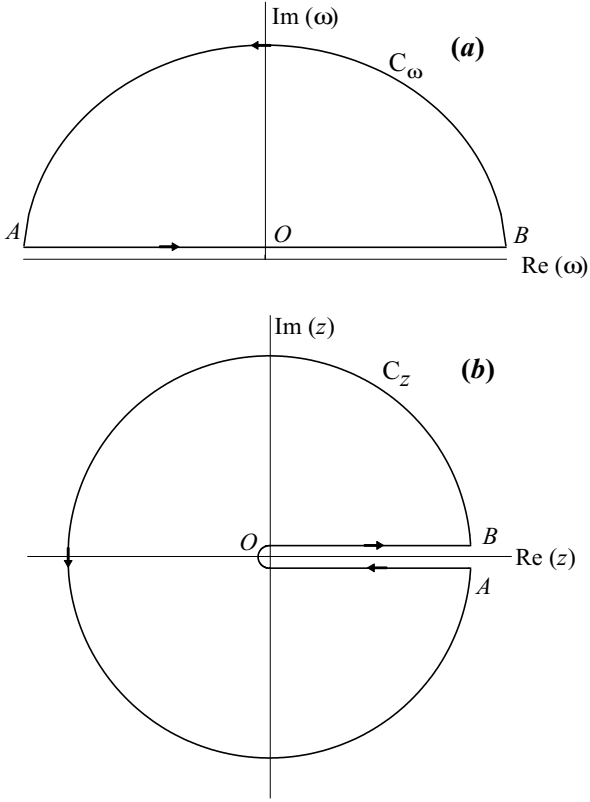}
 \end{center}
\caption{\small Directed contours of integration $C_{\omega}$ and
$C_z$ on complex planes $\omega$ and $z$ respectively.}
  \label{fig12}
 \end{figure}

Following \cite{Pen}, we shall pass from the complex plane $z$ to
the complex plane $Q_c$. For the number of unstable roots $N$ we
obtain  the following expression
$$
N=\frac{1}{2\pi i}\int\limits_{C_z}\frac{d\varepsilon/dz}{\veps}\,dz =
\frac{1}{2\pi i}\int\limits_{C_{\veps}}\frac{d\veps}{\veps}=\frac{1}{2\pi i}\int
\limits_{C_{Q_c}}\frac{dQ_c}{Q_c-Q},
 \eqno {(B2)}
$$
where the directed contour $C_{Q_c}$  is the image of the
directed contour $C_z$ on the
complex plane $Q_c$.

Thus the problem is reduced to constructing a contour $C_{Q_c}$.
The number of times that this contour encloses in anticlockwise
sense the point $Q_c=Q$ (lying on the real positive half axis of
the complex plane $Q_c$) will give us the number of unstable
roots. We need to formulate rules according to which we should
image a contour $C_z$ on the plane $Q_c$ and to establish a
direction of motion along it in its various parts.

\medskip
1. First of all, note that since the entire circle with a big
radius on the plane $z$ is imaged into a unique point $Q_c=0$, the
entire remaining contour $C_{Q_c}$ on the plane $Q_c$ corresponds
to a horizontal part of the contour $C_z$, starting at a point
$A$, and finishing at a point $B$.
\smallskip

2. It is easy to understand that the contour $C_{Q_c}$ is symmetric relative
to the horizontal axis ${\rm Im}\,(Q_c)=0$.
\smallskip

3. Further, we need to understand how the contour $C_{Q_c}$
crosses the horizontal axis at a point $Q_c=0$. Note that the
contour $C_{Q_c}$ should cross the horizontal axis in this point
(at least\footnote {We speak  ``at least'' since it is possible in
principle that for some $z_j$ corresponding to the extrema of the
function $F(x)$, it may be true that $Q_c^j=0$.}) twice. ({\it i})
The first crossing corresponds to the start of the contour from
$Q_c=0$ and its finish at $Q_c=0$ when a point $z$ moves from $A$
to $B$ (that is the crossing corresponds to coincidence  of the
``initial'' and ``final'' points of the closed contour. ({\it ii})
The second crossing corresponds to passage of a contour $C_z$
through a point $O$ (that is $z=0$). It follows directly from
Equation (B1), that $Q_c(0)=0$. Let us examine the manner in which
the contour $C_{Q_c}$ crosses the horizontal axis at $Q_c=0$ in
both cases. The direction -- up- or downwards -- of the crossing
of the real axis at this point is important here.
 \smallskip

\noindent ({\it i}) We  have from (B2) as $|z|\to\infty$
$$
Q_c(z)=z\int\limits_0^{\infty}\frac{F'(x)}{x-z}\,dx$$
$$=
z\int\limits_0^{\infty}\frac{F(x)}{(x-z)^2}\,dx\approx\frac{1}{z}\int\limits_0^{\infty}
F(x)\,dx=\frac{\cal M}{z}, \ \ \ {\cal M}>0.
\eqno {(B3)}
$$
By means of (B3) it is now simple to understand that at a point
$A$ (that is at $z=R-i \epsilon$, $R\to\infty, \ \ \epsilon\to 0$)
${\rm Im}\,(Q_c)\to +0$, and at a point $B$ (that is at $z=R+i
\epsilon$) ${\rm Im}\,(Q_c)\to -0$. It means that this type of
crossing is {\it upwards}. From (B3) it is also visible that
crossing occurs so that $ {\rm Re}\,(Q_c)\to +0$, that is the
contour approaches a point $Q_c=0$ {\it from the right}.
\smallskip

\noindent ({\it ii}) It is more difficult to establish how
crossing occurs that corresponds to passage of a point $z=0$ on
the contour $C_z $. Omitting details, we declare simply that
crossing of the horizontal axis at a point $Q_c=0$ in this case
also occurs {\it upwards} - however, the contour approaches  this
point so that ${\rm Re}\,(Q_c)\to -0$, that is {\it from the
left}.
 \smallskip

4. Now we shall discuss the way the contour crosses the horizontal
axis at points $Q_c^j$, corresponding to the extrema of $F(x)$.
Clearly, each such point is crossed twice: the first crossing
occurs when a point $z$ moves along the positive half axis from
infinity to the center, and the second during subsequent movement
in the opposite direction. We shall show that both crossings occur
in the same direction, namely: {\it downwards} at points $Q_c^j$,
corresponding to the maxima of $F(x)$, and {\it upwards} at points
$Q_c^j$, corresponding  to the minima. For points $z$ on the real
axis $z$ we have
$$
Q_c(z)=z\,\Bigl[\,\int\limits_0^{\infty}\hspace{-10.1pt}-\
dx\,\dfrac{F'(x)}{x-z}+ i \pi F'(z)\,{\rm sign}\,(\omega)\Bigr],
 \eqno {(B4)}
$$
so
 $$
 {\rm Im}\,[Q_c(z)]=\pi z\,F'(z)\,{\rm sign}\,(\omega).
 \eqno {(B5)}
 $$
The origin of the factor ${\rm sign}\,(\omega)$ in (B4) and (B5)
may be  understood from Fig. 12\,({\it b}) where it is seen that
for points $z$, lying a little below the positive half axis $z$,
i.e., at ${\rm Im}\,(z)=-\epsilon$, the pole in the integral over
$x$ in the right part (B2) is indented upwards, and for the points
lying a little above the real axis, i.e., at ${\rm
Im}\,(z)=\epsilon$, it is indented downwards.

Let the point  $z$ pass through the corresponding extremum point
$z_j$  during its first passage, that is from $A$ to $O$ along the
bottom side of the positive real half axis. Then $ {\rm
sign}\,(\omega)=-1$, and the increment  $\Delta z$  is negative,
$\Delta z<0$. From (B5) we have
  \begin{align*}
\Delta {\rm Im}\,[Q_c (z)]=\left\{\frac{d}{dz}\,{\rm Im}\,[Q_c
(z)]\right\}_{z=z_j}\,\Delta z\\
=-\pi\,z_j\,F''(z_j)\,\Delta
z=\pi\,z_j\,F''(z_j)\, |\Delta z|.
  \end{align*}

In its second crossing of this point, that is when the point $z$
moves from $O$ to $B$ along the top side of the positive real half
axis, we have ${\rm sign}\,(\omega)=+1$ and $ \Delta z> 0 $. So we
again have
$$
\Delta\,{\rm Im}\,[Q_c(z)]=\pi\,z_j\,F''(z_j)\,\Delta z.
$$

Thus, indeed, in both crossings the points $Q_c^j$, corresponding
to extrema, pass in the same manner, namely, the minima
($F''(z_j)>0 $) upwards, and the maxima ($F''(z_j)<0$) downwards.

 The above recipes are sufficient to construct
a directed contour $C_{Q_c}$. Having constructed it and found how
many times it winds around the point $Q_c=Q$ lying on the
horizontal axis {\it to the right} from the origin,  we, according
to  Penrose's idea, can draw a conclusion on the
instability/stability of the system under consideration with a
given value of the parameter $Q$. Indeed, from (B2) it is easy to
understand that if the directed contour $C_{Q_c}$ crosses the
horizontal axis from below to the right of the point $Q_c=Q>0$ at
least once, then instability takes place ($N \ge 1 $), since in
this case the point $Q_c=Q$ is enclosed by the contour in an
anticlockwise sense. As the parameter $Q$ can have an arbitrary
positive value, we come to the conclusion that for instability it
is necessary and sufficient that the contour crosses the
horizontal positive half axis {\it upwards} at least once. It
means that the rightmost crossing of the horizontal axis
corresponds {\it to a minimum} of $F(x)$. This leads us to to
formulating the instability criterion cited in subsection 4.8 of
the main text.

Note that the task of constructing a contour can be simplified if
we recall that it involves all unstable roots $z=z_0$ as complex
conjugate pairs (this corresponds to pairs of eigenfrequencies
$\omega_0=\pm\alpha+i\,\beta$, $\beta>0$). Therefore the number of
times the contour $C_{Q_c}$ winds around a point $Q_c=Q$ in the
complex plane $Q_c $ must be {\it even}, $N=2L$. It is simple to
understand that $L$ times occur  when a point $z$ moves in the
complex plane $z$ from infinity to zero (along the bottom side of
the real axis), and $L$ more times occur when it moves in the
opposite direction (from $z=0$ to infinity along  the top side).
Recall that the full contour is symmetric relative to the
horizontal axis, touches itself at point $Q_c=0$, and the
horizontal axis is crossed twice at each point $z_j$ in the same
direction - either both times downwards (maximum), or both times
upwards (minimum). Therefore it is enough to trace the movement of
point $z$ only halfway, say, from $z=0$ to infinity and to image
only the half of the full contour which  is also a closed contour.
For definiteness we shall plot only that  half which corresponds
to movement of point $z $ from the center to infinity (along the
top side of the real half axis). We shall designate this closed
directed ``semi-contour'' as $ {\hat C}_{Q_c} $  and, like the
full contour, it will start and finish at point $Q_c=0 $. Its
beginning will be in the second quarter, and its end in the fourth
quarter of the complex plane $Q_c $.
\smallskip

{\small {\it Note.} The above does not mean that instability
exists for any $0 <Q <Q_c^{({\rm min})}$, where $Q_c^{({\rm
min})}$ is a point of  right-most crossing of a horizontal axis,
related to a minimum $F(x)$. If a positive $Q_c^{({\rm max})}$
exists also for some maximum, instability will take place only
when $ Q_c^{({\rm max})} <Q <Q_c^{({\rm min})}. $ Indeed, from
Fig.\,13 it is seen that in this case the point $Q_c=Q $ will only
be enclosed if $Q$ lies in the specified range.

\begin{figure}
\begin{center}
 \includegraphics[bb=98 80 532 692, width=60mm, clip=true, draft=false]{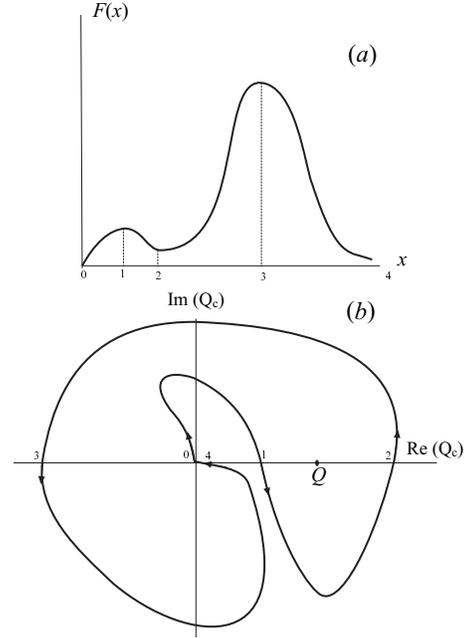}
\end{center}
 \caption{\small Schematic sketch of DF $F(x)$ with two
maxima and one minimum and corresponding arrangement of points
$Q_c^j $ ($j=1, 2, 3 $) on axis ${\rm Im}\,(Q_c)=0$. Numbers 0, 1,
2, 3, 4 correspond to the order in which the directed semi-contour
${\hat C}_{Q_c}$ passes over corresponding points. The situation
when there are two  positive values $Q_c^j$ with $j=1$ and $j=2$,
corresponding to low maximum, and to minimum,  is demonstrated.
The contour encloses a point $Q_c=Q$ in anticlockwise sense once
provided that the point $Q$ is within the range $Q_c^{(1)} <Q
<Q_c^{(2)}$. These values just are boundaries of the range of
instability on parameter $Q$.} \label{fig13}
\end{figure}

 \bsp
 \label{lastpage}
\end{document}